\documentclass[article,aps]{revtex4-1}
\usepackage{graphicx}
\usepackage{color}
\usepackage{multirow}
\begin{document}
\title{ Quasiparticle Renormalization and Pairing Correlations in Spherical Superfluid Nuclei}
\author{A.Idini$^{a,b}$, F.Barranco$^{c}$ and E.Vigezzi$^{b}$}%
\affiliation{
$^a$ Dipartimento di Fisica, Universit\`a degli Studi di Milano, Via Celoria 16, 20133 Milano, Italy.\\
$^b$ INFN, Sezione di Milano, Via Celoria 16, 20133 Milano, Italy.\\
$^c$ Departamento de Fisica Aplicada III, Escuela Superior de Ingenieros, Universidad de Sevilla, 
Camino de los Descubrimientos s/n,  
  41092 Sevilla, Spain.\\
}
\date{\today}

\begin{abstract}
We present a detailed discussion of the solution of Nambu-Gor'kov equations in superfluid nuclei,
which provide a consistent framework to deal with  the interplay  between particle-hole and particle-particle channel,
including the effects of the fragmentation of the quasiparticle strength and of the pairing interaction
induced by the exchange of collective vibrations.
The coupling between quasiparticle and vibrations is determined from the experimental polarizability
of the low-lying collective surface vibrations. This coupling is used to renormalize the properties of 
quasiparticles obtained  from a BCS calculation using the bare nucleon-nucleon interaction.
We apply the formalism to the case of the  nucleus $^{120}$Sn, showing results for the 
low-energy spectrum and the quasiparticle strength distribution in neighbouring odd nuclei 
and for the neutron pairing gap. 

\end{abstract}
\maketitle



















      










\section{\label{introduction} Introduction}

The key role played by pairing correlations in atomic nuclei was recognized soon after 
the development of BCS theory. The number of Cooper pairs involved is small, and one can 
study nuclear superfluidity within BCS in terms of specific orbitals lying close to the Fermi energy, 
interacting through an attractive nucleon-nucleon force.
However, the properties of the particles are considerably renormalized by the coupling to 
the collective vibration of the system, which leads  to an enhancement  of 
the effective mass and of the level density  close to the Fermi energy, 
and to   a fragmentation of the particle strength observed in one-nucleon transfer reactions. 
While these phenomena have been extensively studied in normal nuclei \cite{mah:85,soloviev}, 
their consequences  on pairing correlations have been much less investigated.
Obtaining  an accurate description of these effects requires to go beyond the BCS framework.
A convenient formalism that allows one to  consider the interplay 
between the particle-hole and the particle-particle channel is offered by the Nambu-Gor'kov equations,
used in condensed matter to deal with strong coupling superconductivity \cite{gorkov,nambu,eliashberg}. These
equations imply the calculation of energy-dependent normal and abnormal self-energies,
take into account  the pairing interaction induced by the exchange of  collective vibrations  
between  members of the Cooper pairs and lead to  theoretical 
predictions concerning the low-energy part of the nuclear spectrum;
they also produce the structure elements needed for a consistent  calculation of one- and two-nucleon
transfer reactions \cite{gregory}.

In this paper we extend our previous investigations concerning  the microscopic origin of pairing 
(cf. in particular \cite{prl_noi}-\cite{pastore:08}).
In Section \ref{formalism} we   provide a detailed account of the formalism, comparing 
two different schemes for the dealing with the Nambu-Gor'kov equations, either solving first the
BCS equations with the bare nucleon-nucleon force and then adding renormalization effects (prior scheme,
cf. Section \ref{prior}) , or dealing at the same time
with both sources of pairing (post scheme, cf. Section \ref{post}). The coupling 
between quasiparticle and vibrations is computed according  to the 
basic rules of Nuclear Field Theory (NFT) \cite{nft1,nft2,BM:75}. Use is made of the  
collective model of Bohr and Mottelson to calculate the properties of vibrational states in the
Quasiparticle Random Phase Approximation (QRPA) with a separable force, with a coupling constant 
chosen so as to reproduce the experimental properties of low-lying collective surface modes. 
In this way we avoid the use of free parameters in the calculation, which then only depends on the choice
of the interaction adopted to produce the mean field. In most of the computations we shall use 
the SLy4 effective interaction, which provides a good reproduction of the overall mean field properties, 
and leads to a reasonable level density around the Fermi energy, after including renormalization effects.
We shall derive gap equations  which generalize the usual BCS expression, lend themselves to useful
approximations and allow one to make contact with other studies.
In Section \ref{results} we shall solve the Nambu-Gor'kov equations in the case of $^{120}$Sn, comparing 
the theoretical low-lying spectrum in neighbouring odd nuclei with  experimental  
data derived  from one-neutron transfer reactions.  We shall also compare the 
results obtained with a self-consistent iterative solution of the Nambu-Gor'kov equations with the quasiparticle
approximation. It is possible to clearly separate  the contributions to the pairing gap 
associated with the bare nucleon-nucleon force and  with  the renormalization effects. We find that the two
contributions have comparable magnitudes, confirming previous studies. 
We shall also briefly consider the role played by spin modes, which provide a repulsive contribution to pairing 
in the $^{1}S_0$ channel, whose  magnitude is however very uncertain.
Conclusions and a hindsight  will be presented in Section \ref{conclusions}. 
The sensitivity of our results to various elements of the calculations is discussed in the Appendix.


\section{\label{formalism} The formalism}

\begin{figure*}[h!]
\begin{center}
\includegraphics[width=\textwidth]{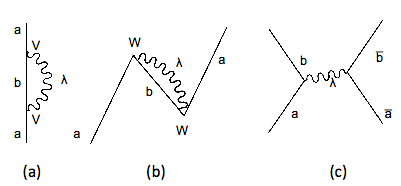}
\end{center}
\caption{ Basic diagrams taken into account in the present study,
which renormalize the normal and abnormal self-energies 
obtained in mean field calculations: polarization (a), correlation (b), induced pairing interaction (c) 
processes.} 
\label{fig:fig1_diagrams}
\end{figure*}

The object of our study are pairing correlations in superfluid nuclei taking into 
account both the contribution of the bare nucleon-nucleon interaction, and the
many-body effects associated with the coupling between particles and vibrations (cf. Fig. \ref{fig:fig1_diagrams}).

Our approach will be based on two basic assumptions:

(A) We shall start from a mean field obtained with a Hartree-Fock (HF) calculation
with an effective force. 
Our quantitative results will then depend to some extent on the choice of the  force adopted
to produce the mean field, because pairing correlations are very sensitive to the position of the single-particle
levels. In principle, one could aim at determining the parameters of a new effective force, by comparing
the results  obtained including renormalization with experimental data. This would require an
extensive investigation, considering several isotopic chains. 
In this work we shall limit ourselves to the study
of renormalization effects in the single nucleus $^{120}$Sn, adopting existing effective forces
which lead to a reasonable description of the available data - although we are aware that 
the quantitative aspects of our findings might be improved by a more careful determination of the mean field. 

\begin{figure*}[h!]
\begin{center}
\includegraphics[width=0.75\textwidth]{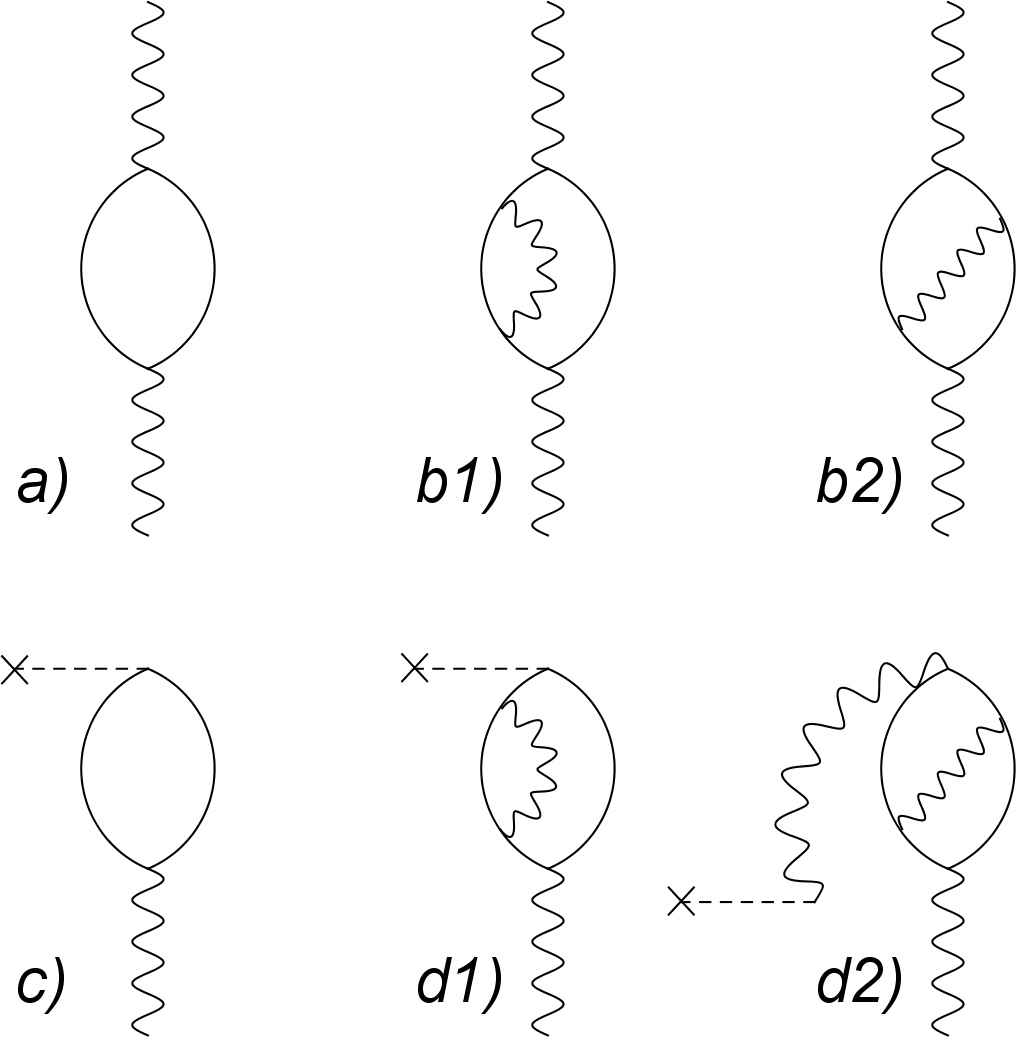}
\end{center}
\caption{ The self-energy (a) of the vibrations calculated in the  
QRPA approximation is renormalized by the basic  diagrams (b1) (leading to a renormalization of the 
particle-hole energies) and (b2) (vertex correction renormalizing the particle-hole interaction).
Similarly, the QRPA transition strength (c) is renormalized by the corresponding diagrams (d1) and (d2).}
\label{fig:fig2_diagrams}
\end{figure*}

(B) We shall not consider explicitly the renormalization processes  affecting the vibrations of the system
 \cite{epj:04,bort:80,broglia:03}. The main diagrams renormalizing the energy
of the phonons are shown in Fig. \ref{fig:fig2_diagrams}(b1)  (self-energy correction
renormalizing the energy of the particle-hole transitions)
and   Fig. \ref{fig:fig2_diagrams}(b2)  (vertex correction renormalizing the particle-hole interaction),
while the diagrams shown in Fig. \ref{fig:fig2_diagrams}(d1) and Fig. \ref{fig:fig2_diagrams}(d2) 
renormalize the transition strength.
The explicit 
inclusion of renormalization effects on phonons, 
on par with those on single-particle levels, 
represents an ambitious program that has been attempted only in a few cases for closed-shell nuclei 
\cite{dickhoff1,dickhoff2,barbieri}. 
We shall instead 
take the view 
that such renormalization effects, if taken properly into account, 
should lead to agreement with experiment. 
In particular, we shall  take the transition densities which are at the basis of the interweaving of single-particle 
(quasiparticle) degrees of freedom and collective modes from the collective macroscopic Bohr-Mottelson model.
We shall determine 
the Particle-Vibration Coupling (PVC) 
strength 
(cf. Fig. \ref{fig:fig2_diagrams}(b2))  from
a QRPA calculation tuned to reproduce the experimental properties of the low-lying phonons, using a mean
field with an effective mass $m^* = m$  that reproduces the experimental single-particle level density
(cf. Fig. \ref{fig:fig2_diagrams}(b1)) (see Section \ref{results} for more details).


Within this frozen-phonon approximation, as a possible alternative,
one might calculate the PVC microscopically making use of the transition density of the RPA phonons
obtained from a self-consistent calculation performed with the same effective force  employed to
obtain the HF mean field. Calculations of this kind have been performed several times in the past
for non-superfluid nuclei, with various degree of approximations \cite{mah:85}; see \cite{colo}
for a recent calculation aiming at a complete self-consistence. The main
drawback in this approach lies in the fact that QRPA leads to a relatively poor reproduction of the
experimental energy and of the transition strength of the low-lying collective vibrational states in semimagic nuclei
(for recent QRPA calculations 
cf. e.g. \cite{ansari:06},\cite{tera:08}), 
which provide the main contribution to renormalization effects. This is consistent with the fact that 
processes beyond QRPA can strongly renomalize the phonon properties. 
We shall instead take the view 
that renormalization effects, if taken properly into account beyond the frozen phonon
approximation, should lead to agreement with experiment. As a consequence, we shall take the PVC from
a QRPA calculation tuned to reproduce the experimental properties of the low-lying phonons, using a mean
field with an effective mass $m^* = m$  that reproduces the experimental level density. 
In other words, we shall assume that vertex corrections 
are effectively included in our PVC.

The actual implementation of assumptions (A) and (B) will be presented below in the next Sections.  
Other, less essential assumptions will be adopted in our calculation in order to reduce the computational
complexity.  First of all, we shall limit the investigation of renormalization
processes  to states lying  close to the Fermi energy. 
We shall then take into account the fact
that the bare interaction - even considering its soft-core or $V_{{\rm low} k}$ versions - couples 
the single-particle states lying close to  the Fermi energy with
states lying up to several hundreds of  MeV, while the phonon-mediated pairing interaction acts between
pair of states separated by a few MeV \cite{noi_schuck,dick:book}.   This suggests the convenience of  
a separate treatment of the two interactions in the pairing problem, by performing a HF+BCS calculation
prior to the calculation of many-body effects.
We shall refer to this approach as the ${\it prior \; scheme}$.
Otherwise, one is forced to consider the bare and the  induced interaction on the same footing
(${\it post \; scheme}$). 
In the following we shall first illustrate in detail the prior scheme, which will be used in most 
calculations presented in this paper.  We shall then outline the main modifications involved in
the post scheme.

\subsection{\label{prior} The prior scheme }

A convenient formalism for the calculation of the properties of 
quasiparticles in superfluid nuclei within the prior scheme  was given by
Van der Sluys et al. \cite{Sluys:93}, although they did not devote particular 
attention to the renormalization of pairing correlations. 
In this approach one first accounts for the action of the 
bare force with a HF+BCS calculation, 
discussed in more detail in Section \ref{baregap} , leading to 
quasiparticle energies $E_a$, quasiparticles amplitudes $u_a$ and $v_a$
and to a pairing gap $\Delta^{BCS}_a$.
One then 
renormalizes the obtained quasiparticles including the coupling to vibrations calculated 
in the QRPA. The derivation of the formalism
was based on the equation of motion method, which  has a very
close relation to the Green's functions 
formalism used to derive the Nambu-Gor'kov equations, commonly used to study superconductivity in
condensed matter physics \cite{Schr:64}. 

In order to calculate the renormalization
of a quasiparticle in spherical nuclei, denoted by its associated quantum numbers $a \equiv \{nlj\}$,  
one has to solve a
system of linear equations  obtained coupling the quasiparticle with more complex 
configurations including phonon states
through the basic vertices $V$ and $W$ shown in Fig. \ref{fig:fig1_diagrams}.
The phonons are characterized by their angular momentum $\lambda$ and by their energy $\hbar \omega_{\lambda,\nu}$.
We shall assume that phonons have natural parity, $\pi = (-1)^{\lambda}$ (see however Section \ref{effspin}). 
For illustration, we write  the diagonalization problem including only two other  quasiparticle
states $b$ and $c$, and  a single phonon ${\lambda, \nu}$:

\begin{eqnarray}
\left( \matrix {
  E_{a} & V(ab\lambda\nu) & V(ac\lambda\nu)  & W(ab\lambda\nu) & W(ac\lambda\nu) & 0 \cr
  V(ab\lambda\nu) & \hbar \omega_{\lambda\nu} + E_{b} & 0  & 0 & 0 &  W(ab\lambda\nu)\cr 
  V(ac\lambda\nu) & 0& \hbar \omega_{\lambda\nu} + E_{c} & 0 & 0  &  W(ac\lambda\nu)\cr 
  W(ab\lambda\nu) & 0 & 0  & -\hbar \omega_{\lambda\nu} - E_{b} & 0 &  -V(ab\lambda\nu)\cr
  W(ac\lambda\nu) & 0 & 0  & 0& -\hbar \omega_{\lambda\nu} - E_{c} & -V(ac\lambda\nu)\cr
  0  & W(ab\lambda\nu) & W(ac\lambda\nu)  & -V(ab\lambda\nu) & -V(ac\lambda\nu) & -E_{a}} \right )
\left ( 
\matrix {
    x_{a(n)}\cr
 C_{a(n),b,\lambda\nu}\cr
 C_{a(n),c,\lambda\nu}\cr
 -D_{a(n),b,\lambda\nu}\cr
 -D_{a(n),c,\lambda\nu}\cr
 -y_{a(n)}} \right ) = \nonumber
\end{eqnarray}

\begin{eqnarray}
\tilde E_{a(n)}
\left ( 
\matrix {
    x_{a(n)}\cr
 C_{a(n),b,\lambda\nu}\cr
 C_{a(n),c,\lambda\nu}\cr
 -D_{a(n),b,\lambda\nu}\cr
 -D_{a(n),c,\lambda\nu}\cr
 -y_{a(n)}} \right )
\label{MasterMatrix}
\end{eqnarray}
Many eigenvalues $\tilde E_{a(n)}$ and eigenstates $a(n)$ are obtained from the diagonalization of the matrix 
(\ref{MasterMatrix}),  
giving rise to a fragmentation of  the associated quasiparticle strength. 
For a given eigenvalue $\tilde E_{a(n)} >0$ there exists a corresponding eigenvalue $\tilde E_{a(-n)}= - \tilde E_{a(n)}$.
As in  standard BCS theory,  we will keep only positive energy solutions, $ n > 0$. 

The amplitudes obey the normalization condition
\begin{equation}
 x_{a(n)}^2  + \sum_{b,\lambda,\nu} \left [C_{a(n),b,\lambda\nu}^2\right ] + y_{a(n)}^2 + \sum_{b,\lambda,\nu} \left [D_{a(n),b,\lambda\nu}^2\right ]  = 1,
\label{norm}
\end{equation}
$C_{a(n),b,\lambda\nu}$ and  $D_{a(n),b,\lambda\nu}$ being the components on the complex $1qp \otimes 1ph$  states,
while $ x_{a(n)}$ and $y_{a(n)}$ are the components on the original $1qp$ state $a$.
The   fragment $a(n)$  carries a fraction of the strength 
\begin{equation}
N_{a(n)} = x_{a(n)}^2 + y_{a(n)}^2 < 1 
\label{norm_strength}
\end{equation}
which is to be compared with the experimental quasiparticle strength, as determined for example in 
one-particle-transfer reactions.


The excitation operator 
appearing in Eq.(1)
$$
\tilde O^\dagger_{a(n)} = \tilde \alpha^\dagger_{a(n)_{qp}} + 
C_{a(n)b\lambda\nu} \alpha^\dagger_b \Gamma^\dagger_{\lambda\nu}-
D_{a(n)b\lambda\nu} \alpha_{\bar b} \Gamma_{\lambda \bar \nu}
$$
contains the quasiparticle component
$$
\tilde \alpha^\dagger_{a(n)_{qp}} = x_{a(n)} \alpha^\dagger_a - y_{a(n)} \alpha_{\bar a}.
$$
Taking into account the BCS Bogoliubov transformation,
$$
\alpha^\dagger_a = u_a a^\dagger_a + v_a a_{\bar a}
$$
one can write
$$
\tilde \alpha^\dagger_{a(n)_{qp}} = \tilde u_{a(n)} a^\dagger_a + \tilde v_{a(n)} a_{\bar a},
$$
where the quantities
\begin{eqnarray}
\tilde u_{a(n)} = x_{a(n)} u_a - y_{a(n)} v_a\cr
\tilde v_{a(n)} = x_{a(n)} v_a + y_{a(n)} u_a.
\label{newUVS}
\end{eqnarray}
represent the  new quasiparticle  amplitudes  associated with a given fragment $a(n)$:
their squares give the spectroscopic factors associated with one-nucleon transfer reactions.

We can then calculate the renormalized pairing gap as (cf. below Eq. (\ref{gap_prior})) 
\begin{equation}
\tilde \Delta_{a(n)} = \frac {2 \tilde E_{a(n)} \tilde u_{a(n)} \tilde v_{a(n)}} {\tilde u_{a(n)}^2 + \tilde v_{a(n)}^2}.
\label{deltaprior}
\end{equation}






The quasiparticle-phonon matrix elements $V(ab\lambda\nu)$ and $W(ab\lambda\nu)$ 
in Eq. (\ref{MasterMatrix}) are
given by 
\begin{eqnarray}
& V(ab\lambda\nu)=\left[\frac{2\lambda+1}{2j_a+1}\right]^{1/2} \sum_{c\leq d} (1+\delta_{cd})^{-1/2} \times & \cr
& \left[X_{cd}(\lambda \nu)V(cd\lambda b;a)
+(-1)^{j_a - j_b +\lambda}Y_{cd}(\lambda\nu)V(cd\lambda a;b)\right] &
\end{eqnarray}
and 
\begin{eqnarray}
& W(ab\lambda\nu)=\left[\frac{2\lambda+1}{2j_a+1}\right]^{1/2} \sum_{c\leq d} (1+\delta_{cd})^{-1/2} \times & \cr
& \left[X_{cd}(\lambda \nu)R(cd\lambda b;a)
+(-1)^{j_a - j_b +\lambda}Y_{cd}(\lambda \nu)Q(cd\lambda a;b)\right], &
\end{eqnarray}
where $X_{cd}$ and $Y_{cd}$ are the forward and backward amplitudes resulting from the QRPA calculation.

The terms $V(cd \lambda b;a)$, $Q(cd \lambda b;a)$ and $R(cd \lambda b;a)$ are given by

\begin{eqnarray}
V(cd \lambda b;a)=-(u_a v_b u_c u_d - v_a u_b v_c v_d)G(abcd \lambda)\cr
+(u_a u_b u_c v_d - v_a v_b v_c u_d)F(abcd \lambda) \cr
+(-1)^{j_c - j_d + \lambda}(u_a u_b v_c u_d - v_a v_b u_c v_d)F(abdc \lambda)
\label{VGedFV}
\end{eqnarray}

\begin{eqnarray}
Q(cd\lambda b;a)=(u_a u_b u_c u_d + v_a v_b v_c v_d)G(abcd \lambda) \cr
+(u_a v_b u_c v_d + v_a u_b v_c v_d)F(abcd \lambda) \cr
+(-1)^{j_c - j_d + \lambda}(u_a v_b v_c u_d + v_a u_b v_c u_d) F(abdc \lambda)
\label{VGedFQ}
\end{eqnarray}

\begin{eqnarray}
R(cd \lambda b;a)=-(u_a u_b v_c v_d + v_a v_b u_c u_d)G(abcd \lambda) \cr
+(u_a v_b v_c u_d + v_a u_b u_c v_d)F(abcd \lambda) \cr
+(-1)^{j_c - j_d + \lambda}(u_a v_b u_c u_d + v_a u_b v_c v_d)F(abdc \lambda),
\label{VGedFR}
\end{eqnarray}
where $F(abcd \lambda)$ and $G(abcd \lambda)$ denote the angular momentum coupled antisymmetrized
particle-hole and particle-particle $<ab \lambda |V|cd \lambda>_{as}$  matrix element respectively.
Note that in the limit of non-superfluid nuclei (all $u,v$ terms equal to 0 or 1), 
the matrix elements $V(ab\lambda\nu)$ connect pairs of states $a,b$ above or 
below the Fermi energy through the $F-$terms, and pairs of states on opposite parts of the Fermi energy
through the $G-$terms; while the opposite is true for the $W(ab\lambda\nu)$ matrix elements.
In the following we shall not take into account 
the coupling with pair vibration modes, neglecting the $G$ terms in Eqs. (\ref{VGedFV})-(\ref{VGedFR}). 
While this coupling is known to be relevant for closed shell
nuclei, it is expected to be much less important for the superfluid case, since most of the two-particle
transfer strength is already incorporated in the BCS ground state (gauge space deformed)  wave function
\cite{broglia:advances}.

The QRPA amplitudes could be obtained from a calculation performed using the same force adopted in
the HF+BCS calculations. In our approach, the QRPA calculation 
is instead decoupled from the renormalization process, in keeping with the main assumption
(B) discussed above.
In  our QRPA calculation we shall use the separable force
\begin{equation}
 V(\vec r_1,\vec r_2) = -\kappa_{self} \; r_1\frac{\partial U}{\partial r_1}r_2\frac{\partial U}{\partial r_2}
\sum_{\lambda \mu} \chi_{\lambda} Y^*_{\lambda \mu}(\theta_1)Y_{\lambda \mu}(\theta_2)
\label{separable}
\end{equation}
where $U(r)$ is 
a potential that gives a good reproduction of the experimental levels. 
In practice, we adopt the Woods-Saxon parametrization given in \cite{BM:69} (cf. Eq. (2-182))
together with an empirical pairing coupling constant adjusted to reproduce the pairing gap deduced
from the experimental odd-even mass difference. 
The parameters $\chi_{\lambda}$ 
are determined so as
to get a good agreement with the observed properties (energy and transition strength)
of the low-lying surface modes. More precisely, we shall reproduce the polarizability $\beta^2_{\lambda 1}/\hbar \omega_{\lambda 1}$
of the low-lying modes, where $\beta_{\lambda\nu}$ denotes the experimental nuclear deformation parameter. In fact, the matrix
elements of the phonon-induced pairing interaction for levels close to the Fermi energy are approximately proportional
to the polarizability of the mode (cf. Eq. (\ref{VIND_prior}) below).
The resulting values of $\chi_{\lambda}$  turn out to be close to 1 (cf. Section \ref{results}), 
indicating that the QRPA coupling constant  is close to the Bohr-Mottelson self-consistent
coupling constant $\kappa_{self}= \left[{\int r\frac{\partial \rho}{\partial r}r\frac{\partial U}{\partial r}r^2 dr}\right]^{-1} $. 


This scheme then reduces to the collective particle-shape vibration (phonon) coupling 
scheme given by Bohr and Mottelson
\cite{BM:75} (cf. Eqs. 6-207- 6-209).
In fact  the particle-hole  matrix elements, neglecting the exchange terms (cf. on this point \cite{rowe},
Eq. (14.54) and Chap.16),
are given by 
\begin{equation}
F(abcd \lambda ) = -\kappa_{self}\chi_{\lambda} <ab \lambda \mu| r_1\frac{\partial U}{\partial r_1}Y^*_{\lambda \mu}(\theta_1)|0>
<0| r_2\frac{\partial U}{\partial r_2}Y^*_{\lambda \mu}(\theta_2)|cd \lambda\mu>
\end{equation}
where $\mu$ is any of the $z$-projections of the angular momentum $\lambda$.
In this expression the QRPA-like single-particle indices $(c,d)$ and the scattered particle indices $(a,b)$
appear in separated factors, so that one gets the  angular momentum reordering property 
$F(abdc \lambda)=(-1)^{j_c-j_d+ \lambda}F(abcd \lambda) =(-1)^{j_a-j_b+ \lambda}F(bacd \lambda)$,
and
\begin{eqnarray}
V(ab \lambda\nu)=-\kappa_{self}\chi_{\lambda} (u_au_b-v_av_b)<ab\lambda\mu| r_1\frac{\partial U}{\partial r_1}Y^*_{\lambda \mu}(\theta_1)|0>
\left[\frac{2\lambda+1}{2j_a+1}\right]^{1/2} \cr 
\times \sum_{c\leq d} (1+\delta_{cd})^{-1/2} 
\left[
(X_{cd}(\lambda\nu)+Y_{cd}(\lambda\nu))(u_cv_d+v_cv_d)<0| r_2\frac{\partial U}{\partial r_2}Y^*_{\lambda \mu}(\theta_2)|cd\lambda\mu>
\right] 
\end{eqnarray}
The quantity in the summation is precisely the transition amplitude $M(\lambda\nu)$ of the 
$\hat M = r_2\frac{\partial U}{\partial r_2}Y^*_{\lambda \mu}(\theta_2)$
operator, which is usually expressed in terms of the so-called collective deformation parameter as $M(\lambda\nu)=\alpha^o_{\lambda\nu}/\kappa_{self}$,
assuming a collectively deformed density 
$\delta\rho=-r\frac{\partial\rho}{\partial r}\sum_{\lambda \mu} Y^*_{\lambda \mu}(\theta)\alpha_{\lambda \mu}$

In this way we can write
\begin{equation}
V(ab\lambda\nu)=-\chi_{\lambda} (u_au_b-v_av_b)<ab\lambda\mu| r_1\frac{\partial U}{\partial r_1}Y^*_{\lambda \mu}(\theta_1)|0>
\left[\frac{2\lambda+1}{2j_a+1}\right]^{1/2} \alpha^o_{\lambda\nu}.
\end{equation}
Finally, following the notation in \cite{BM:75}, Eqs.(6-207 to 6-209) 
using the reduced matrix element $<j_b||Y_{\lambda}||j_a>= (-1)^{j_a-j_b}<j_aj_b;\lambda\mu|Y_{\lambda\mu}|0> \sqrt{2\lambda+1}$ and
the relation $\alpha^o_{\lambda\nu}=\beta_{\lambda\nu}/\sqrt{2\lambda+1}$,
we can write
\begin{equation}
V(ab\lambda\nu) = h(ab\lambda\nu)(u_a u_b - v_a v_b),
\end{equation}
where
\begin{eqnarray}
 h(ab\lambda\nu) =-(-1)^{j_a-j_b} \beta^{eff}_{\lambda\nu} <a| r_1\frac{\partial U}{\partial r_1}|b><j_b||Y_{\lambda}||j_a> \cr
\left[\frac{1}{(2j_a+1)(2\lambda+1)}\right]^{1/2},
\label{hvertex}
\end{eqnarray}
which is the basic vertex in \cite{BM:75}  corrected by our effective deformation parameter  $\beta^{eff}_{\lambda}=
\chi_{\lambda\nu} \beta_{\lambda\nu}$.

Analogously one finds
\begin{equation}
 W(ab\lambda\nu) = h(ab\lambda\nu)(u_a v_b + v_a u_b).
\end{equation}


A self-consistent  renormalization procedure by iterating the diagonalization process,
using the previously renormalized quasiparticles to build the complex $ 1qp \otimes 1ph $ states. 
This means that the basic $V,W$ matrix elements are now calculated as 
\begin{eqnarray}
& V(ab(m)\lambda\nu) =& h(ab\lambda\nu)(u_a \tilde u_{b(m)} - v_a \tilde v_{b(m)})\cr
& W(ab(m)\lambda\nu) =&h(ab\lambda\nu)(u_a \tilde v_{b(m)} + v_a \tilde u_{b(m)})
\label{eq:VW}
\end{eqnarray}
where the $u_a$ and $v_a$ amplitudes obtained from the initial HF+BCS calculation are kept fixed
in the iteration process; on the other hand the amplitudes $\tilde u_{b(m)}$ and $\tilde v_{b(m)}$ refer to the 
amplitudes associated with the $m-$fragment resulting from the renormalization of the state $b$ in the previous iteration step.
In this way  a consistent fragmentation of the different states is 
constructed through the iterative procedure.

\begin{figure*}[t!]
\begin{center}
\includegraphics[width=0.8\textwidth]{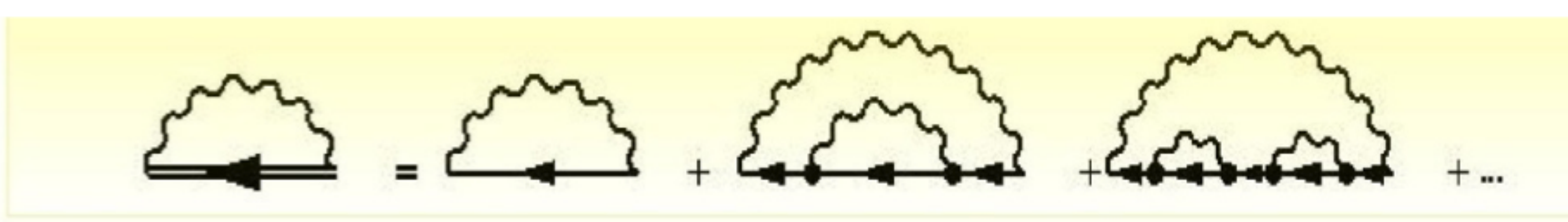}
\end{center}
\caption{ The coupling of the quasiparticle to many-phonon states, corresponding to the  rainbow diagrams
shown in the figure, is included in our approach through the iteration of the self-consistent diagonalization
of the Nambu-Gor'kov matrix.}
\label{fig:fig3_diagrams}
\end{figure*}

The iteration process  gives rise to the so called no-line crossing
rainbow series of the self-energy (see below), in which all orders  are summed up coherently, 
and  which is expected to play an important role in the limit of strong coupling (cf. Fig.\ref{fig:fig3_diagrams}). 
The size of the matrix
to be diagonalized increases exponentially at each iteration, and some sort of numerical approximation
is needed, as will be discussed in Section \ref{results}.

\subsubsection{\label{energy} Energy dependent, BCS-like formulation}

The diagonalization of the  eigenvalue problem  (\ref{MasterMatrix}) is equivalent to 
the $2\times2 $ energy-dependent eigenvalue problem derived from an approach based on  Green's function formalism
(cf. e.g. \cite{Ter:02}):

\begin{eqnarray}
\label{Matrice2x2}
 \left ( \matrix{ 
  E_{a}+ \Sigma^{11pho}_{a(n)} & \Sigma^{12pho}_{a(n)}\cr
 \Sigma^{12pho}_{a(n)}  & - E_{a}+ \Sigma^{22pho}_{a(n)}  } \right )
\left ( \matrix{ 
x_{a(n)}\cr
y_{a(n)}
  } \right ) = \tilde E_{a(n)}
\left ( \matrix{ 
x_{a(n)}\cr
y_{a(n)}
}  \right ) 
\end{eqnarray} 
where one has introduced the energy-dependent, normal self-energies $\Sigma^{11pho}_{a(n)}$ and
$\Sigma^{22pho}_{a(n)}$
given by
\begin{eqnarray}
\Sigma^{11pho}_{a(n)} = \sum_{b,m,\lambda,\nu} \frac{ V^2(a b(m)\lambda\nu)}
{\tilde E_{a(n)} - \tilde E_{b(m)} - \hbar \omega_{\lambda \nu}} +
\sum_{b,m,\lambda,\nu} \frac{ W^2(a b(m)\lambda\nu)}
{\tilde E_{a(n)} + \tilde E_{b(m)} + \hbar \omega_{\lambda\nu}} \cr
\Sigma^{22pho}_{a(n)} = \sum_{b,m,\lambda,\nu} \frac{ W^2(a b(m)\lambda\nu)}
{\tilde E_{a(n)} - \tilde E_{b(m)} - \hbar \omega_{\lambda \nu}} +
\sum_{b,m,\lambda,\nu} \frac{ V^2(a b(m)\lambda\nu)}
{\tilde E_{a(n)} + \tilde E_{b(m)} + \hbar \omega_{\lambda\nu}}
\label{Sig11}
\end{eqnarray}
and  the abnormal self-energy  
\begin{eqnarray}
\Sigma^{12pho}_{a(n)} = - \sum_{b,m,\lambda,\nu} V(a b(m)\lambda \nu) W(a b(m) \lambda \nu) \cr
\left [ \frac{1}{\tilde E_{a(n)}-\tilde E_{b(m)}-\hbar\omega_{\lambda \nu}}- 
  \frac{1}{\tilde E_{a(n)}+\tilde E_{b(m)}+\hbar\omega_{\lambda \nu}} \right ].
\label{Sig12prior}
\end{eqnarray}

We note that $\Sigma^{11}$  evaluated at a given energy $E$ is equal to $-\Sigma^{22}$ evaluated at $-E$, that is,
$\Sigma^{11pho}_{a(n)} = - \Sigma^{22pho}_{a(-n)}$.
The normalization of the 
quasiparticle strength of the $n$-fragment (cf. Eq. (\ref{norm})) is given by  \cite{Bloch}
\begin{equation}
x_{a(n)}^2 + y_{a(n)}^2 
-  \frac{\partial \Sigma^{11pho}_{a(n)}}{\partial \tilde E_{a(n)}} x^2_{a(n)}
-  \frac{\partial \Sigma^{22pho}_{a(n)}}{\partial \tilde E_{a(n)}} y^2_{a(n)}
- 2 \frac{\partial \Sigma^{12pho}_{a(n)}}{\partial \tilde E_{a(n)}} x_{a(n)} y_{a(n)} = 1.
\label{normalization}
\end{equation}

 In this form one can easily make contact with the formalism based on the Green's function
for superfluid systems,
\begin{equation}
 \hat G_a (\tilde E_a + i\delta) = [ (\tilde E_a  + i \delta){\bf 1} - E_a \tau_3 - \hat \Sigma^{pho}_{a}(\tilde E_a+i\delta) ]^{-1},
\label{green1}
\end{equation}
where $\tau_3$ denotes a Pauli matrix, and where 
\begin{equation}
\hat \Sigma^{pho}_{a}(\tilde E_a+i\delta) =
 \left ( \matrix{ \Sigma^{11pho}_{a}(\tilde E_a+i\delta)& \Sigma^{12pho}_{a}(\tilde E_a+i\delta)\cr
 \Sigma^{12pho}_{a}(\tilde E_a+i\delta) & \Sigma^{22pho}_{a}(\tilde E_a+i\delta) } \right )
\label{green2}
\end{equation}
is our phonon-mediated self-energy matrix evaluated at the complex energy $ \tilde E_a+i\delta $, which coincides with
 the matrix introduced  by Nambu and extensively used in condensed matter
to deal with strong coupling superconductivity \cite{Schr:64}.
Eqs. (\ref{green1}),(\ref{green2}) must be solved self-consistently due to the fact that
the quasiparticle strengths needed for  the evaluation of 
 $\hat \Sigma^{pho}_{a}(\tilde E_a+i\delta) $ through the $V$ and $W$ matrix elements are obtained from the imaginary part
of $ \hat G_a (E+i\delta)$ (cf. \cite{Sluys:93}, Eqs.(46-47)).


In order to get more insight concerning  the respective 
 contributions of the bare and of the phonon-induced interaction to the pairing gap,
it is useful to rewrite the 2x2 eigenvalue problem (\ref{Matrice2x2}) 
in terms of the amplitudes $\tilde u,\tilde v$, instead of  
$x,y$, by inverting  the relation (\ref{newUVS}), obtaining

\begin{eqnarray}
\label{Matrice2x2uv}
 \left ( \matrix{ 
  (\epsilon_{a}-\epsilon_F)+\tilde \Sigma^{11}_{a(n)}&\tilde \Sigma^{12}_{a(n)}\cr
 \tilde \Sigma^{12}_{a(n)}&-(\epsilon_{a}-\epsilon_F)+\tilde \Sigma^{22}_{a(n)}  } \right )
\left ( \matrix{ 
\tilde u_{a(n)}\cr
\tilde v_{a(n)}
  } \right ) = \tilde E_{a(n)}
\left ( \matrix{ 
\tilde u_{a(n)}\cr
\tilde v_{a(n)}
}  \right ), 
\label{eq:inverted}
\end{eqnarray}
where $\epsilon_a$ denotes the HF single-particle energy, while  
the new normal self-energies are given by
\begin{eqnarray}
 \tilde \Sigma^{11}_{a(n)}= u_a^2 \Sigma^{11pho}_{a(n)} +v_a^2 \Sigma^{22pho}_{a(n)}
-2u_av_a \Sigma^{12pho}_{a(n)}\cr
 \tilde \Sigma^{22}_{a(n)}= u_a^2 \Sigma^{22pho}_{a(n)} +v_a^2 \Sigma^{11pho}_{a(n)}
+2u_av_a \Sigma^{12pho}_{a(n)}.
\end{eqnarray}
One can separate the abnormal self-energy into two terms, writing
\begin{equation} 
\tilde \Sigma^{12}_{a(n)} =  \Delta^{BCS}_a +
 \tilde \Sigma^{12,pho}_{a(n)}.
\label{sigma_decomp}
\end{equation}

The first term, $\Delta^{BCS}_a $, is the pairing 
gap associated with the bare interaction obtained in the HF+BCS calculation, while 
the second term is associated with the phonon-induced interaction and is given by 
\begin{equation}
\tilde \Sigma^{12,pho}_{a(n)}= \Sigma^{12pho}_{a(n)}(u^2_a-v_a^2)+
u_av_a(\Sigma^{11pho}_{a(n)}-\Sigma^{22pho}_{a(n)}).
\end{equation}
Using Eqs. (\ref{eq:VW}), (\ref{Sig11}) and (\ref{Sig12prior}) this expression can be simplified
and the abnormal self-energy can be rewritten as
\begin{equation}
\tilde \Sigma^{12,pho}_{a(n)} = -\sum_{b,m} \frac{(2j_b+1)}{2} V_{ind}(a(n)b(m))  \tilde u_{b(m)} \tilde v_{b(m)},
\label{Sig12phoprior}
\end{equation}
where we have  introduced the induced pairing interaction:

\begin{eqnarray}
&V_{ind}(a(n)b(m)) = \sum_{\lambda,\nu}\frac{2 h^2(ab\lambda\nu)}{(2j_b+1)} \times & \cr
&\left[ \frac{1}{\tilde E_a(n)-\tilde E_b(m)-\hbar\omega_{\lambda\nu}}- \frac{1}{\tilde E_a(n)+\tilde E_b(m)+\hbar\omega_{\lambda\nu}} \right]. &
\label{VIND_prior}
\end{eqnarray}
Furthermore, we can symmetrize the matrix (\ref{Matrice2x2uv}) in order to get   
a $2 \times 2$  eigenvalue equation which is formally identical to 
the BCS eigenvalue equation.
This can be achieved multiplying Eq. (\ref{eq:inverted}) by  the $Z_{a(n)}$ energy-dependent function, 

\begin{equation}
 Z_{a(n)}= \left( 1-\frac{\tilde \Sigma^{odd}_{a(n)}}{\tilde E_{a(n)}} \right)^{-1},
\end{equation}
where $\tilde \Sigma^{odd}$ is the odd part of $\tilde \Sigma^{11}_{a(n)}$
\begin{equation}
\tilde \Sigma^{odd}_{a(n)} = \frac{\tilde \Sigma^{11}_{a(n)} + \tilde \Sigma^{22}_{a(n)}}{2}=
\frac{ \Sigma^{11pho}_{a(n)} +  \Sigma^{22pho}_{a(n)}}{2}.
\end{equation}
We note that  according to the definition above, $Z$ is the inverse of the correspondent  quantity
as defined in \cite{Ter:02},\cite{Schr:64}. We also note that the symbol $Z$ is often used instead of $N$ to define
the quasiparticle strength. In fact the two quantities tend to take similar values close to the Fermi energy
(cf. Fig. \ref{fig:zeta_sly4} below).  
This similarity can be explained, on the one hand, by noting that for the lowest pole, 
 the finite difference in the expression of  $Z$ can be approximated by a derivative: 
\begin{equation}
Z_{a(n)}  = \left (1 - \frac{\Sigma^{11}_{a(n)} - \Sigma^{11}_{a(-n)}}{2 \tilde E_{a(n)}} \right )^{-1} \approx  \left ( 1 -  \frac {\partial \Sigma^{11}_{a(n)}}
{\partial \tilde E_{a(n)}} \right )^{-1}.
\end{equation}
On the other hand, in the normalization of the strength of quasiparticles close to the Fermi energy, $x >> y$ so that, using
Eq. (\ref{normalization}),
\begin{equation}
N_{a(n)}  = x_{a(n)}^2 + y_{a(n)}^2  \approx x_{a(n)}^2 \approx \left ( 1 -  \frac {\partial \Sigma^{11}_{a(n)}}{\partial \tilde E_{a(n)}} \right )^{-1}.
\end{equation}

It is possible to  rewrite Eq. (\ref{Matrice2x2uv}) as
\begin{eqnarray}
\label{eq.Secular_prior}
 \left ( \matrix{ 
  \tilde \epsilon_{a(n)}- e_F & \tilde \Delta_{a(n)}\cr
 \tilde \Delta_{a(n)} & - (\tilde \epsilon_{a(n)}-e_F) }\right )
\left ( \matrix{ 
\tilde u_{a(n)}\cr
\tilde v_{a(n)}
  } \right ) = \tilde E_{a(n)}
\left ( \matrix{ 
\tilde u_{a(n)}\cr
\tilde v_{a(n)}
}  \right ) 
\end{eqnarray} 
where 
\begin{equation}
 \tilde \epsilon_{a(n)}- e_F = Z_{a(n)}\left[ (\epsilon_{a}- e_F) +\tilde \Sigma^{even}_{a(n)} \right],
\label{ean_prior}
\end{equation}
and where $\tilde \Sigma^{even}_{a(n)}$ is the even part of $\tilde \Sigma^{11}_{a(n)}$:

\begin{equation}
\tilde \Sigma^{even}_{a(n)} = \frac{\tilde \Sigma^{11}_{a(n)} -\tilde \Sigma^{22}_{a(n)}}{2}=
(u_a^2-v_a^2)\frac{\Sigma^{11pho}_{a(n)} - \Sigma^{22pho}_{a(n)}}{2}-2u_av_a  \Sigma^{12pho}_{a(n)}.
\end{equation}
The term $\tilde \epsilon_{a(n)}$ in Eq. (\ref{eq.Secular_prior}) represents the renormalized single-particle energy,
and one can now identify the pairing gap with the term $\tilde \Delta_{a(n)} $ \cite{Schr:64}:
\begin{eqnarray}
\tilde \Delta_{a(n)}= Z_{a(n)} \tilde \Sigma^{12}_{a(n)} =  Z_{a(n)} \left( \Delta^{BCS}_a + \tilde \Sigma^{12,pho}_{a(n)} \right)
\equiv \tilde \Delta_{a(n)}^{bare}+ \tilde \Delta_{a(n)}^{pho}.
\label{gap_prior}
\end{eqnarray}
This quantity coincides with the expression \ref{deltaprior} introduced above,
which can be used when solving the energy independent problem (\ref{MasterMatrix}), because it involves 
only quasiparticle energies and amplitudes.
Note also that the quasiparticle energy relates to the new gap and single-particle
energy as in the usual BCS equations:

\begin{equation}
\tilde E_{a(n)}= \sqrt{(\tilde \epsilon_{a(n)}- e_F)^{2}+\tilde \Delta^{2}_{a(n)} }
\end{equation}

It is now possible to  write a generalized gap equation.
Introducing the total quasiparticle strength for a given fragment,
\begin{equation}
 \tilde u^{2}_{a_{n}} + \tilde v^{2}_{a_{n}} = N_{a(n)},
\end{equation}
the product $\tilde u_{a(n)} \tilde v_{a(n)}$ may be obtained in the usual way from the 2x2 secular equation as 

\begin{equation}
\tilde u_{a(n)} \tilde v_{a(n)}=  N_{a(n)} \frac{\tilde \Delta_{a(n)}} {2\sqrt{(\tilde \epsilon_{a(n)}- e_F)^{2}+\tilde \Delta^{2}_{a(n)} }}.
\end{equation}

\begin{equation}
\tilde u_{a(n)} \tilde v_{a(n)}=  N_{a(n)} 
\frac{Z_{a(n)} \tilde \Sigma^{12}_{a(n)}} 
{2\sqrt{Z^{2}_{a(n)}(\epsilon_{a}-e_F+\tilde \Sigma^{even}_{a(n)})^{2}+( Z_{a(n)} \tilde \Sigma^{12}_{a(n)})^2 }}.
\end{equation}
Substituting in Eq. (\ref{Sig12phoprior}) one obtains

\begin{equation}
\tilde \Sigma^{12}_{a(n)}= \Delta_{a} -  \sum_{b(m)}V_{ind}(anbm) N_{b(m)} 
\frac{\tilde \Sigma^{12}_{b(m)}} {2\sqrt{(\epsilon_{b}-e_F+\tilde \Sigma^{even}_{b(m)})^{2}+(\tilde \Sigma^{12}_{b(m)})^2 }}. 
\label{delta_noi}
\end{equation}
The second term in this equation is a generalization of the usual BCS gap equation, and 
clearly demonstrates  how the action of the different fragments of the original quasiparticles
is modulated by their quasiparticle strength $ N_{b(m)}$.
The equation, however, is of little  practical use as it stands because 
it involves the energy dependent interaction $ V_{ind}$
which contains a "dangerous" denominator (cf. Eq. (\ref{VIND_prior})).
The formula  will be further discussed in Section \ref{quasiparticle},  
and we  shall present a similar expression in Section \ref{post} (cf. Eqs. \ref{delta_noi_post}-\ref{delta_noi2}).

%






 The approach  presented so far (and in \cite{Sluys:93}) 
neglects the $1qp$-exchange interaction between the complex
$1qp \otimes 1ph$  states (see fig. 6.10 c-d in \cite{BM:75}). 
In fact the associated matrix elements are set to 0 in the matrix (\ref{MasterMatrix}). 
The ignored processes would account for some violation of the Pauli principle arising from the
microscopic structure of the QRPA phonons, which may imply double occupation of the quasiparticle
state in the complex $1qp \otimes 1ph$  state. These violations  are anyway
very small for the calculation reported in this paper (cf. Section \ref{bubble}).



Those processes account also for vertex renormalization
terms in the self-energy, which are not taken into account  in the rainbow series we have considered.
While in condensed matter they are usually neglected based on the Migdal theorem,
in nuclear physics they are usually considered to be of minor importance because
their contribution to the self-energy implies a recoupling of angular momenta 
that when summed up over all possible intermediate states 
is expected to lead to a rather strong cancellation \cite{belyaev,avdeenkov}.  
As we have mentioned discussing approximation (B) above, 
we shall assume that they are implicitely included
in our  effective PVC, computed
making  use of phenomenological  phonons and single-particle levels.

\begin{figure*}[t!]
\begin{center}
\includegraphics[width=7cm]{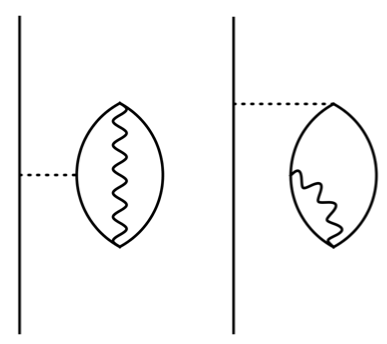}
\end{center}
\caption{ Lowest order diagrams representing the renormalization of the quasiparticle self-energy
by zero-point fluctuations.}
\label{fig:tadpole}
\end{figure*}

\begin{figure*}[t!]
\begin{center}
\includegraphics[width=0.6\textwidth]{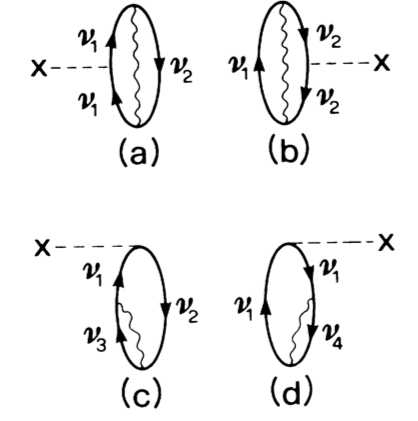}
\end{center}
\caption{ Lowest order diagrams representing the renormalization of  the density operator (represented by the cross)
by zero-point fluctuations in closed shell nuclei.}
\label{fig:tadpole_prl}
\end{figure*}

Although within NFT tadpole diagrams should be included \cite{nft3}, 
we will neglect the energy-independent contributions associated with them (cf. Fig. \ref{fig:tadpole}), 
which take into account the effect of zero-point fluctuations on the quasiparticle energy \cite{khodel}.
This kind of diagrams modifies the nuclear  density (cf. Fig. \ref{fig:tadpole_prl}) and
plays a particular important role in the calculation of nuclear radii \cite{barranco_radii,khodel_radii,fayans},  
representing the leading correction beyond mean field for closed shell nuclei, and producing sizeable isotopic effects 
\cite{barranco_plb}.
However, they lead to relatively small changes in the mean-field potential. The shifts of the single-particle 
energies for $A = 120$ can be estimated to be  of the order of 150 keV  \cite{madurga}, and, being of static nature, we shall assume that 
they are effectively taken into account in the mean field.  
The tadpole diagrams can also influence the 
abnormal density an the calculation of the bare pairing gap \cite{saperstein:08}, 
but no calculations have been performed for superfluid nuclei. We have
estimated that the effect of  diagrams (a) and (b) in Fig. \ref{fig:tadpole_prl} changes the abnormal
density $\kappa$  by about 5\%. This value 
can be obtained  from  the expression
\begin{equation}
\delta {\kappa}(r) = \sum_{ab \lambda}   (2\lambda +1) Y_{ab\lambda}^2   \kappa _a(r)/(2j_a+1),
\end{equation}
where $Y_{ab\lambda}$ denotes a backward amplitude calculated  in QRPA. 
The diagrams (c) and (d), in the normal case,
give a contribution to the renormalization of the radius which is about three times larger
than the one produced by  (a) and (b) \cite{barranco_radii}. This can be considered as an upper
limit since the strong $r^2$ dependence tend to enhance their relative importance.

Thus we estimate a total tadpole effect to the pairing field of less than 20/
value, that is less than 200 keV, which is a number rather consistent with that estimated above for the
normal single-particle energy.

Finally, we note that the formalism presented in this Section  is similar to the one adopted by 
Avdeenkov and Kamerdzhiev \cite{Avd:1,Avd:2}. However, a precise comparison of their results with ours  
is difficult, because  they have followed a different approach, trying to extract 'bare' single-particle levels and 
pairing gaps from the experimental levels, instead of renormalizing the levels obtained from
an effective mean field.

\subsection  {\label{post} The post scheme }

In the previous Section we have outlined the prior scheme  in the calculation of the pairing 
properties of superfluid nuclei. That is, we have started from a HF+BCS calculation which accounts 
for the bare pairing interaction and we have then 
added the effects of the PVC diagonalizing the matrix in Eq.(\ref{MasterMatrix}).
However one could also adopt a different scheme (which we will call the {\it post}-scheme)
in which one starts from the HF solution and
then  one includes simultaneously both 
renormalization  effects
and 
the bare pairing interaction through the iterative procedure.
This approach is particularly appropriate  when the 
phonon induced pairing provides the leading contribution
to pairing correlations, and is certainly needed  when the bare interaction alone is not able to produce
a superfluid solution. This is the case, for instance, in calculations of halo nuclei based on the 
PVC coupling \cite{noi:11li,noi:11be}.
On the other hand,  the prior-scheme is more natural, in the framework of  general schemes 
based on the corrections to mean field properties.

In the following we shall apply the post scheme to the calculation of pairing properties 
in $^{120}$Sn, which is a well bound, superfluid nucleus. 
In this case  we expect that the two schemes should give similar results 
concerning  physically relevant quantities, namely quasiparticle energies, spectroscopic factors and pairing gaps.

The calculation in the post scheme is
performed by introducing the  bare pairing field $\Sigma^{12bare}$ associated with the bare 
interaction in Eq.(\ref{MasterMatrix}),
by writing 

\begin{eqnarray}
\left( \matrix
{
  E_{a} & V(ab\lambda\nu) & V(ac\lambda\nu)  & W(ab\lambda\nu) & W(ac\lambda\nu) & \mp \Sigma^{12, bare}\cr
  V(ab\lambda\nu) & \hbar\omega_{\lambda\nu} + E_{b} & 0  & 0 & 0 &  W(ab\lambda\nu)\cr 
  V(ac\lambda\nu) & 0& \hbar\omega_{\lambda\nu} + E_{c} & 0 & 0  &  W(ac\lambda\nu)\cr 
  W(ab\lambda\nu) & 0 & 0  & -\hbar\omega_{\lambda\nu} - E_{b} & 0 &  -V(ab\lambda\nu)\cr
  W(ac\lambda\nu) & 0 & 0  & 0& -\hbar\omega_{\lambda\nu} - E_{c} & -V(ac\lambda\nu)\cr
 \mp \Sigma^{12bare}& W(ab\lambda\nu) & W(ac\lambda\nu)  & -V(ab\lambda\nu) & -V(ac\lambda\nu) & -E_{a}} \right )
\left ( 
\matrix {
    x_{a(n)}\cr
 C_{a(n),b,\lambda\nu}\cr
 C_{a(n),c,\lambda\nu}\cr
 -D_{a(n),b,\lambda\nu}\cr
 -D_{a(n),c,\lambda\nu}\cr
 -y_{a(n)}} \right ) = \nonumber
\end{eqnarray}

\begin{eqnarray}
\tilde E_{a(n)}
\left ( 
\matrix {
    x_{a(n)}\cr
 C_{a(n),b,\lambda\nu}\cr
 C_{a(n),c,\lambda\nu}\cr
 -D_{a(n),b,\lambda\nu}\cr
 -D_{a(n),c,\lambda\nu}\cr
 -y_{a(n)}} \right )
\label{PostMatrix}
\end{eqnarray}
where now the unperturbed "quasiparticle" energy $E_{a} $ contains no contribution
from pairing and is simply equal to the difference between the HF single-particle energy
and the Fermi energy,
$$
E_a = | {\epsilon_a - \epsilon_F}|,
$$
where $\Sigma^{12bare}$ obeys the equation
\begin{equation}
\Sigma^{12bare} = - \sum_{b,n}V_{bare}(ab) \frac {(2j_b+1)} {2} \tilde u_{b(n)} \tilde v_{b(n)},
\label{Delta}
\end{equation}
and the -(+) sign is to be used for particle-(hole-) states, that is $\epsilon_{a} > e_{F}$ ($\epsilon_{a} < e_{F}$).

The amplitudes $\tilde u_{a(n)}, \tilde v_{a(n)}$ are given by Eq. (\ref{newUVS}), 
taking into account the fact that  the initial $u_a$ and  $v_a$ factors to be inserted in the iterative solution of
Eq.(\ref{MasterMatrix}) are now equal to 1 or 0, depending on whether the level is above or below
the Fermi energy:
\begin{equation}
 \tilde u_{a(n)} = x_{a(n)} \quad   \tilde v_{a(n)} = y_{a(n)}
\label{newUVparticle}
\end{equation}
and 
\begin{equation}
 \tilde u_{a(n)} = -y_{a(n)} \quad  \tilde v_{a(n)} = x_{a(n)}
\label{newUVhole}
\end{equation}
respectively in the case of a particle- and a hole-state.

Note that, putting the 
$V-W$ matrix elements to zero, Eq. (\ref{Delta}) reduces to the standard BCS equation for the pairing gap.
The solution must be obtained through a self-consistent iterative procedure which in the general case involves simultaneously the
bare interaction and the PVC vertices $V$ and $W$. In general this is not equivalent to the two-step diagonalization
performed in the prior scheme, in which one considers
first only $ \Sigma^{12bare}$ (which is exactly equivalent to solve the bare BCS problem), and then
takes into account the 
matrix elements $V$'s and $W$'s. Nevertheless we expect that the two schemes lead to similar results,
as long as the first diagonalization leads to a finite value of the order parameter, the pairing gap.

As in the prior scheme one can use the $2 \times 2 $ energy dependent BCS-like matrix to solve the problem.
The only difference with  respect to Eq. (\ref{eq.Secular_prior}) is that now 
the renormalized pairing gap must include a term $\Sigma^{12bare}$ representing the contribution 
of the bare pairing interaction:
\begin{eqnarray}
\tilde \Delta_{a(n)}= Z_{a(n)} \left( \Sigma^{12bare}_{a(n)} +\tilde \Sigma^{12,pho}_{a(n)} \right)
= \tilde \Delta^{bare}_{a(n)}  + \tilde \Delta^{pho}_{a(n)} .
\label{gap_schr_post}
\end{eqnarray}

This  expression can be rewritten in the more appealing way
\begin{eqnarray}
\tilde \Delta_{a(n)}=  - Z_{a(n)} \sum_{b,m} \frac{(2j_b+1)}{2} V_{eff}(a(n)b(m)) \tilde u_{b(m)} \tilde v_{b(m)} =
\label{gap_schr_post2}
\end{eqnarray}
where one has introduced the effective interaction $V_{eff}$, which is the sum
of the bare and phonon-induced interactions:
\begin{equation}
 V_{eff}(a(n)b(m)) = V_{bare}(ab) + V_{ind}(a(n)b(m))
\end{equation}

Furthermore one can obtain a very closed form for the gap equation by eliminating the
amplitudes $u$ and $v$ using  Eq. (\ref{eq.Secular_prior}):
\begin{equation}
\tilde u_{a(n)} \tilde v_{a(n)}=  N_{a(n)} \frac{\tilde \Delta_{a(n)}}{2 \tilde E_{a(n)}} 
= N_{a(n)}  \frac{Z_{a(n)} \tilde \Sigma^{12}_{a(n)}} 
{2\sqrt{Z^{2}_{a(n)}(\epsilon_{a}-e_F+\tilde \Sigma^{even}_{a(n)})^{2}+( Z_{a(m)} \tilde \Sigma^{12}_{a(n)})^2 }}.
\end{equation}
leading to 
\begin{eqnarray}
&\tilde \Delta_{a(n)} = Z_{a(n)} \tilde \Sigma^{12}_{a(n)}= \cr
& - Z_{a(n)} \sum_{b(m)}   \frac{2 (j_b+1)}{2} V_{eff}(a(n)b(m)) N_{b(m)} 
\frac{\tilde \Sigma^{12}_{b(m)}} {2\sqrt{(\epsilon_{b}-e_F+\tilde \Sigma^{even}_{b(m)})^{2}+ (\tilde \Sigma^{12}_{b(m)})^2 }}. 
\label{delta_noi_post}
\end{eqnarray}

We note that for levels near to the Fermi energy the  $Z-$factor is very close to the quasiparticle strength $N$.
A more compact expression, bearing a direct resemblance to the standard BCS gap equation,
may be obtained by reintroducing the $Z-$function both  in the numerator and in the denominator:
\begin{equation}
\Delta_{a(n)}=   - Z_{a(n)} \sum_{b(m)} V_{eff}(a(n)b(m)) N_{b(m)} 
\frac{\Delta_{b(m)}} {2 \tilde E_{b(m)}}. 
\label{delta_noi2}
\end{equation}
This equation is an exact consequence of the Nambu-Gor'kov energy-dependent problem and can be used as an useful starting point for
approximate gap equations (see Section \ref{quasiparticle}  ). In particular, restricting the sum to the main peak $m=1$ and neglecting the difference
between $N$ and $Z$, Eq. (\ref{delta_noi2}) 
formally reduces  to an approximate gap equation previously presented in the case of uniform matter (\cite{lombardo} (cf. Eq.(12)), 
\cite{baldo}).

\section{\label{results} Results}


In the following we present the solution of the Nambu-Gor'kov equations for $^{120}$Sn in various approximations.
In all cases, we shall limit our attention to the five neutron levels belonging to the shell around the Fermi energy
(namely, the 1$g_{7/2}$, the 2$d_{5/2}$, the 3$s_{1/2}$, the 2$d_{3/2}$ and the 1$h_{11/2}$ orbitals).
Most of the calculations will be performed in the prior scheme, making use of the Argonne  nucleon-nucleon interaction as the
bare pairing force in the $^1S_0$ channel, which gives by far  the dominant contribution to $T=1, J=0$ pairing 
\cite{baroni_gap}. In Appendix \ref{barepair} we shall also show a few results obtained with the $V_{{\rm low} k}$ potential. 

In general we shall iterate the renormalization process. This requires
looking for a self-consistent solution, either  by successive diagonalizations
of the matrix (\ref{MasterMatrix}) or (\ref{PostMatrix}), or looking for the solutions of the equivalent 
energy-dependent $2 \times 2$  problem  (\ref{Matrice2x2}). Nearly all of the results  presented 
in this work have been obtained from the the diagonalization procedure. However, we have verified in several instances 
the numerical agreement between the two approaches. 
In order to control the iteration process, one must introduce a cutoff  procedure, or perform
some averaging,  in order to avoid  the exponential
increase of the number of solutions, retaining at the same time the essential information \cite{vanneck:91},\cite{vanneck:93}. 
We shall make use of two simple numerical procedures. 
A cutoff procedure can be adopted  when one solves the energy-dependent problem. 
In this case one can limit the number of solutions kept at each iteration, by selecting only those fragments carrying
a quasiparticle strength larger than a given cutoff $N_{cut}$. An extreme case is represented by the so-called one-quasiparticle
approximation, in which one keeps only the most important  pole for each orbital $a$.
We shall instead  make use of the averaging procedure when we solve the energy-independent problem 
diagonalizing the matrices (\ref{MasterMatrix}) or (\ref{PostMatrix}). In this case we define a number
$N_{zones}$  of energy zones. The zones are generally not of the same size, but are chosen so as to reflect the main features of the 
quasiparticle strength  function.  After each iteration we collect all the strength  obtained within a given energy zone
into a single fragment, placed at the calculated average energy position. In this way,  the number of solutions 
associated with a given orbital $a$  is  kept fixed to
$N_{dim} =  2 \times  \left [N_{phon} \times N_{sps} \times N_{zones} +2 \right]$, 
where $N_{phon}$ is the number of RPA
solutions retained in the diagonalization and  $N_{sps}$ is the number of single-particle levels considered. 

We shall compute the PVC coupling by first performing a QRPA calculation with 
the separable force (\ref{separable})
for the multipolarities  and parities $\lambda^{\pi}= 2^+,3^-,4^+$ and $5^-$. In each case, the coupling constant
as determined so as to reproduce the experimental value of the ratio $B(E \lambda)/\hbar \omega_{\lambda,1}$ (cf. the discussion
about the approximation (B) in Section \ref{formalism} above). Experimental data are taken from \cite{expdata_phon}. The quasiparticle
states used in the QRPA calculation are obtained from a BCS calculation with a monopole pairing interaction $- G_0 P^{\dagger}P$ 
based on  the levels of a Woods-Saxon potential parametrized as in  ref. \cite{BM:69},
Eq. (2-182).  The pairing  coupling constant $G_0$ is adjusted so as to reproduce the value $\Delta_{exp} \approx 1.4$ MeV derived from the experimental odd-even mass difference. The PVC matrix elements are then obtained from Eq. (\ref{hvertex}).

Concerning the QRPA spectrum, we adopt an averaging procedure similar to that adopted for the quasiparticle strength.
We include explicitly the strong collective low-lying vibrational states and we collect the remaining strength in a
small number of peaks, reflecting the gross structure of the response for each multipolarity \cite{Ter:02}.
We have verified that the results are not sensitive to the details of the high-lying phonons.
The properties of the low-lying phonons employed in the calculations are listed in Table \ref{table_phonons},
where they are compared with the available experimental data. We remark that the values 
of $\chi_{\lambda}$ (cf. Eq. (\ref{separable})) are close to 0.9, reflecting  the rather good accuracy of the collective model.

\begin{table}
\begin{center}
\begin{tabular}{|c|c|c|c|c|c|c|c|}
\hline
&\multicolumn{3}{|c|} {Exp.}  
&
\multicolumn{4}{|c|} {Theory} 
\\ 
\hline
\textbf{$\lambda^{\pi}$} &$\hbar \omega_{\lambda 1}  $&$\beta_{\lambda 1}$&$\beta_{\lambda 1}^2/\hbar \omega_{\lambda 1} $
&$\hbar \omega_{\lambda 1}  $&$\beta^{eff}_{\lambda 1} $&$(\beta^{eff}_{\lambda 1})^2/\hbar \omega_{\lambda 1} $
& $\chi_{\lambda}$
\\
\hline
\multirow{1}{*}{$2^+$} &  $1.17 $ &  $0.13$ & $0.014$ & $1.20$ &
$0.13$ & 0.013 & 0.86  \\
\hline
\multirow{1}{*}{$3^-$} & $2.40$ &  $0.16$ & $0.011$     & $2.71$ &
$0.16$  &  0.010 & 0.95  \\
\hline
\multirow{1}{*}{$4^+$} & $3.10$ &  $0.11$    & $0.004$     & $2.33$  & $0.10$ &
0.004& 0.91 \\
\hline
\multirow{1}{*}{$5^-$} & $2.27$  &  $0.08$& $0.003$ & $2.48$  & $0.09$   
& 0.003 & 0.91\\
\hline
\end{tabular}
\caption{\protect In the first columns we list the experimental energies 
$\hbar \omega_{\lambda 1}$ in (MeV), deformation parameters $\beta_{\lambda 1} $
and polarizabilities $\beta^2_{\lambda 1}/ \hbar \omega_{\lambda 1}$
of the low-lying states associated with the $2^+$, $3^-$, $4^+$ and $5^-$ multipolarities. 
They are compared  with the   
corresponding quantities calculated in QRPA, making use of the effective deformation 
parameter $\beta^{eff}_{\lambda\nu} = \chi_{\lambda} \beta_{\lambda\nu}$. In the last column
we give the  values of  $\chi_{\lambda}$. 
In the case of $4^+$,  the experimental low-lying strength is fragmented in four peaks  lying between 
2.2 and 3.8 MeV; the numbers in the table refer to an average weighted with the transition strength of each peak.  }
\label{table_phonons}
\end{center}
\end{table}

\subsection {\label{bare} Calculations with bare pairing potentials}

In this Section, we shall discuss solutions of the Nambu-Gor'kov equations making use of bare nucleon-nucleon potentials
as pairing interactions. As we have remarked above, we shall adopt the prior scheme. 
The practical difficulty with the post scheme is that the basis needed to account for the realistic bare interactions
must span a broad energy interval  
(about 1 GeV in the case of the Argonne $v_{14}$ interaction, and about 200 MeV in the case of $V_{{\rm low} k})$. 
This  implies a large numerical   effort in order to
handle Eq.(\ref{PostMatrix}), in particular because the matrix associated with a given angular momentum must include
states with different number of nodes. We shall further consider the relation between the prior and post schemes 
presented in 
Section \ref{formalism}, making use of a simplified bare interaction.

\subsubsection{\label{baregap} Bare pairing gap}

The first step in our calculation is represented by the 
solution of the HF+BCS equation with the bare force in the pairing channel; we shall not consider the influence 
of pairing on the mean field.

We notice that the
calculation with the bare force is in fact an extended BCS calculation, 
because for a hard-core interaction it is essential to 
consider the coupling between pairs of levels with different number of nodes, in order to take properly into 
account the coupling between bound and continuum levels \cite{noi_plb,pizzo}. We take into account states up to 1 GeV. 
As a consequence, for a given
set of quantum numbers $a = \{lj\}$ one obtains  a set of quasiparticles $\{E_k\}$; to each quasiparticle
is associated  an array of quasiparticle amplitudes $u_{nk}$ and $v_{nk}$, 
which are the components of the quasiparticle states over the HF basis states $\phi_{nlj}$. 
Going to the canonical basis, where the density matrix  takes a diagonal form \cite{ring_schuck},
we look for the state  having the largest value of the abnormal density, $u_{max} v_{max}$. As a rule, for
a stable nucleus like $^{120}$Sn, this
canonical state is dominated by the quasiparticle state having the lowest value of the quasiparticle energy,
$E_{min}$. We then approximate the extended BCS calculation associating the  value of $v_{max}$ and $E_{min}$,
treating them as proper BCS quantities; in particular we derive an associated pairing gap as
$\Delta^{BCS}= 2 u_{max} v_{max} E_{min}$, which is very close to the diagonal value of
the gap in the original basis. The value of $v_{max},E_{min}$ and $\Delta^{BCS}$ are then employed
as input values for the solution of the Nambu Gor'kov equations in the prior scheme described in Section \ref{prior}
(cf. Eqs. (\ref{MasterMatrix}) and (\ref{newUVS}), where they are denoted $u_a,v_a$ and $E_a$).
This approximation leads to a substantial simplification of our numerical scheme. For nuclei close to the 
line of stability, where the canonical states have significant weight on continuum state and the PVC easily
connects bound and continuum states, this approximation would not be justified  
and one should rather consider different numerical schemes, for example based on the continuum Green's function. 

The mean field will be taken from a HF calculation with the SLy4 interaction \cite{chabanat}. 
This interaction gives a good reproduction of the bulk properties of nuclei;
moreover the resulting level density
close to the Fermi energy (associated with an effective mass $m_k \approx 0.7 m$), 
once increased by  renormalization effects, is in reasonable agreement with the experimental
one (the associated effective mass increasing to $m^* = m_k  m_{\omega} \approx m$).
The energies of the five single-particle levels lying closest to the Fermi energies are shown in Appendix \ref{meanfield} 
(cf. Fig. \ref{spart_levels}). 
The detailed features of our calculations are of course influenced by the specific properties of the 
mean field, and in particular by its effective mass.   In order to 
have some insight about the dependence on the adopted mean field, in the 
Appendix \ref{meanfield} we provide results obtained making use of two  different Skyrme interactions,
having effective masses smaller and larger than the SLy4 interaction. 
One should distinguish two different effects that influence the pairing gap calculated with the bare pairing interaction: 
on the one hand, the pairing gap is sensitive to the detailed 
position of the levels close to the Fermi energy, that could be influenced by static contributions not considered here,
as those produced from tensor correlations  \cite{duguet_tensor,colo_tensor} and from tadpole diagrams 
\cite{saperstein:08} (for the latter, cf. the comments at the end of Section \ref{prior}); on the other hand,
the pairing gap is also sensitive to the momentum dependence of $m_k$ for large momenta.

In a previous work  \cite{epj:04} we presented a solution of  the HFB equations  using the Argonne $v_{14}$ equation in the pairing channel, 
which led to a pairing gap  of the order of 700 keV. There we used a modified SLy4 mean field,  
reducing the spin-orbit coupling strength by about 15\% (furthermore we included in the Hamiltonian 
the terms proportional to  the square of the spin density, which will not be included in the following,
and used a slightly larger PVC strength).
This was done, in order to improve the agreement of the results obtained after renormalization 
with  the experimental  spectrum and pairing gap.  Here we prefer to take the HF field 
obtained with the original SLy4 interaction. This is associated with a larger level density around the Fermi energy. 
As a consequence, the obtained pairing gap, shown in Fig. \ref{fig:renogap_sly4} (full dots), 
is larger,  being equal to about 1.1 MeV. Similar calculations have  
been performed with the $V_{{\rm low} k}$ potential using the same mean field \cite{hebeler}. In that case, the result
depends on the cutoff $\Lambda$ adopted to obtain $V_{{\rm low} k}$: 
for values of $\Lambda $ smaller than about 4 fm$^{-1}$, the gap is close to 1.4 MeV,
(cf. Fig. \ref{fig:renogap_sly4_vlowk}), while for increasing values of $\Lambda$  the $V_{{\rm low} k}$ 
potential reduces to the Argonne potential and one obtains 
the result shown in  Fig. \ref{fig:renogap_sly4}. We notice that our  numerical results with the Argonne potential
are in very good agreement with those reported in ref. \cite{hebeler} for very large values of $\Lambda$.
The reason for the difference between the results obtained with the Argonne and $V_{{\rm low} k}$ interaction has been discussed in 
detail \cite{hebeler,saperstein}:
using the Argonne potential implies using the Skyrme effective mass $m_k$ (equal to about 0.7$m$ inside the nucleus) up to very high momenta, while 
the renormalization process leading to $V_{{\rm low} k}$ implies that $m_k$ goes to the free mass for momenta larger than the cutoff $\Lambda$.
The lower value of $m_k$ at high momenta  leads to a smaller level density and to a 
reduction of about 300 keV in the value of the gap with the Argonne interaction with respect
to the values obtained with $V_{{\rm low} k}$ for cutoffs $\Lambda$ up to about  $ 4$fm$^{-1}$.  
An analogous difference shows up in calculation of the gap in uniform matter. 
While the lack of momentum dependence of the effective mass associated with  Skyrme interactions is certainly unrealistic,
the proper momentum dependence is still to be established;  
and the density dependence of $m_k$ associated with the SLy4 interaction in nuclear matter  
is not far from that  resulting at the Fermi energy from 
calculations based on Br\"uckner theory  \cite{saperstein}. 
Furthermore, a precise determination of the value of the bare  gap within the $V_{{\rm low} k}$ scheme requires the consideration of 
the effects of the three-body force, which is expected to provide a repulsive contribution \cite{lombardo_three}. 
A calculation of three-body effects within the $V_{{\rm low} k}$ renormalization scheme \cite{duguet_three,duguet_threea}
leads in fact to a reduction of the gap, down to values close to those obtained with the Argonne interaction 
in Fig. \ref{fig:renogap_sly4}. An average value of the bare gap close to 1 MeV was also derived in the analysis 
of refs. \cite{Avd:1,Avd:2}.

We have brought circumstantial evidence which testifies to the fact that a  
pairing gap $\Delta^{BCS} \approx $ 1.1 MeV for the levels around the Fermi energy 
obtained  using the  $v_{14}$ potential as bare pairing interaction and the effective mass 
associated with SLy4  represents a reasonable starting point, 
being well aware that the determination of the mean field and of the associated effective mass 
is one of the most important issues which remains to be fully clarified, 
for a quantitative and microscopic  calculation of the gap in finite nuclei.  In Section \ref{meanfield} we 
investigate the dependence of the results on the adopted mean field, whille in Section \ref{barepair} we 
provide some 
results obtained adopting $V_{\rm{low} k}$ as bare pairing interaction.


\subsubsection{\label{solution} Solution of the Nambu-Gor'kov equations}

\begin{figure*}[t!]
\begin{center}
\includegraphics[width=0.8\textwidth]{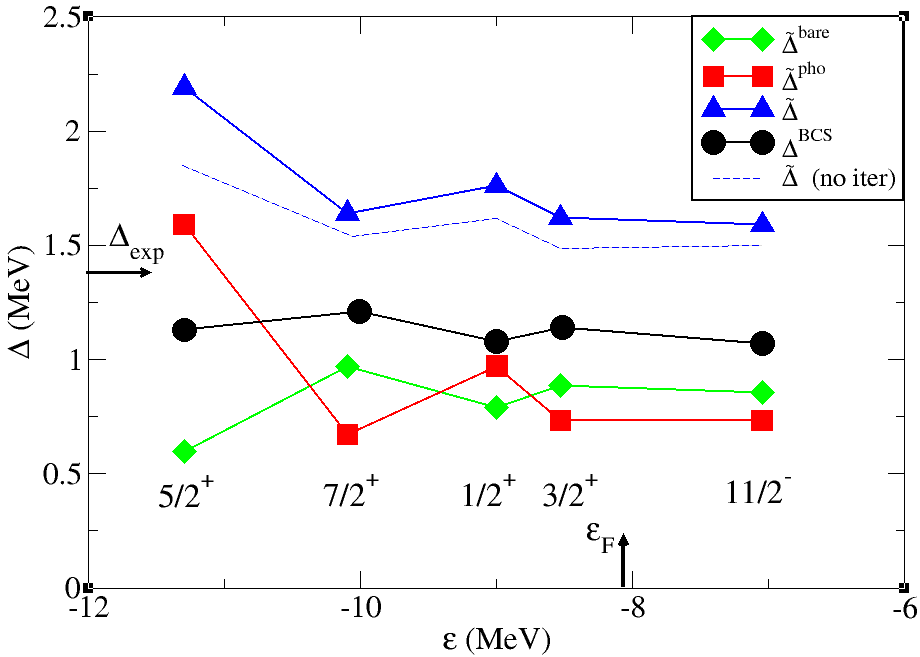}
\end{center}
\caption{ The state-dependent neutron paring gap $\Delta^{BCS}$
calculated in BCS with the bare $v_{14}$ interaction  is shown as a function of the 
SLy4 HF single-particle energy of the five valence orbitals (cf. Fig. \ref{spart_levels}), and is  compared to
the renormalized gap $\tilde \Delta$ (cf. Eq. (\ref{gap_prior}) )
obtained solving the Nambu-Gor'kov equations by iteration; also shown (dashed line) is the  gap 
obtained at the first iteration. 
We also show the decomposition of $\tilde \Delta$
into the bare and phonon contributions $\tilde \Delta^{bare}$ and $\tilde \Delta^{pho}$. 
The value of the Fermi energy $\epsilon_F$ and of the gap obtained from the experimental odd-even
mass difference $\Delta_{exp}$ are also indicated.} 
\label{fig:renogap_sly4}
\end{figure*}

\begin{figure*}[b!]
\begin{center}
\includegraphics[width=0.8\textwidth]{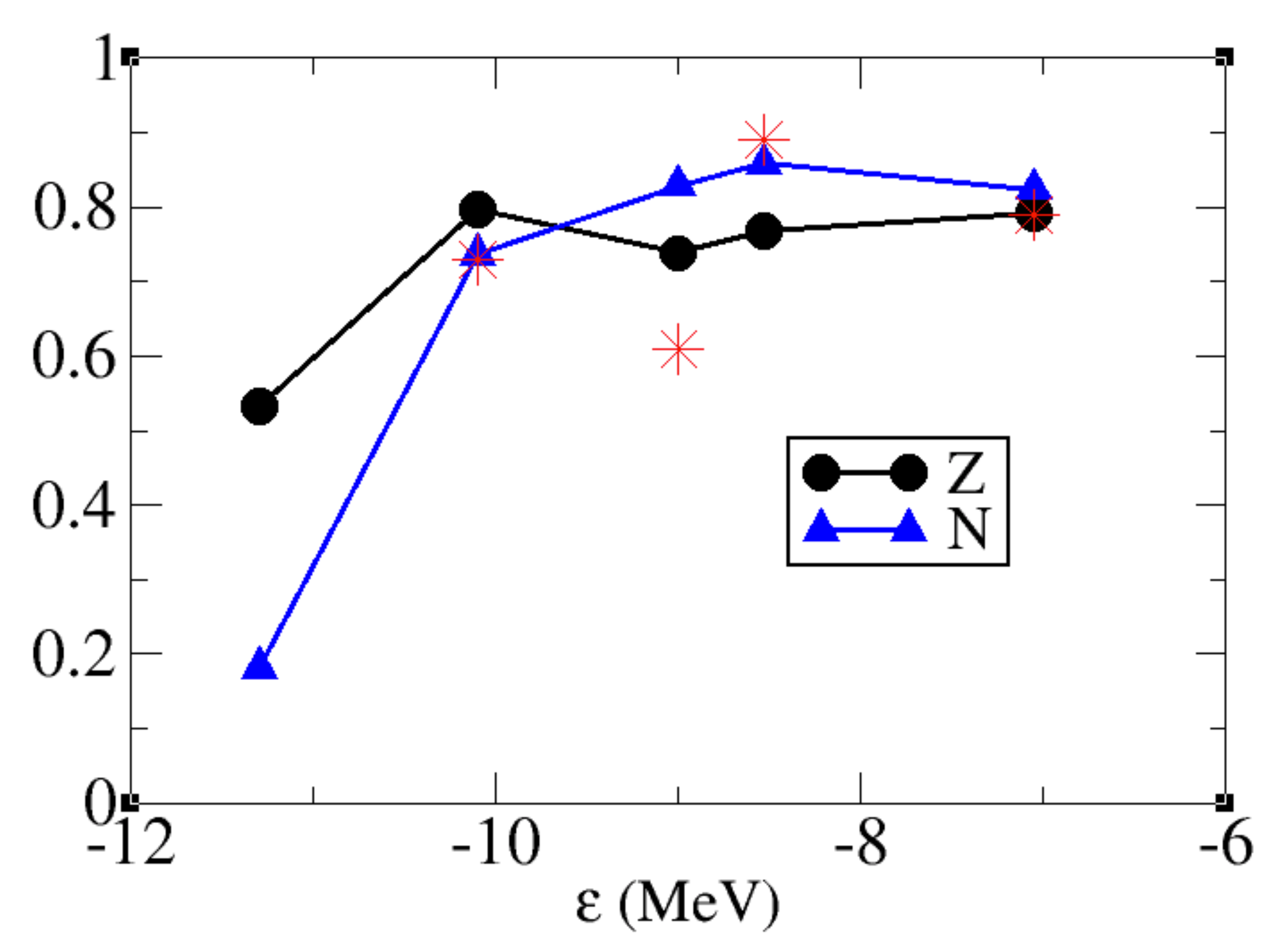}
\end{center}
\caption{ Comparison of the $N-$ and $Z-$ factors associated with the lowest quasiparticle peaks
in the  Nambu-Gor'kov calculation shown in Fig. \ref{fig:renogap_sly4}. Also shown by stars are the values of the experimental
quasiparticle strength \cite{exp_transf2}, except for the $d_{5/2}$ orbital which shows a pronounced fragmentation (cf. text).}
\label{fig:zeta_sly4}
\end{figure*}

The state-dependent pairing gap obtained from the solution of the Nambu-Gor'kov equations is compared 
to the  bare gap in Fig. \ref{fig:renogap_sly4}. Renormalization effects lead to a total gap $\tilde \Delta$ 
 about 600 keV larger than $\Delta^{BCS}$. Most of the effect is obtained already with the first diagonalization of the Nambu-Gor'kov matrix. 
The self-consistent iteration process leads to a further  increase of the gap by about 10\%. 
 We recall that one can distinguish two contributions to the renormalized gap 
$\tilde \Delta  = Z [ \Delta^{BCS} + \tilde \Sigma^{12,pho}] = \tilde \Delta^{bare} + \tilde \Delta^{pho}$,
associated with  the bare and  with the phonon-induced
interaction (cf. Eq. (\ref{gap_prior})).  
They are also shown in Fig. \ref{fig:renogap_sly4} and turn out to be of similar magnitude. 
This confirms that the phonon-induced pairing interaction is as important as the bare interaction in determining
pairing properties of heavy nuclei.  Notice that a proper comparison between the role of these two sources of pairing
can only be made analyzing their contribution to the  final physical result, and not just by comparing the BCS and complete
pairing gaps. In fact, 
processes beyond mean field act in a complex way, not only giving rise to the induced pairing interaction,
but also reducing the effect of the bare interaction through the $Z-$factors.
The values of $Z$ for the 
five orbitals are shown in Fig. \ref{fig:zeta_sly4} where it can be seen that they are close to 0.7,
bringing the bare contribution from the BCS result (about 1.1 MeV) down to about 0.8 MeV, which is
about one half of the total renormalized gap. The other half is provided by the pairing induced interaction.
The values of $Z$ are similar  to the values  of the quasiparticle strength $N= U^2 +V^2$, except for the orbital $d_{5/2}$. 
The values of $N$ are in rather good agreement with the  
quasiparticle strength measured in one-nucleon transfer reactions, shown by stars in Fig. \ref{fig:zeta_sly4} (cf. also
below Figs. \ref{fig:strength_h112} and \ref{fig:strength_d52} with the  related  discussion), although one has to consider that 
the experimental values are affected by a large error, estimated to be about 30\%.

\begin{figure*}[t!]
\begin{center}
\includegraphics[width=0.8\textwidth]{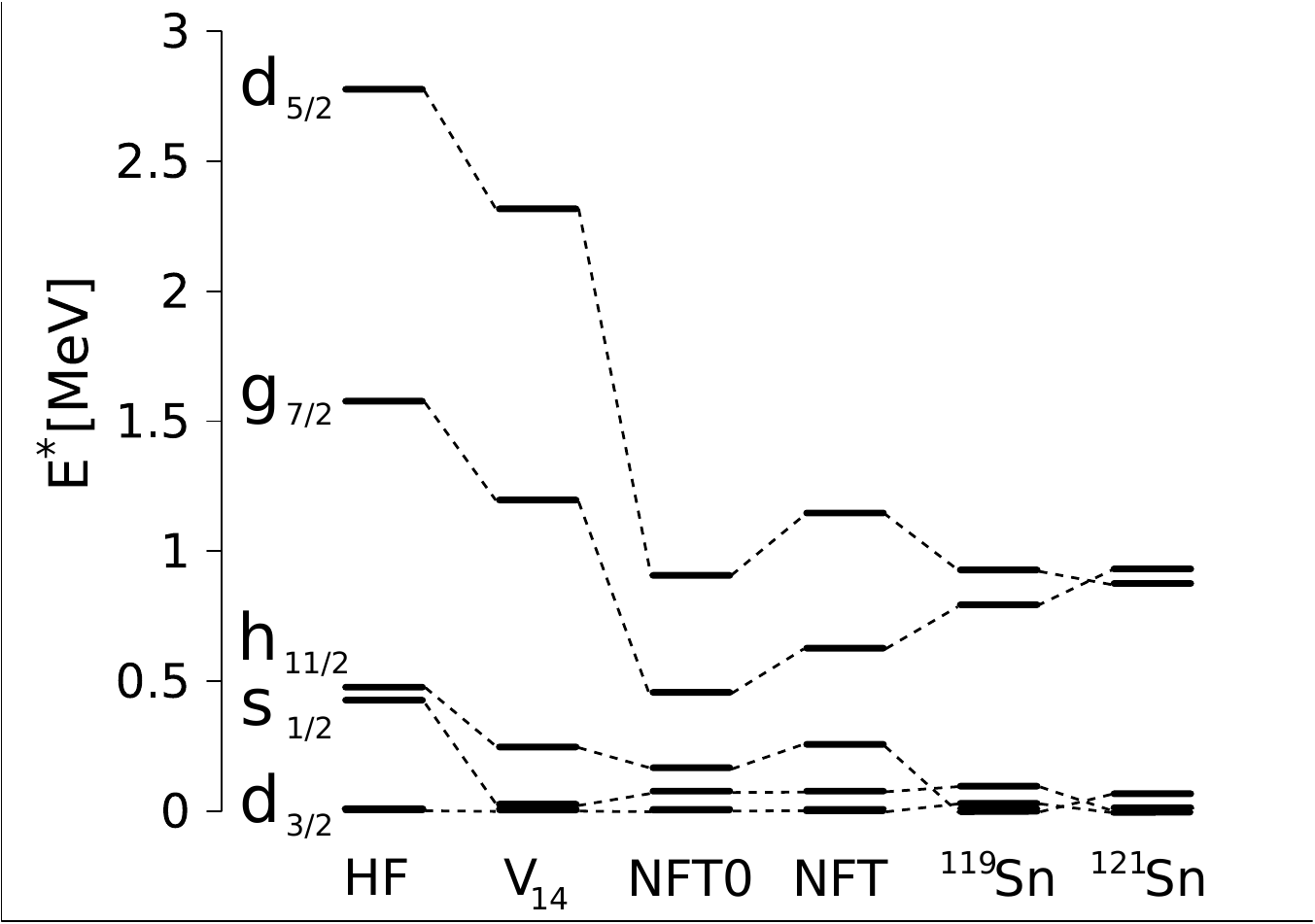}
\end{center}
\caption{ The theoretical quasiparticle spectra obtained at the various steps of the calculation are compared
to the experimental data. See the text for more explanation.} 
\label{fig:renospectra_sly4}
\end{figure*}

The renormalized pairing gap exceeds the experimental value obtained from the odd-even mass difference by about 300 keV,
the value of $\Delta_{exp}$ lying in between $\Delta^{BCS}$ and $\tilde \Delta$. We remark that 
the estimate of $\Delta_{exp}$ varies by  about $\pm$  100 keV depending on which odd-even mass formula
is used \cite{bender,duguet_oddeven} ( the five-point formula $\Delta^{(5)}$ yielding a value of 1.39 MeV), 
and that a more consistent comparison would imply a  theoretical calculation of the binding  energies. 
Such a  precise comparison of the gaps is probably not very significant at the present stage of
the theory, given the  uncertainties which affect both the BCS calculation  - like the dependence on the 
adopted HF mean field and  the effect of three-body forces mentioned above - and the renormalization process - 
in particular the exchange of spin modes, not included in this calculation  (see the discussion in Section \ref{effspin}
and Section \ref{Hindsight}).
However, the fact that the value of the renormalized pairing gap is found to be relatively close to experiment is satisfactory.
The necessity of going beyond mean field turns out  anyway to be clearer considering  other physical quantities,
in particular the energy spectrum of neighbouring odd nuclei with  the  associated strength functions and
spectroscopic factors.   
This can be seen  in Fig. \ref{fig:renospectra_sly4} where we compare the features of the 
quasiparticle spectrum obtained at the different steps of the calculation.   In the first column (HF), we report the 
absolute value of the difference $| \epsilon_a - \epsilon_F|$ obtained  in the HF calculation with the SLy4 force,
referred to the value of this difference for the level $d_{3/2}$, which lies closest to the Fermi energy $\epsilon_F$ = -8.05 MeV
(cf. Fig. \ref{spart_levels}). 
In the second column (BCS), we give the values of the quasiparticle energies obtained in the HF+BCS calculation with the Argonne force,
referred to the lowest quasiparticle.
In the third column (NFT0) we show the spectrum obtained taking into account processes beyond mean field 
calculated with the Nambu-Gor'kov equations without iterating,
while in the fourth (NFT) we show the self-consistent solution: in these cases
we show the energy of the lowest fragment for each quantum number (this is also the one collecting the largest quasiparticle
strength, except for the $d_{5/2}$ orbital, which is very fragmented and whose case is discussed below): 
Finally in the fifth and sixth columns we give the position of the lowest peaks measured in $^{119}$Sn and $^{121}$Sn. 
The main discrepancy of the renormalized spectrum concerns the $7/2^+$ state, whose position is brought too low  by
the PVC  by about 300 keV ( this is probably related to the initial position of the $g_{7/2}$ orbital in the HF spectrum,
which lies too close to the Fermi energy,  cf. Fig. \ref{spart_levels}). 
The experimental energies of the lowest three  states are very close to each other, being separated by less than 100 keV,   and are 
well separated by the other two  levels. This gross structure is already present in the HF result, which, however, greatly underestimates
the density of levels. This remains the case 
including pairing correlations at the BCS level (see second column), which lead to a limited improvement. 
The PVC leads to a denser spectrum, considerably improving the agreement with experiment,
although one should not expect a detailed agreement in the order of three lowest levels. The main effect is already obtained
with a single diagonalization (NFT0), but the self-consistent treatment leads to an appreciable rearrangement of the spectrum,
somewhat reducing the initial compression of the levels, slightly improving the agreement with experiment.
The increase of level density can be expressed as an increase of the neutron effective mass from $m^* \approx 0.7 m$ to $m^* \approx m$.  One could argue
that a mean field BCS  calculation in a potential with $m^*=1$ 
could lead to agreement with experiment in a simpler and more direct way. 
However, such a calculation (still performed  with the bare interaction) would greatly overestimate the gap \cite{esb}.   Thus, the
simultaneous consideration of gap and low-energy spectra clearly favours a description that includes 
renormalization effects on both 
quantities.


\begin{figure*}[t!]
\begin{center}
\includegraphics[width=0.8\textwidth]{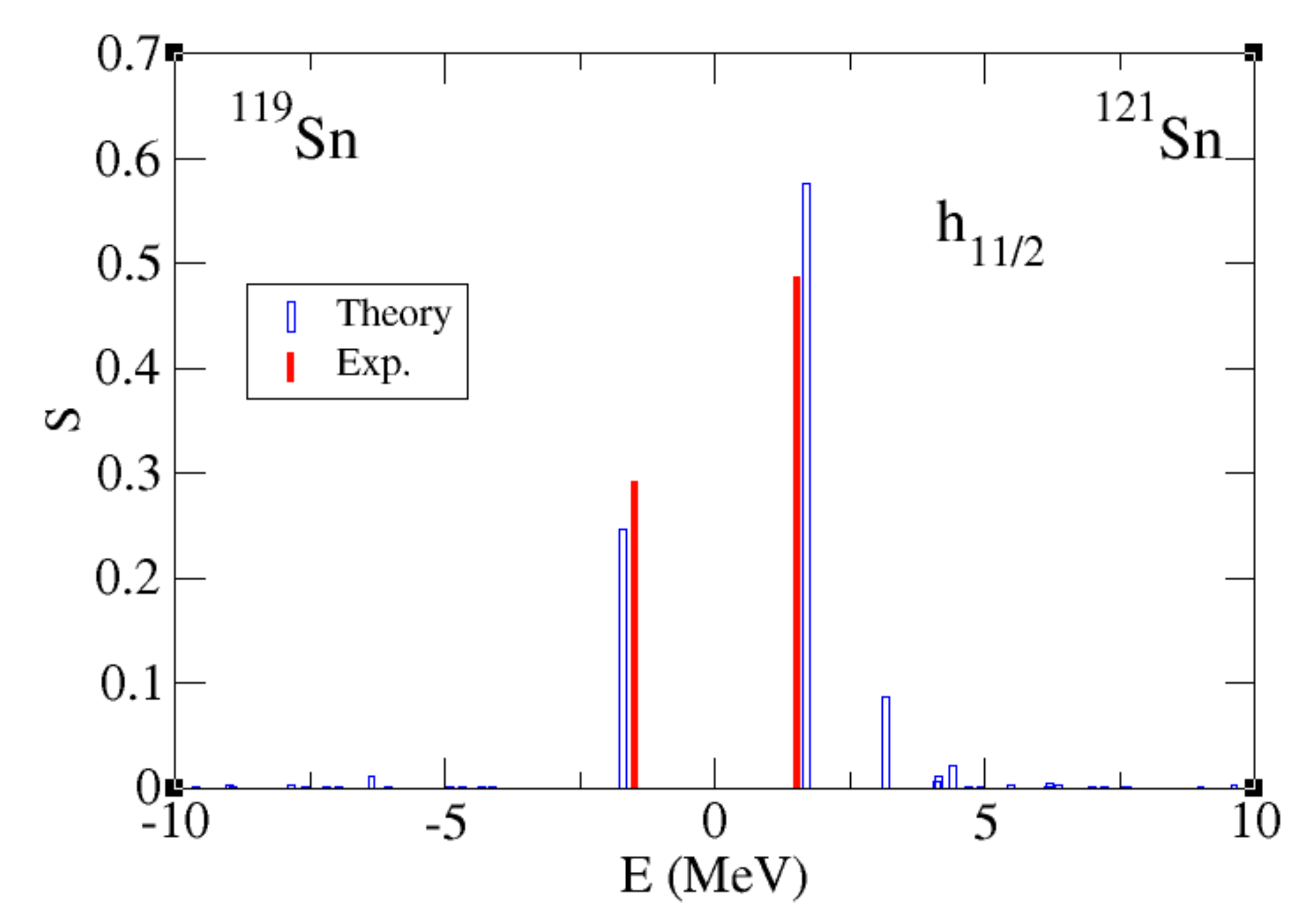}
\end{center}
\caption{ The theoretical strength function calculated for the $h_{11/2}$ orbital 
is compared to the spectroscopic factors associated with experimental levels detected in one-neutron transfer reactions.} 
\label{fig:strength_h112}
\end{figure*} 

\begin{figure*}[h!]
\begin{center}
\includegraphics[width=0.8\textwidth]{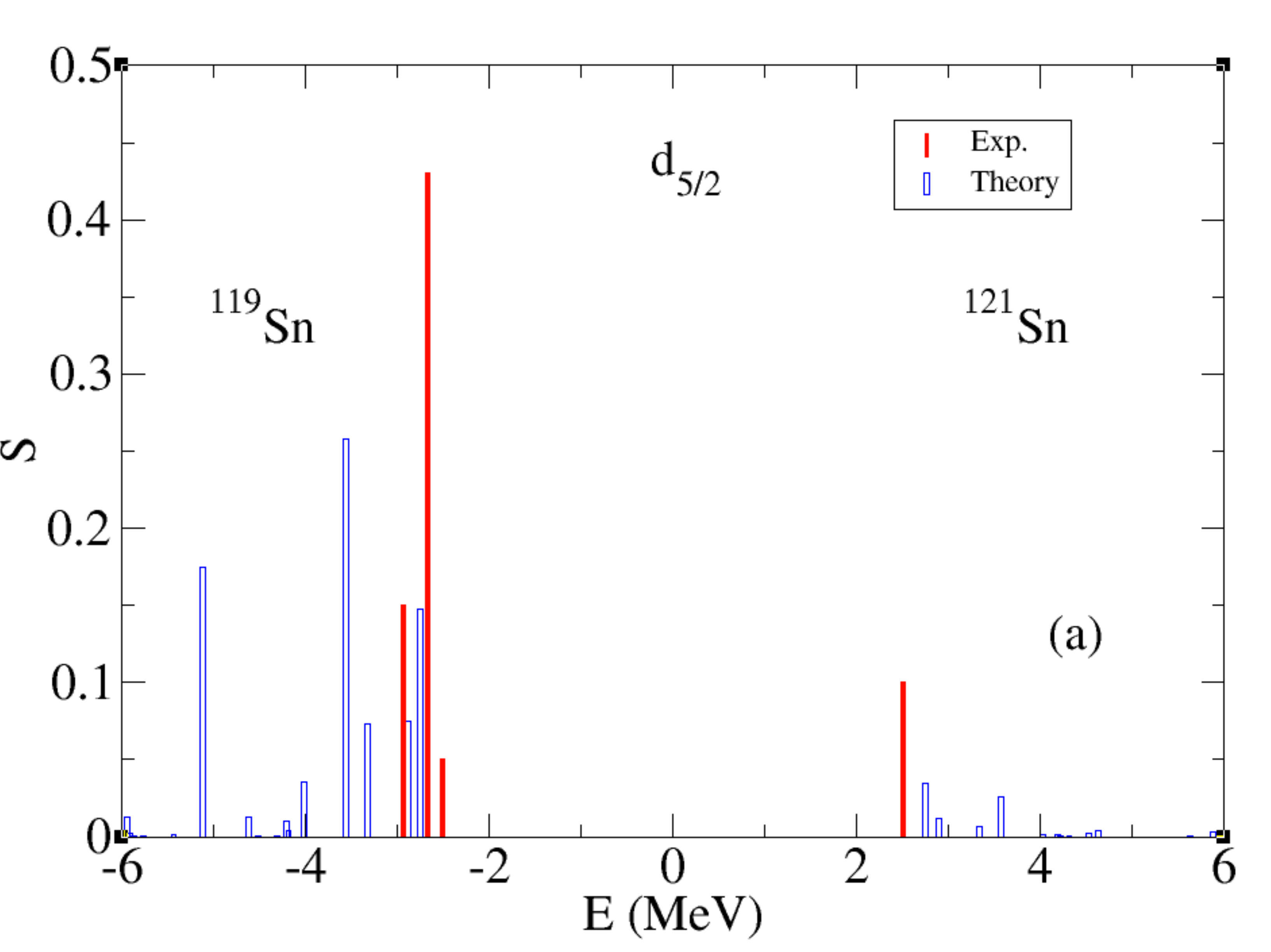}
\end{center}
\caption{ The theoretical strength function calculated for the $d_{5/2}$ orbital 
is compared to the spectroscopic factors associated with experimental levels detected in one-neutron transfer reactions.
} 
\label{fig:strength_d52}
\end{figure*} 

The most specific fingerprint of the renormalization processes is represented by the fragmentation of the quasiparticle
strength, that in our approach is due to  the coupling with $1qp \otimes 1ph$  states. Experimentally, the effects of such
a fragmentation leads to a breaking  of the observed quasiparticle strength into several peaks, and consequently to a reduction 
of the strength observed in the main peaks, as compared to the full strength expected in mean field calculations. 
In the present case, strength functions measured in 
one-neutron transfer experiments involving  $^{119}$Sn or $^{121}$Sn
\cite{exp_transf1,exp_transf2,exp_transf3} show a strong, isolated  peak for the quantum numbers
$J^{\pi} = 1/2^+,3/2^+,7/2^+,11/2^-$.
The case of $11/2^-$ is shown in Fig. \ref{fig:strength_h112}, where we compare the theoretical values of $\tilde v^2$ ($\tilde u^2)$
with the observed  peak in $^{119}$Sn ($^{121}$Sn).
Here, and in similar figures, in order to compare the experimental to the theoretical strength functions, 
we have added the  energy of the lowest calculated quasiparticle to  the experimental excitation energy in the odd nuclei
$^{119}$Sn and $^{121}$Sn. 
The summed strength of the two main peaks is equal to  0.79 in experiment
and to 0.8 in theory. The remaining theoretical strength is calculated to lie at higher energy in $^{121}$Sn. A good agreement 
between theory and experiment is found also in the case of the other three quantum numbers, as was shown in 
Fig. \ref{fig:zeta_sly4}. 

\begin{figure*}[h!]
\begin{center}
\includegraphics[width=0.7\textwidth]{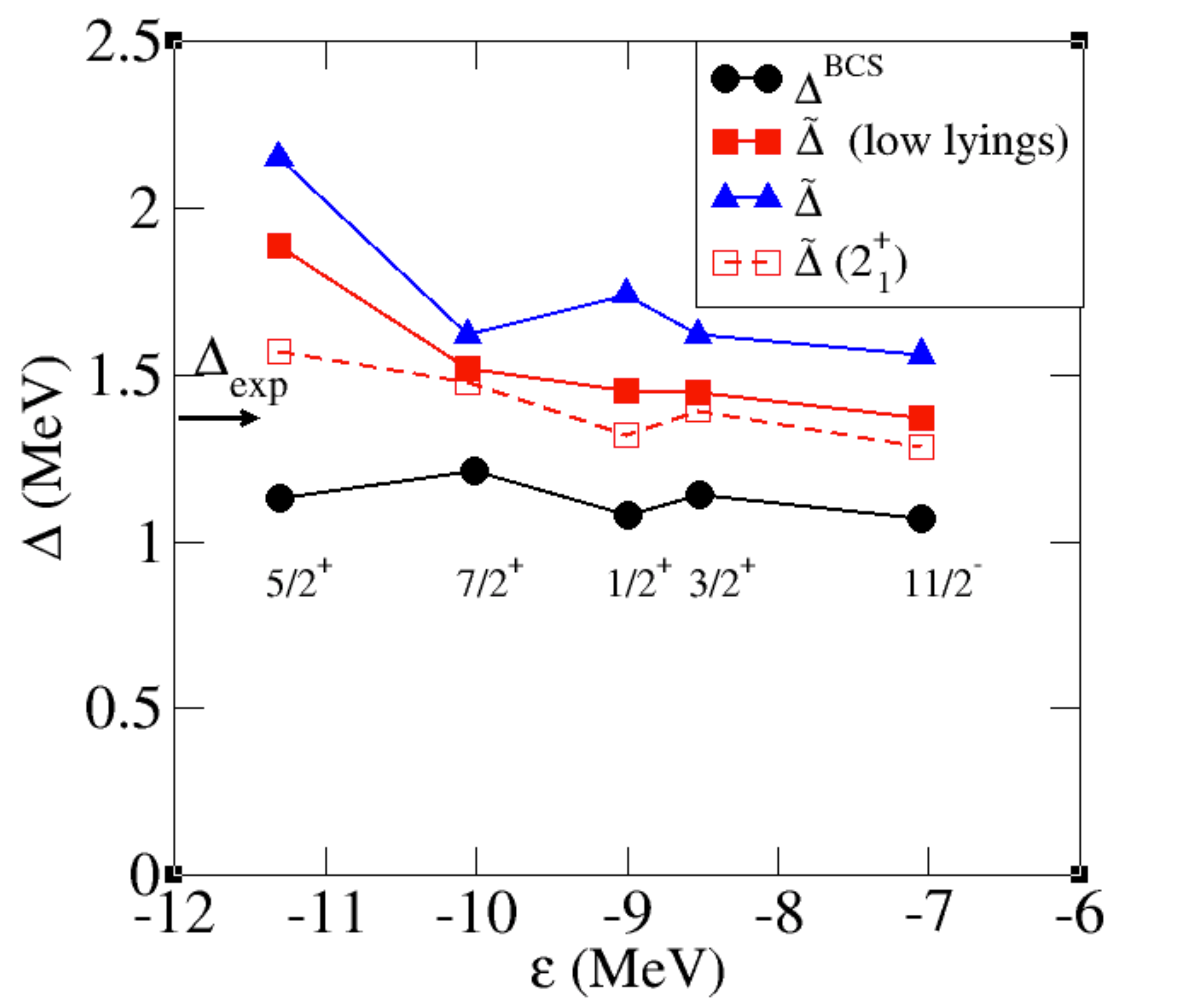}
\end{center}
\caption{ The pairing gaps $\Delta^{BCS}$
obtained in the BCS calculation (dots) and $\tilde \Delta$ including the renormalization 
effects (triangles) are compared to those obtained including only the renormalization 
due to  the coupling to the lowest  $2^+,3^-,4^+$ and $5^-$ QRPA modes (filled squares), 
or only  the lowest $2^+$ mode (open squares).}  
\label{fig:gaps_lowlying}
\end{figure*}

It is interesting to notice that  the main contribution to the renormalization effects 
originates from the coupling to the lowest vibrational state of each multipolarity : this
demonstrates  the key role played by the  
interweaving of collective and single-particle modes which is at the basis of the NFT treatment
of the elementary modes of nuclear excitation.
This is clearly seen in Fig. \ref{fig:gaps_lowlying}, where we compare the 
full calculations with results obtained including only couplings to the lowest vibrational states.
It is seen that the coupling to the $2^+$ low-lying mode is the dominant one, providing half of the increase
of the gap due to the renormalization effects. The other low-lying modes  
provide about one-third of the remaining increase of the gap, while higher energy modes 
(mainly giant resonances) complete the picture. The strength of the various PVC couplings
is reported in Table \ref{table_mat}, where for each pair of orbitals $(a,b)$ 
we list the value of the squares of the PVC matrix elements  summed over the phonons of a given multipolarity, 
$\sum_{\nu} h^2(ab\lambda\nu)/(2j_b+1)$. 
We observe that although the deformation parameter
associated with the lowest $3^-$ state is the largest one (cf. Table \ref{table_phonons}) its influence is hindered 
with respect to the $2^+$ by its higher energy and by the fact that it acts efficiently only between the
$d_{5/2}$ and the $h_{11/2}$ orbitals, which are the only ones having opposite parity and the same spin alignment.
We also note that the dominance of the coupling to low-lying vibrational states found
in the present calculation  makes our approach distinctly different from other approaches  
based on microscopic calculations of  the PVC with zero-range forces, 
which need an energy cutoff to avoid an ultraviolet divergence
in the self-energy \cite{colo:10}.

The strength of the remaining orbital, $5/2^+$, is much more fragmented both in theory and experiment,
as is shown in Fig. \ref{fig:strength_d52}.   
The experiment shows a number of low-energy peaks, which exhaust about 63\% of the strength. The theoretical
distribution is more fragmented than observed in the data; moreover most of the strength lies about 1 MeV
above the experimental findings. 
In the calculation, the energy of the $d_{5/2}$ quasiparticle is lowered by its strong coupling with 
$s_{1/2}$ state (through the $2^+$ vibrations) and with the $h_{11/2}$ state (through the $3^-$ vibrations),
as can be seen in Table \ref{table_mat}. 
These couplings bring the energy of the $d_{5/2}$ state 
close to  the energies of the $s_{1/2} \otimes 2^+_1$ and $d_{3/2} \otimes 2^+_1$ configurations.
The matrix elements of the induced interaction $V_{ind} $ (cf. Eq. (\ref{VIND_prior})) are listed 
in Table \ref{table_vind}. They are calculated using the values of the renormalized quasiparticle energies
$\tilde E_a(1)$ of the lowest fragment for each orbital. 
Their values can be compared to the typical value of the matrix
elements of the bare interaction, which is about $G_0 = - 0.22$ MeV (cf. below Section \ref{priorpost}).  
They are rather well correlated with the PVC matrix elements reported in Table \ref{table_mat}, with the 
remarkable exception of $d_{5/2}$: in the latter case, the induced interaction with the orbitals
$s_{1/2}$ and $d_{3/2}$ takes large values, associated with the almost degeneracy of the energy of 
the $d_{5/2}$ quasiparticle with the configurations $s_{1/2}\otimes 2^+_1$ and $d_{3/2}\otimes 2^+_1$,
which leads to the fragmentation of the $d_{5/2}$ strength.

These results may indicate that the  unperturbed quasiparticle energy of the 
$d_{5/2}$ orbital lies too high. In fact, moving the energy of the single-particle energy by 600 keV
towards the Fermi energy,
and leaving everything else unchanged, and then solving the BCS and Nambu-Gor'kov equations, one  
obtains a  better agreement with experiment, as will be shown below in Section \ref{Hindsight} 
(cf. Fig. \ref{fig:hindsight}(d)).
This shows that a detailed study of the fragmentation process can give important indications about how to improve
the mean field (cf. also Section \ref{meanfield}).

\begin{table}
\begin{center}
\begin{tabular}{|c|c|c|c|c|c|}
\hline
                          &$d_{5/2}$&$g_{7/2}$&$s_{1/2}$&$d_{3/2}$&$h_{11/2}$
\\
\hline
\multirow{2}{*}{$d_{5/2}$} &  $0.190 $ &  $0.016$ & $0.467$ & $0.071$ &
$0.340$   \\
\cline{2-6}
                          &  $0.088$ &  $0.041$ & $0.$     & $0.266$ &
$0.138$   \\
\hline
\hline
\multirow{2}{*}{$g_{7/2}$} & $0.016$ &  $0.130$ & $0.$     & $0.236$ &
$0.020$   \\
\cline{2-6}
                          & $0.041$ &  $0.071$ & $0.168$   & $0.081$ &
$0.043$   \\
\hline
\hline
\multirow{2}{*}{$s_{1/2}$} & $0.467$ &  $0.$    & $0.$     & $0.461$  & $0.$   \\
\cline{2-6}
                          &  $0.$ &  $0.168$ & $0.$     & $0.$     &
$0.389$   \\
\hline
\hline
\multirow{2}{*}{$d_{3/2}$} & $0.071$  &  $0.236$& $0.461$ & $0.253$  & $0.$   \\
\cline{2-6}
                          &  $0.266$ &  $0.081$& $0.$     & $0.$     &
$0.092$   \\
\hline
\hline
\multirow{2}{*}{$h_{11/2}$}& $0.340$ &  $0.020$& $0$     & $0.$     &
$0.169$   \\
\cline{2-6}
                          &  $0.138$&  $0.043$& $0.389$  & $0.092$  &
$0.100$   \\
\hline
\hline
\end{tabular}
\caption{\protect For each pair $(a,b)$  of the five valence orbitals, we list the value of 
the sum $\sum_{\nu} h^2(ab\lambda\nu)/(2j_b+1)$, in MeV$^2$. 
For pairs of orbitals of the same parity, the numbers in the upper (lower)
row give the contribution associated with $\lambda= 2^+$ ($\lambda= 4^+$) phonons; for pairs of orbital of different parity,
the numbers in the upper (lower) row give the contribution associated with  $\lambda= 3^-$ ($ \lambda = 5^-$) phonons.}
\label{table_mat}
\end{center}
\end{table}

\begin{table}
\begin{center}
\begin{tabular}{|c|c|c|c|c|c|}
\hline
                          &$d_{5/2}$&$g_{7/2}$&$s_{1/2}$&$d_{3/2}$&$h_{11/2}$
\\
\hline
\multirow{2}{*}{$d_{5/2}$} &  $-0.223 $ &  $-0.031$ & $-1.701$ & $-1.230$ &
$-0.309$   \\
\cline{2-6}
                          &  $-0.054$ &  $-0.030$ & $0.$     & $-0.245$ &
$-0.075$   \\
\hline
\hline
\multirow{2}{*}{$g_{7/2}$} & $-0.015$ &  $-0.157$ & $0.$     & $-0.513$ &
$-0.016$   \\
\cline{2-6}
                          & $-0.023$ &  $-0.045$ & $-0.117$   & $-0.062$ &
$-0.021$   \\
\hline
\hline
\multirow{2}{*}{$s_{1/2}$} & $-0.383$ &  $0.$    & $0.$     & $-0.686$  & $0.$   \\
\cline{2-6}
                          &  $0.$ &  $-0.100$ & $0.$     & $0.$     &
$-0.177$   \\
\hline
\hline
\multirow{2}{*}{$d_{3/2}$} & $-0.055$  &  $-0.219$& $-0.503$ & $-0.319$  & $0.$   \\
\cline{2-6}
                          &  $-0.139$ &  $-0.047$& $0.$     & $0.$     &
$-0.041$   \\
\hline
\hline
\multirow{2}{*}{$h_{11/2}$}& $-0.183$ &  $-0.012$& $0$     & $0.$     &
$-0.212$   \\
\cline{2-6}
                          &  $-0.051$&  $-0.017$& $-0.167$  & $-0.042$  &
$-0.067$   \\
\hline
\hline
\end{tabular}
\caption{\protect Matrix elements of the induced interaction $V_{ind(a(1)b(1))}$ (cf. Eq. (\ref{VIND_prior})) between the
lowest fragments of the five valence orbitals, in MeV. For pairs of orbitals for the same parity, the numbers in the upper (lower)
row give the contribution of the matrix elements involving  $2^+$ ($4^+$) phonons; for pairs of orbital of different parity,
the numbers in the upper (lower) row give the contribution of the matrix elements involving $3^-$ ($5^-$) phonons.}
\label{table_vind}
\end{center}
\end{table}


\clearpage

\subsubsection{\label{effspin} Effects of  spin modes}

\begin{figure*}[t!]
\begin{center}
\includegraphics[width=0.8\textwidth]{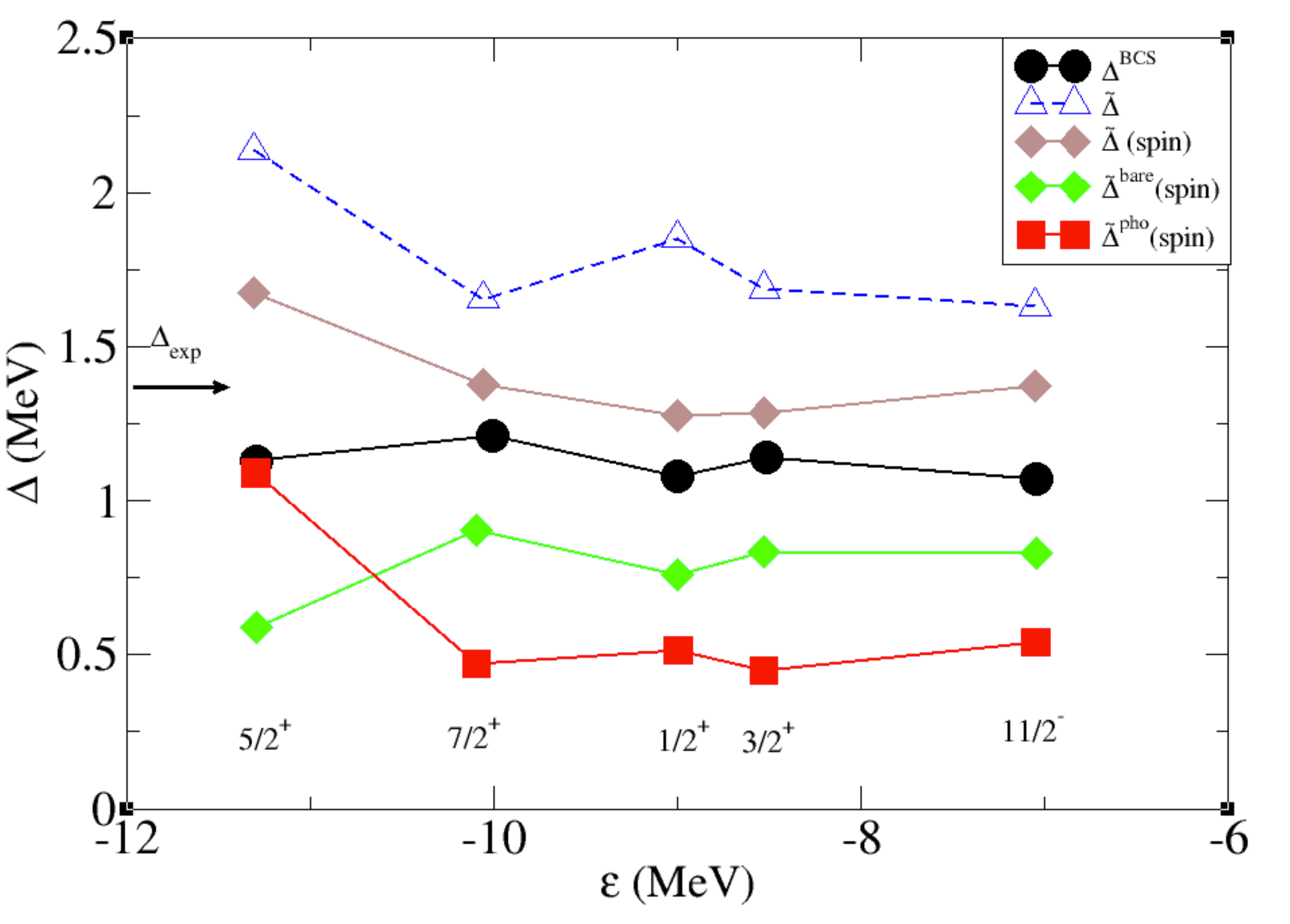}
\end{center}
\caption{ The BCS (dots) and renormalized pairing gaps associated with the 
lowest quasiparticle peaks obtained solving the Nambu-Gor'kov equations (triangles, cf. Fig. \ref{fig:renogap_sly4})
are compared to the gaps obtained including the schematic spin-induced interaction discussed in the text (diamonds),
and to their decomposition into $\tilde \Delta^{bare}$ and $\tilde \Delta^{pho}$.} 
\label{fig:gaps_spin}
\end{figure*}

\begin{figure*}[b!]
\begin{center}
\includegraphics[width=0.8\textwidth]{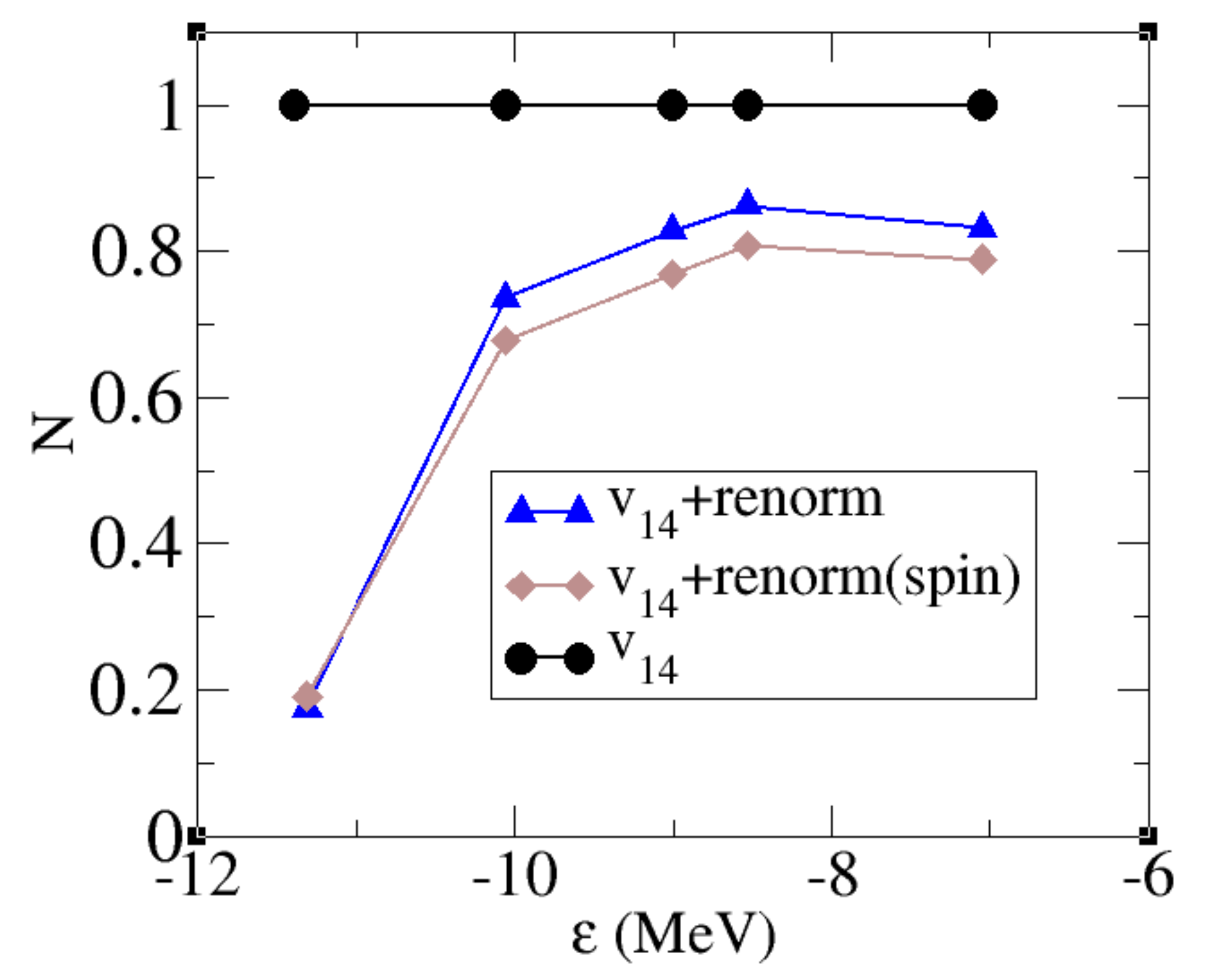}
\end{center}
\caption{ The strength $N$ associated with the main quasi particle peaks in the bare HF+BCS calculation (dots) and 
solving the Nambu-Gor'kov equations (triangles) is compared to the result obtained 
including the schematic spin-induced interaction discussed in the text (diamonds).}
\label{fig:zeta_spin}
\end{figure*}


It is interesting to speculate about the effect of $S=1$ modes, that should be included in a more complete calculation.
Their contribution represents the leading renormalization effect in calculations in uniform neutron matter, where  it induces a
repulsive contribution on the pairing interaction that wins over the attractive contribution associated with density modes - thanks to the
larger multiplicity of spin modes -  leading to
a quenching of the  pairing gap \cite{chen}-\cite{viverit}. On the other hand, one expects that the relative weight of  spin modes should be considerably less
important in finite nuclei, where there is no spin equivalent to low-lying surface collective vibrations.

The repulsive character of spin modes, as opposed to density modes,  arises from their different transformation properties  
under time reversal. This is reflected in a sign change in the basic  $V,W$ vertices 
(cf. Eq. (6-207) in \cite{BM:75}),
which become (cf. Eq. (\ref{eq:VW})) 
\begin{eqnarray}
& V(ab(m)\lambda\nu) =& h_{S=1} (ab\lambda\nu)(u_a \tilde u_{b,m} + v_a \tilde v_{b,m})\cr
& W(ab(m)\lambda\nu) =&h_{S=1} (ab\lambda\nu)(-u_a \tilde v_{b,m} + v_a \tilde u_{b,m}).
\label{eq:VWspin}
\end{eqnarray}
As a consequence. the product $VW$ which determines the abnormal self-energy $\Sigma^{12pho}$ (cf. Eq. (\ref{Sig12prior}))  also changes its sign, 
leading to a repulsive contribution. On the other hand the normal self-energies depend on the square of $V$ and $W$ so that
their value is increased by the action of spin modes.
There is furthermore a change in the expression of the basic vertex $h$, which is given by 
\begin{equation}
h_{S=1}(abJ\nu) =-(-1)^{j_a-j_b} \beta_{J\nu} <a|
f_{J}(r)|b><j_b||(Y_{\lambda}\sigma)_{J}||j_a> 
\left[\frac{1}{(2j_a+1)(2J+1)}\right]^{1/2},
\label{hvertexS1}
\end{equation}
where $ f_{J}(r)$ is the form factor associated to the S=1 modes
transition density.


 The  contribution of spin modes to pairing correlations in $^{120}$Sn  was estimated in ref. \cite{Gori:05}, based on a QRPA calculation
performed with the SkM$^*$ interaction. There it was found that, performing a calculation of the pairing gap  including only 
the induced interaction, using the simplified expression discussed in Section \ref{quasiparticle} (cf. Eq. (\ref{VIND_vecchia})), the spin modes
decreased  the gap arising from the coupling with  $S=0$ modes by about 25\%.
We have incorporated this effect in our present calculation by introducing  a schematic $S=1$ response function, consisting of a single
peak at an energy of about $ 1 \hbar \omega_{osc}$, in keeping with the fact that there are no  low-lying collective $S=1$ modes in finite nuclei.
We have used an average value for the product $\beta_{J\nu} $  $<a| f_{J}(r)|b>$, treating it as an 
adjustable parameter in order to reproduce the 25\% reduction in $\tilde \Sigma^{12pho}$ mentioned above,
when no bare interaction is included.
In this way we can estimate the effects of $S=1$ modes on the total gap and on the quasiparticle energies and strengths calculated in the present paper.

The values obtained for $\tilde \Delta$ and for the quasiparticle strength $N$ are shown respectively 
in Figs. \ref{fig:gaps_spin} and \ref{fig:zeta_spin}. In keeping with their contribution to the normal self-energy,
the action of spin modes tends to reduce, although slightly,
the value of $Z$ and $N$, not modifying significantly  the agreement  with experiment.
Spin modes also lead to a reduction of $\tilde \Delta^{pho}$ by about 25\% (150 keV), which reflects directly the
reduction introduced  in $\tilde \Sigma^{12pho}$, while the value of $\tilde \Delta^{bare}$ is practically unaffected.
Altogether, this brings the  value of $\tilde \Delta$ closer to the experimental value of $ \Delta_{exp}$= 1.4 MeV,
without deteriorating the agreement with the experimental quasiparticle strength.


\clearpage

\subsubsection {\label{quasiparticle} One-quasiparticle approximation and the gap equation}

\begin{figure*}[h!]
\begin{center}
\includegraphics[width=0.6\textwidth]{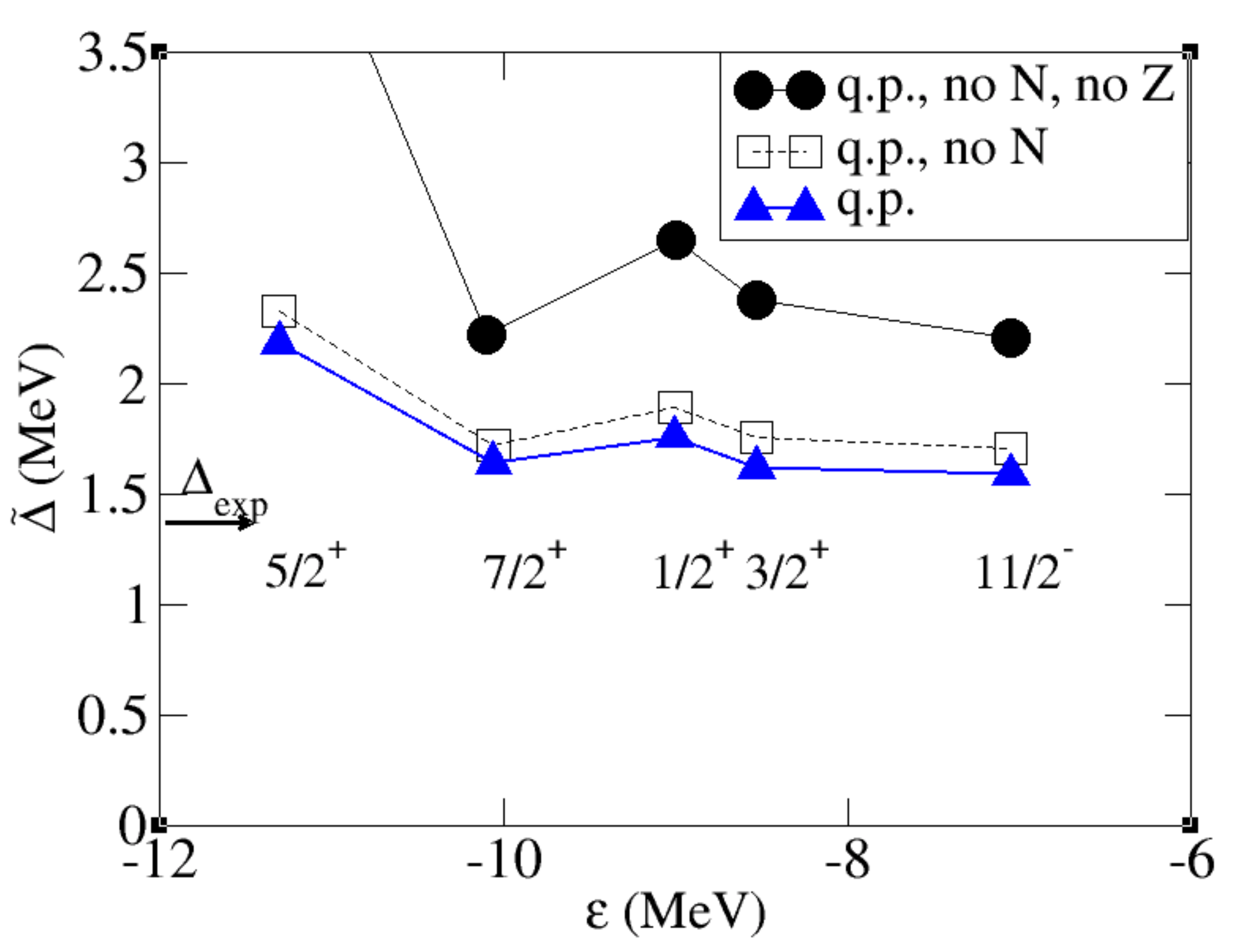}
\end{center}
\caption{ The pairing gaps obtained in the quasiparticle calculation (triangles) are compared to the 
results obtained neglecting the normalization factors $N$, and  neglecting both $N$ and $Z$.} 
\label{fig:gapvsesp1qp}
\end{figure*}

We have seen that the quasiparticle strength function for the 
states close to the Fermi energy displays a limited amount of fragmentation. 
If one is only interested
in the properties of the main quasiparticle peaks, it may be  interesting to solve the 
Nambu-Gor'kov problem by keeping only the main quasiparticle for each orbital
in the iteration process. 
One then gets an excellent agreement with the complete calculation for the quasiparticle 
gap $\tilde \Delta_a^{q.p.}= Z_a \tilde \Sigma_a^{12,q.p.}$, as shown in Fig. \ref{fig:gapvsesp1qp}:
the result cannot be distinguished from the calculation for $\tilde \Delta$  shown in Fig. \ref{fig:renogap_sly4}.

In order to have more insight on these results, we recall the gap equation Eq.(\ref{delta_noi}):
\begin{equation}
\tilde \Sigma^{12}_{a(n)}= \Delta_{a}^{BCS}
 -  \sum_{b(m)}V_{ind}(anbm) N_{b(m)} 
\frac{\tilde \Sigma^{12}_{b(m)}} {2\sqrt{(\epsilon_{b}-e_F+\tilde \Sigma^{even}_{b(m)})^{2}+(\tilde \Sigma^{12}_{b(m)})^2 }}.
\label{delta_noi1}
\end{equation}

While the implementation of the quasiparticle approximation is 
straightforward, it is important to notice that  in order to get good  agreement one must include 
the factors $N$ and $Z$ associated with  the quasiparticle peaks. Neglecting them, that is,
attributing the full quasiparticle strength to the single  peak retained in the calculation,  leads to
an increase of the gap by about 50\%, in keeping with the typical values $Z \approx 0.7$, $N \approx 0.8$.
Neglecting $N$ produces a  smaller effect than neglecting $Z$, 
because $N$ affects only the induced part of the gap 
in Eq. (\ref{delta_noi1}). The error is especially pronounced for the case of $d_{5/2}$, where the 
one-quasiparticle approximation selects the lowest peak, which is associated with a small value of $N$ and $Z$ 
(cf. Fig. \ref{fig:zeta_sly4}). From this result, one concludes that the gap receives only very small contributions from the fragments 
with $ m > 1$, which, on the other hand, account for more than 20\% of the quasiparticle strength. This is mostly due to 
the energy dependence of $V_{ind}$, which is strongly peaked at the Fermi energy 
\cite{noi_schuck}.


A gap equation  equivalent to Eq. (\ref{delta_noi1})  has been obtained in the post scheme (cf. Eq. (\ref{delta_noi2})): 
\begin{equation}
\Delta_{a(n)}=   - Z_{a(n)} \sum_{b(m)} V_{eff}(a(n)b(m)) N_{b(m)} 
\frac{\Delta_{b(m)}} {2 \tilde E_{b(m)}}. 
\label{delta_noi2_bis}
\end{equation}
This expression  lends itself to approximations, which are useful  in order to make a connection with previous  works 
about renormalization effects on pairing.
The most important feature of these approximations
is the use of simplified expressions for the induced interaction $V_{ind}$ (cf. Eq. (\ref{VIND_prior})),
\begin{eqnarray}
&V_{ind}(a(n)b(m)) = \sum_{\lambda,\nu}\frac{2 h^2(ab\lambda\nu)}{(2j_b+1)} \times & \cr
&\left[ \frac{1}{\tilde E_a(n)-\tilde E_b(m)-\hbar\omega_{\lambda\nu}}- \frac{1}{\tilde E_a(n)+\tilde E_b(m)+\hbar\omega_{\lambda\nu}} \right], &
\label{VIND_nuova}
\end{eqnarray}
based on the Feynman diagram (1c) in Fig. \ref{fig:fig1_diagrams} and neglecting fragmentation:
\begin{equation}
V_{ind}(ab) \approx \sum_{\lambda,\nu}\frac{2 h^2(ab\lambda\nu)}{(2j_b+1)}
 \frac{2}{E_0 - \tilde E_a -\tilde E_b -\hbar\omega_{\lambda\nu}},
\label{VIND_vecchia}
\end{equation}
in which $ \tilde E_a = | \epsilon_a - \epsilon_F| $ and $E_0$ is the pairing  energy per Cooper pair (equal to about - $ 2 \Delta$) \cite{prl_noi}.
This expression was inserted in the BCS  gap equation, without taking explicitly into account the renormalization
effects on the single-particle density, namely, setting the $N$- and $Z-$ factors equal to one, and using a mean field potential
characterized by an effective mass $m^* = m$. One then found gaps about 20\%
larger than those obtained solving the Nambu-Gor'kov equations \cite{Ter:02},\cite{Ter:navi_spaziali}.
The simplified expression for $V_{ind}$ was also employed in a more elaborate calculation \cite{pastore:08} 
which used HF level from the SLy4 interaction and a constant value $N = 0.7$. 
However, $\tilde \Sigma_{12}$ was identified with the pairing gap, which then 
tended to be overestimated by a factor $1/Z$.

\subsection{\label{priorpost} Comparison of prior and post scheme with the monopole pairing force}

We have already noticed that solving the Nambu-Gor'kov equations in the post scheme 
is numerically very demanding in the case of realistic nucleon-nucleon interactions. 
In this Section we want to compare the prior and the post schemes using 
a simple 
monopole pairing force - $G_0 P^{\dagger}P$ acting only in our valence space.
The present calculation is similar to that presented in ref. \cite{Ter:02}, except for details in the mean field (here obtained from a SLy4 
interaction instead of a Woods Saxon potential with effective mass $m^*$= 0.7), in the QRPA spectrum and in the determination of the 
PVC coupling.  We shall however present new results which will extend our previous investigation.

\begin{figure*}[t!]
\begin{center}
\includegraphics[width=0.6\textwidth]{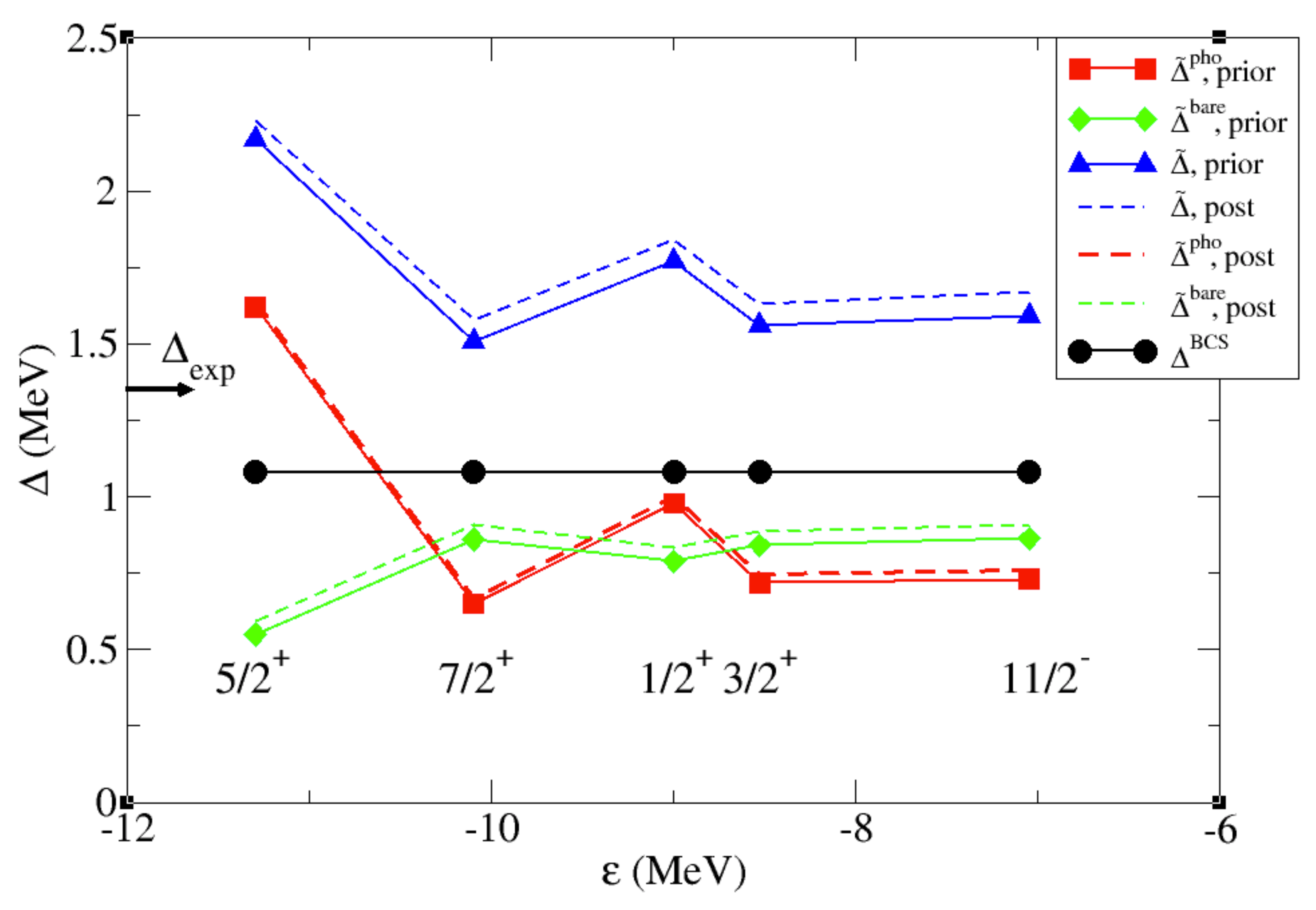}
\end{center}
\caption{Comparison of the renormalized  pairing gaps $\tilde \Delta$  associated with the 
lowest energy fragments, obtained solving the Nambu-Gor'kov equations
in the prior (solid line with triangles) and in the post (dashed line) scheme. 
We also show the bare ($\tilde \Delta^{bare}$) and the phonon 
($\tilde \Delta^{pho}$) components of the renormalized 
gaps in the prior (solid lines with diamonds and squares) and in the post (dashed lines) scheme.
The pairing interaction is a monopole force with coupling  constant $G_0= 0.22$ MeV, acting in the valence shell
around the Fermi energy in $^{120}$Sn, which produces the gap  $\Delta^{BCS}$ (solid lines with dots) obtained in the BCS calculation. 
The symbols refer to the position of the various valence 
orbitals in the SLy4 HF potential. 
The value of the gap $\Delta_{exp}$ obtained from the experimental odd-even mass difference is also indicated.} 
\label{fig:delta_g022}
\end{figure*}

\begin{figure*}[h!]
\begin{center}
\includegraphics[width=0.6\textwidth]{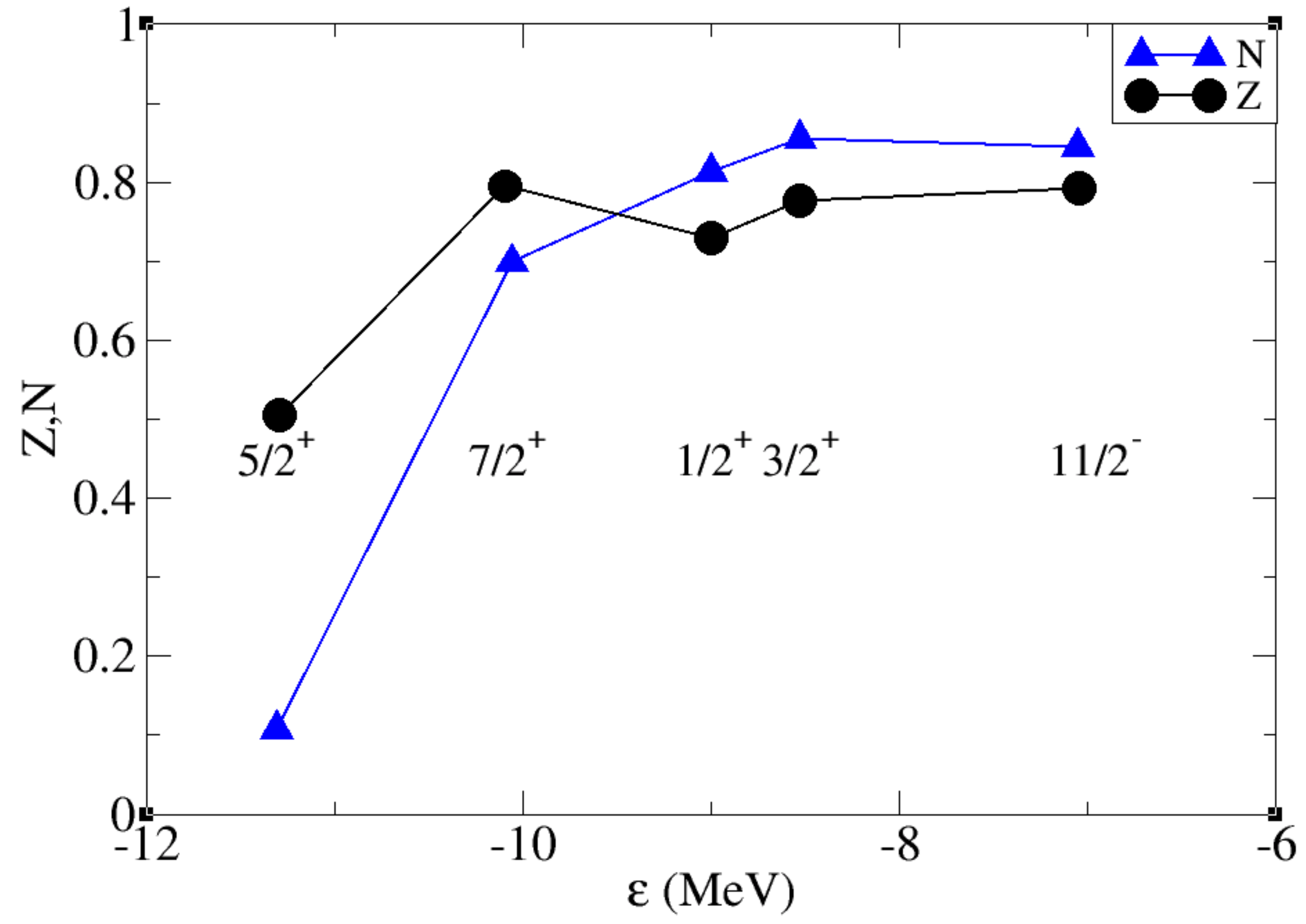}
\end{center}
\caption{$Z-$ and $N-$factors associated with the various valence orbitals, for the calculation
presented in Fig. \ref{fig:delta_g022}. Prior and post values are indistinguishable at the scale of the 
figure.}
\label{fig:zeta_g022}
\end{figure*}

\begin{figure*}[b!]
\begin{center}
\includegraphics[width=0.6\textwidth]{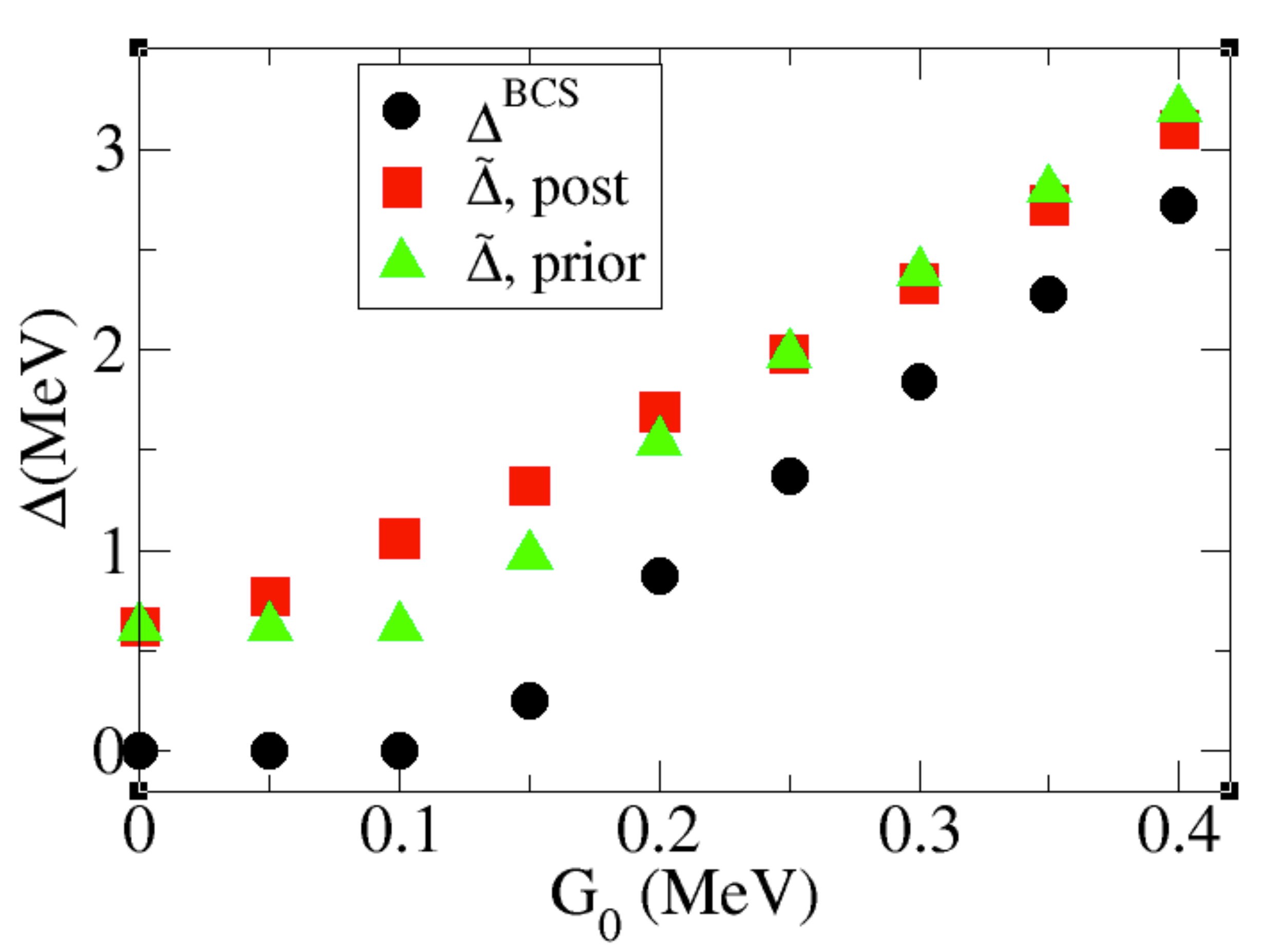}
\end{center}
\caption{ Renormalized gaps $\tilde \Delta$ obtained solving the Nambu Gor'kov equations in the prior
and in the post scheme with the monopole pairing force as a function of the pairing constant $G_0$, averaged over the five valence orbitals. 
Also shown is the gap $\Delta^{BCS}$ obtained solving the BCS equation.} 
\label{fig:deltavsg}
\end{figure*}

In Fig. \ref{fig:delta_g022} we compare the results obtained in the prior and in the post  schemes, for a value $G_0$ = 0.22 MeV, corresponding to a bare pairing gap
$\tilde \Sigma^{12bare} = $ 1.08 MeV.
We have chosen this particular value of $G_0$ in order to reproduce the average value of the gap $\Delta^{BCS}$ 
obtained previously with the $v_{14}$ interaction (cf. Fig. \ref{fig:renogap_sly4}).
 We see that the renormalized pairing gaps are very similar in the
prior and post schemes. Moreover the results are also similar to those obtained with the Argonne interaction, shown in Fig. \ref{fig:renogap_sly4},
not only for the total gap $\tilde \Delta$ but also concerning the decomposition into bare and phonon contribution.
The similarity extends also to the $Z-$ and $N-$factors, shown in Fig. \ref{fig:zeta_g022} (cf. Fig. \ref{fig:zeta_sly4}).
The fact that the results turn out to be 
close to those obtained using the Argonne interaction is easy to understand
in the prior scheme, which is based on the quasiparticle and occupation amplitudes obtained with the bare interaction. 
In fact, the realistic nucleon-nucleon interaction displays a limited state dependence over the valence orbitals,
 with an average value $\Delta^{BCS} =1.1 $ MeV,  very close to the constant
BCS gap obtained with the monopole interaction  with $G_0= 0.22$ MeV.

In Fig. \ref{fig:deltavsg} we show the value of the gap, averaged over the five valence orbitals, as a function of
the pairing strength $G_0$. 
In the prior scheme, one starts from the $u_a,v_a$ amplitudes obtained in a previous BCS calculation with the bare force, 
that for $G_0$ smaller than 0.1 MeV, produces  no superfluid solution. On the other hand, the induced interaction
$V_{ind}$ is able to produce a pairing gap by itself. As a consequence,  in the prior scheme
only $\tilde \Sigma^{12}_{phon}$ contributes to the gap, which is therefore underestimated and independent of $G_0$
for $G_0 < 0.1$.
Instead, using the 
post scheme, the gap grows as a function of $G_0$ since the bare  interaction 
can provide a contribution to pairing correlations even for small values of $G_0$
when it is added to the induced interaction
through the effective interaction $V_{eff} = V_{bare} + V_{ind}$. 
For values of $G_0$ larger than 0.1 MeV, the gaps in the prior and post scheme 
become closer and closer, until the prior scheme becomes 
satisfactory for values of the order of $G_0$ = 0.2 MeV.  It is interesting to notice that the calculations for the
bare and renormalized gaps  performed with $G_0 = 0.25$ MeV reproduce quite well the results  
obtained using the $V_{{\rm low} k}$ interaction as bare pairing force, as
discussed in Section \ref{barepair}. This indicates that Fig. \ref{fig:deltavsg} 
can be used to assess in a simple  way the  effect of renormalization processes on pairing correlations,
for the adopted QRPA spectrum.



\section{\label{conclusions} Conclusions}



We have presented a convenient  formalism to deal with the basic renormalization processes
induced by the coupling between quasiparticle and collecive vibrations
in superfluid spherical nuclei. We have solved the  Nambu-Gor'kov equations
determining the normal and abnormal energy-dependent self-energies self-consistently. 
This allows a detailed calculation of the low-energy part of the nuclear spectrum in odd nuclei taking into
account the fragmentation of the quasiparticle strength, as well as the calculation of the pairing gap
of the system including the pairing interaction induced by the exchange of collective modes.
The mean field is based on a Hartree-Fock calculation with the effective SLy4 interaction, while the coupling
between quasiparticles and vibrations is determined by a QRPA calculation that reproduces the empirical polarizability of the 
low-lying 
vibrational modes. In the  solution of Nambu-Gor'kov equations there are then no free parameters, when
we use the bare $v_{14}$ Argonne nucleon-nucleon potential as bare pairing force.
We have  obtained a gap equation (cf. Eq. (\ref{delta_noi})) which takes into account renormalization
effects in a compact way, and allows one to make contact with previous studies. 

We have  
discussed in detail the practical implementation of the formalism, 
introducing two  calculational schemes,
depending on whether the bare pairing interaction and the renormalization effects are taken into
account simultaneously (post scheme) or in two steps (prior scheme), performing first a BCS calculation
with the bare force. 
We have shown that the two schemes lead to similar results for states close to Fermi energy 
in the case of a simple monopole pairing force with a realistic coupling strength.
The prior scheme is computationally much simpler
in the case of realistic nucleon-nucleon forces which can scatter particles to very high energies. 

We have  presented results for 
the semimagic nucleus $^{120}$Sn, which are  in rather good agreement with those extracted 
from one-neutron transfer reactions. 
The formalism allows also for a clear separation between the contribution 
of the bare force and of renormalization effects: their contributions  to the pairing gap turn out to
be of similar magnitude, confirming previous studies.

As we have already summarized in Section \ref{Hindsight}, we get an overall agreement with the experimental
spectra. This agreement appears to be stable, respect to reasonable changes in the ingredients of our
calculations (cf. the Appendix), and the contribution of the pairing interaction induced by surface vibration   
appears to be well established within our scheme.

Several elements remain to be further investigated, before one can reach firm quantitative  theoretical results,
in particular concerning the absolute value of the total pairing gap.
On the one hand, the mean field should be either derived microscopically, or at least refitted comparing theory and experiment
taking into account renormalization effects; on the other hand, contributions which are expected to provide a repulsive contribution to the 
pairing interaction, associated with three-body effects and with the influence of spin modes, have started only 
recently to be examined.


\subsection{\label{Hindsight} Hindsight}

\begin{figure*}[h!]
\centering
\includegraphics*[width=0.4\textwidth]{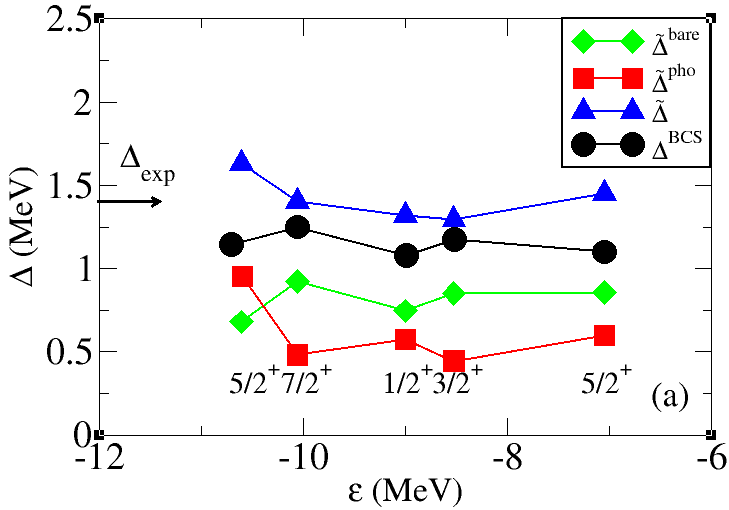}
\includegraphics*[width=0.4\textwidth]{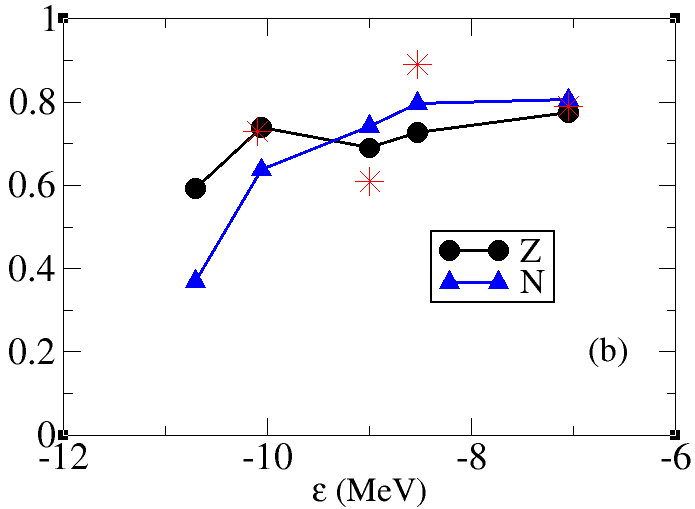}
\vspace{3mm}
\includegraphics*[width=0.4\textwidth]{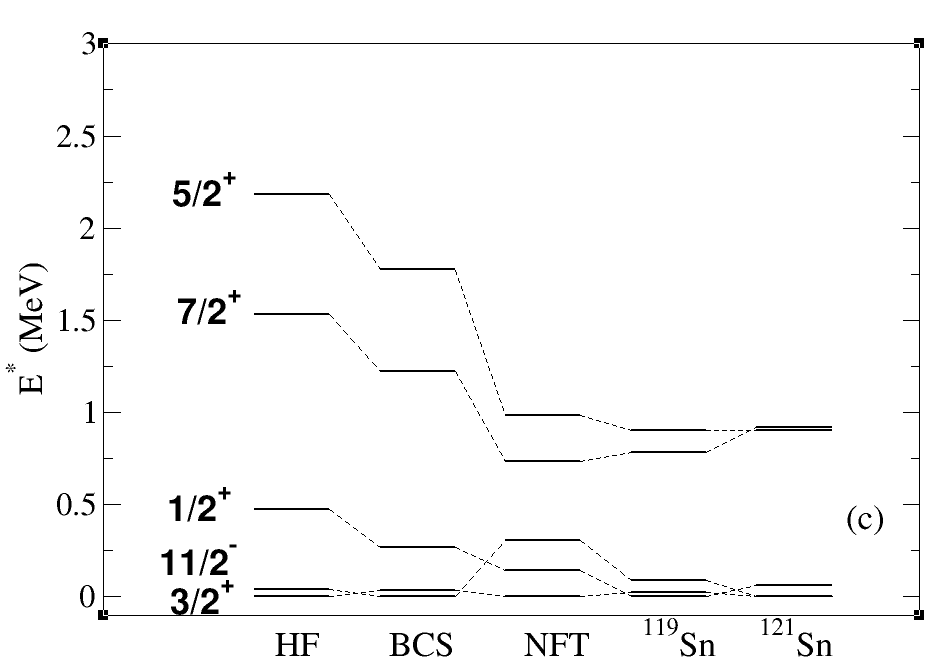}
\includegraphics*[width=0.4\textwidth]{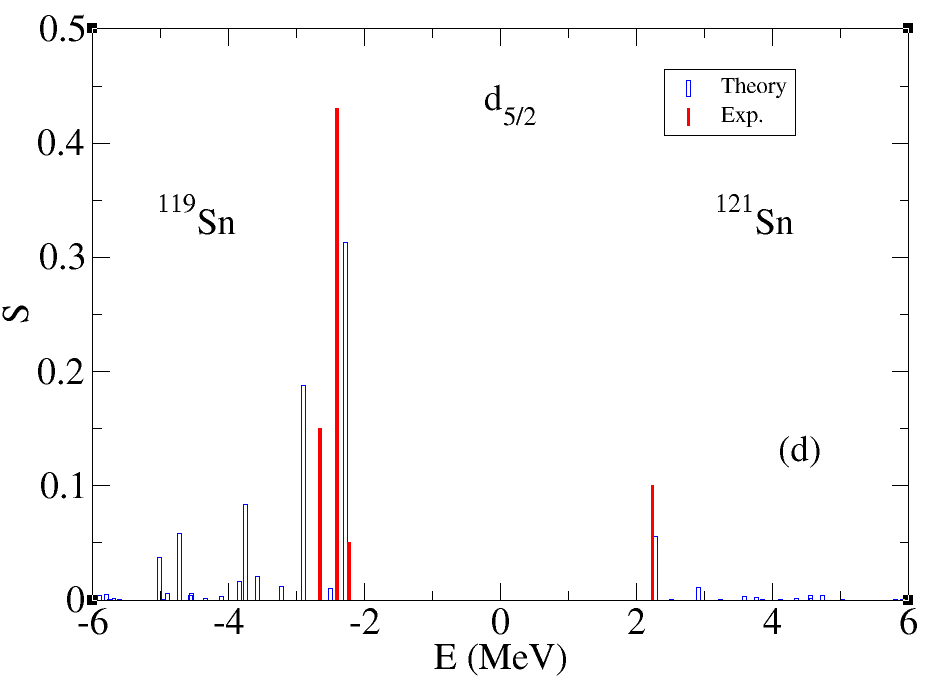}
\caption{ Results obtained solving the Nambu-Gor'kov equations including PVC to spin modes with the SLy4 mean field,
shifting the energy of the $d_{5/2}$ orbital by 600 keV towards the Fermi energy:
(a) renormalized gaps, (b) $Z$- and $N-$ factors, (c) quasiparticle spectrum, (d)  
$5/2^+$ strength function.}
\label{fig:hindsight}
\end{figure*}

\begin{figure*}[t!]
\begin{center}
\includegraphics[width=0.7\textwidth]{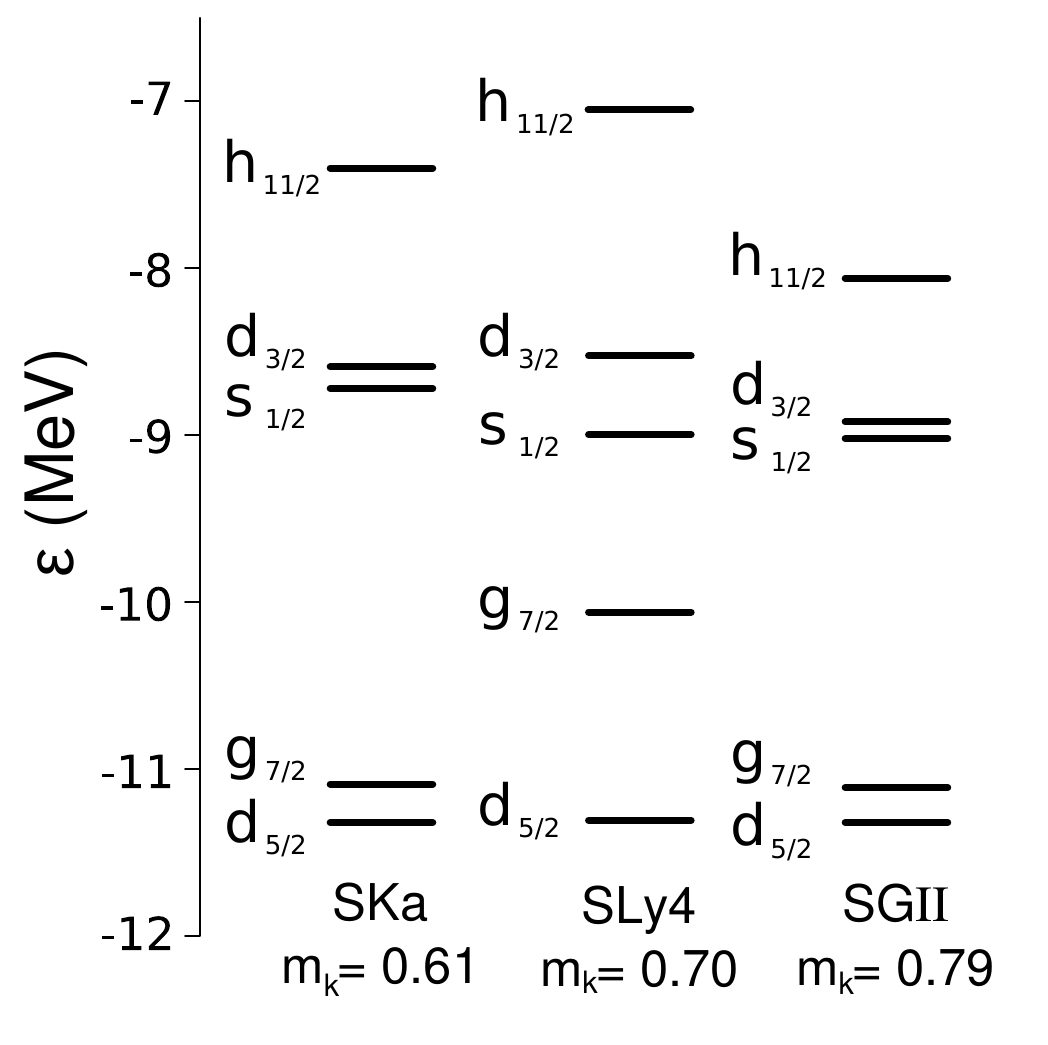}
\end{center}
\caption{ Energies of the five HF single-particle levels lying close to the Fermi energy in $^{120}$Sn, calculated 
with the effective interactions SKa, SLy4 and SGII. The effective mass associated with these forces at saturation density 
is also indicated.} 
\label{spart_levels}
\end{figure*}

One can conclude that it is possible, at least in the case of 
$^{120}$Sn, to draw a picture that is consistent with the available experimental data concerning the quasiparticle
properties close to Fermi energy. This is  summarized in Fig. \ref{fig:hindsight}, which shows the results of a 
calculation with the bare pairing interaction, including all the elements discussed in the previous sections.

The spectroscopic factors and the induced pairing gap  are not much affected by the details of the 
mean field or of the effective mass associated with the effective forces we have examined.
These quantities are  essentially determined by the interweaving with 
low-lying collective surface vibrations, 
which shifts the  quasiparticle energies, increasing the level density close to the Fermi energy, and
leads to spectroscopic factors ($Z-$ and $N-$ factors) in the range 0.6-0.8 (cf. Fig.\ref{fig:hindsight} (b))   
and to an induced  pairing gap $\Delta^{pho}$ of about 0.8 MeV.
The coupling of spin modes, due to their higher energy and smaller collectivity, does not affect much the spectroscopic results. It reduces the induced pairing gap by about 0.2 MeV, leading to a final induced pairing gap
$\Delta^{pho}$  close to  0.6 MeV (cf. Fig. \ref{fig:hindsight} (a)). 

The calculated strength function of fragmented levels (the $d_{5/2}$ in the present case) depends on the detailed
position of the levels in the mean field potential,   and the comparison with experiment 
provides important information on it.  In the present case, we have 
explicitly verified that it is possible to reach a good overall agreement between theoretical and experimental spectra,
both for the fragmented $5/2^+$ strength and for the other valence orbitals for which the quasiparticle approximation is valid,  
simply by adjusting the position of the $d_{5/2}$ orbital in the SLy4 HF potential (cf. Figs. \ref{fig:hindsight} (c),(d)) .

We also find good agreement between the total pairing gap $\tilde \Delta$ and  the value 
derived from the experimental odd-even mass difference, starting from the value $\Delta^{BCS}$ = 1.1 MeV obtained
with the value of $m_k$ associated with the SLy4 interaction (cf. Fig. \ref{fig:hindsight} (a)). 
In fact $\tilde \Delta = \tilde \Delta^{pho} + \Delta^{BCS} \times 
Z = $ 0.6 MeV + 0.8 MeV = 1.4 MeV.

We remark  that microscopic pairing forces lead to a weak state dependence for the orbitals close to the Fermi energy, 
so that in this region we obtain essentially the same results - both at the BCS level and after the renormalization process -  
with a simple monopole interaction, using a suitable value of the pairing constant.
Furthermore, different values of $\Delta^{BCS}$,
associated with different momentum dependences of $m_k$ far from the Fermi energy, 
essentially lead to a shift of the final gap $\tilde \Delta$, but do not alter significantly the spectroscopic 
results shown in Fig. \ref{fig:hindsight}(b-d) (cf. Appendix \ref{meanfield}).
  
Unfortunately, a conclusive theoretical calculation of $m_k$ is not yet available. Furthermore, one should add
the effects of the three-body force, which is expected to lead to a repulsive contribution to the 
pairing interaction.    Interestingly, a recent calculation of  $\Delta^{BCS}$ using the $V_{{\rm low} k} $ pairing  interaction 
and including three-body effects within the renormalization group approach, has produced a value for $\Delta^{BCS}$ close to
1.1 MeV in $^{120}$Sn \cite{duguet_three,duguet_threea}.

\clearpage

\section{Acknowledgment}
Most of the work reported in this paper has been carried out in close collaboration with R.A. Broglia. 
F.B. acknowledges financial support from the Ministry of Science and Innovation of Spain grants FPA2009-07653 and ACI2009-1056. 
\clearpage

\section {Appendix}

In this Appendix  we consider the sensitivity of our results to some of the prescriptions and ingredients of the calculations.

\subsection{\label{meanfield} Mean field}

\begin{figure*}[ht!]
\begin{center}
\includegraphics[width=0.7\textwidth]{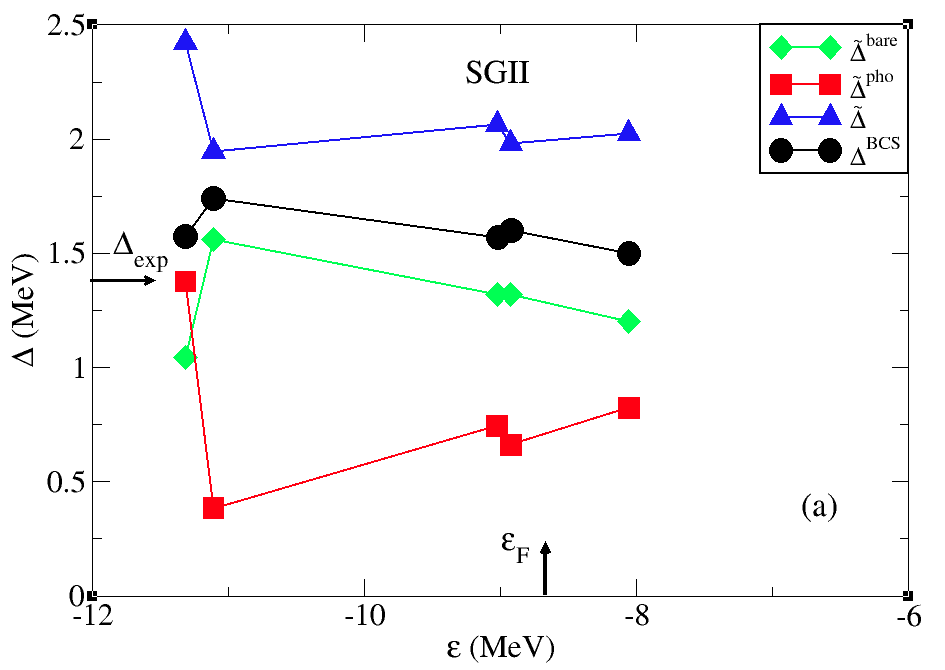}
\includegraphics[width=0.7\textwidth]{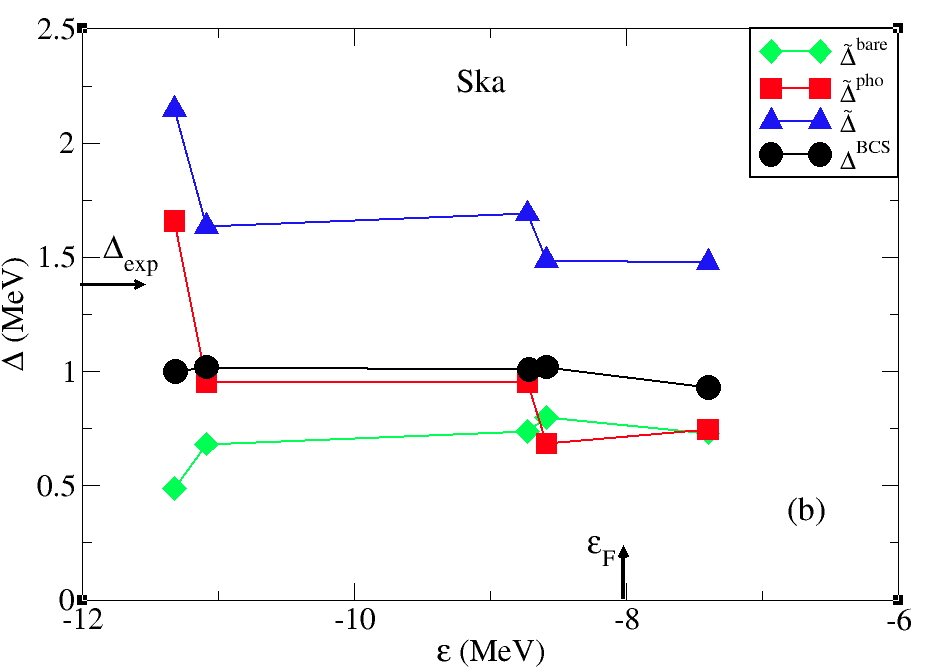}
\end{center}
\caption{ The state-dependent paring gap $\Delta^{BCS}$
calculated in BCS with the bare $v_{14}$ interaction  is compared to
the renormalized gap $\tilde \Delta$ (cf. Eq. (\ref{gap_prior}) )
obtained solving the Nambu-Gor'kov equations. 
We compare results obtained with a  mean field produced with
the SGII interaction (a) and with the SKa interaction (b) 
(cf. Fig. \ref{fig:renogap_sly4} for the corresponding 
calculation with the SLy4 mean field) \cite{micene}.
The symbols refer to the position of the various valence 
orbitals in the SGII HF potential. We also show the decomposition of $\tilde \Delta$
into the bare and phonon contributions $\tilde \Delta^{bare}$ and $\tilde \Delta^{pho}$. 
The value of the Fermi energy $\epsilon_F$ and of the gap obtained from the experimental odd-even
mass difference $\Delta_{exp}$ are also indicated.} 
\label{fig:renogap_sgii}
\end{figure*} 

\begin{figure*}[t!]
\begin{center}
\includegraphics[width=0.7\textwidth]{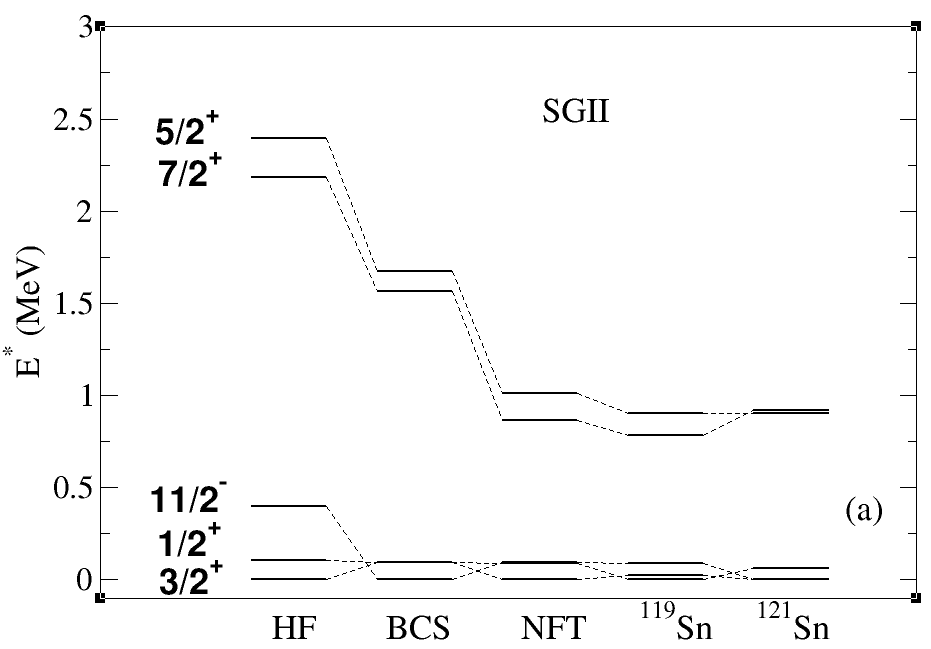}
\includegraphics[width=0.7\textwidth]{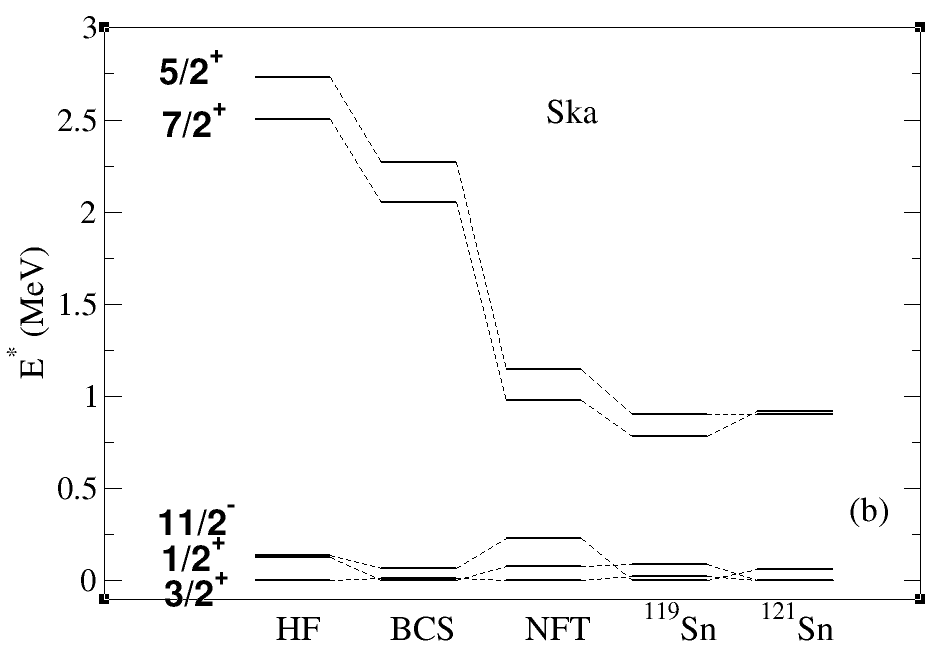}
\end{center}
\caption{ The theoretical quasiparticle spectra obtained at the various steps of the calculation are compared
to the experimental data.
We compare results obtained with a  mean field produced with
the SGII interaction (a) and with the SKa interaction (b) 
(cf. Fig. \ref{fig:renospectra_sly4} for the corresponding 
calculation with the SLy4 mean field).} 
\label{fig:renospectra_sgii}
\end{figure*}

The results obtained in the main text have been obtained with a HF mean field produced by the 
SLy4 interaction, whose  effective mass at saturation is $m_k = 0.7 m$. In this Section 
we show results obtained with two different interactions of the Skyrme type: the SKa \cite{kohler}  and 
the SGII \cite{sagawa},
having respectively a lower ($m_k = 0.61 m$) and a larger ($m_k = 0.78 m$) effective mass.  
All the other features of the calculations are the same as for SLy4. In particular, the PVC matrix 
elements  are the same, because in our approach they only depend on the properties of the phenomenological phonons
and are calculated with the wavefunctions of a fixed potential with effective mass equal to one.
As expected, the SGII interaction produces a mean field associated with a higher level density,
leading  to a larger pairing gap in the BCS calculation with the $v_{14}$ potential, as 
shown in Fig. \ref{fig:renogap_sgii}(a): the value of $\Delta^{BCS}$ is equal to 
about 1.6 MeV, to be compared with the value 1.1 MeV previously obtained with 
the SLy4 interaction (cf. Fig. \ref{fig:renogap_sly4}). The value obtained with the SKa interaction
is instead equal to about 1 MeV (cf. Fig. \ref{fig:renogap_sgii}(b)).

The renormalization processes act in a very similar way as was previously found in the case of 
SLy4. The average value of the phonon induced gap $\Delta^{pho}$ is in all cases equal to about 0.8 MeV: 
in the case of SGII $\Delta^{pho}$ accounts for about 30\% of the total gap, while in the case
of SKa $\Delta^{pho}$ and $\Delta^{bare} $ are of similar magnitude, as in the case of SLy4.

\begin{figure*}[h!]
\begin{center}
\includegraphics[width=0.7\textwidth]{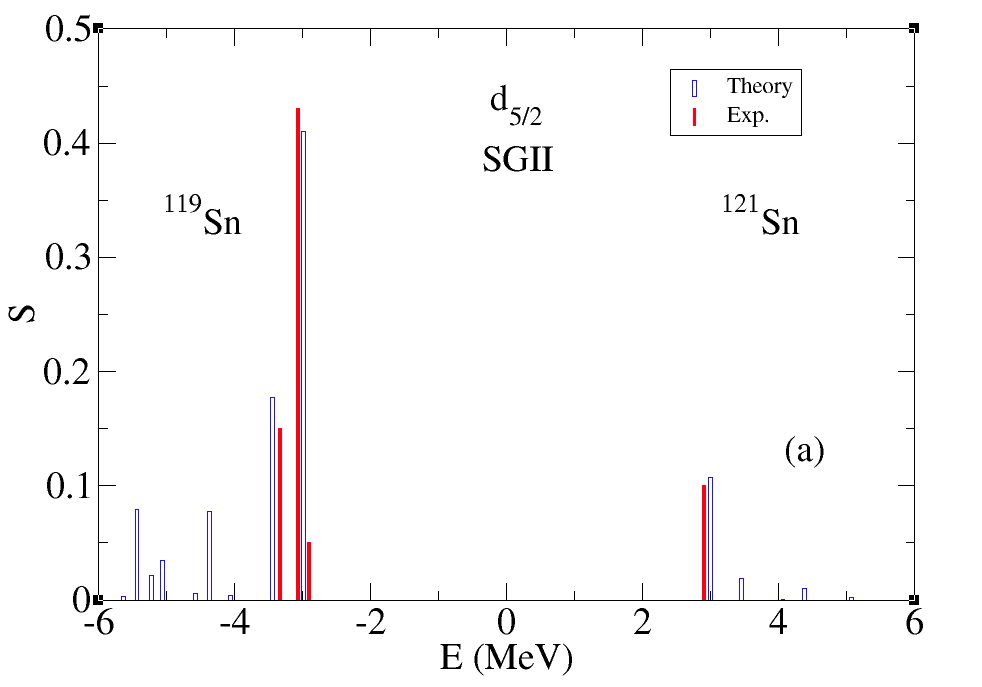}
\includegraphics[width=0.7\textwidth]{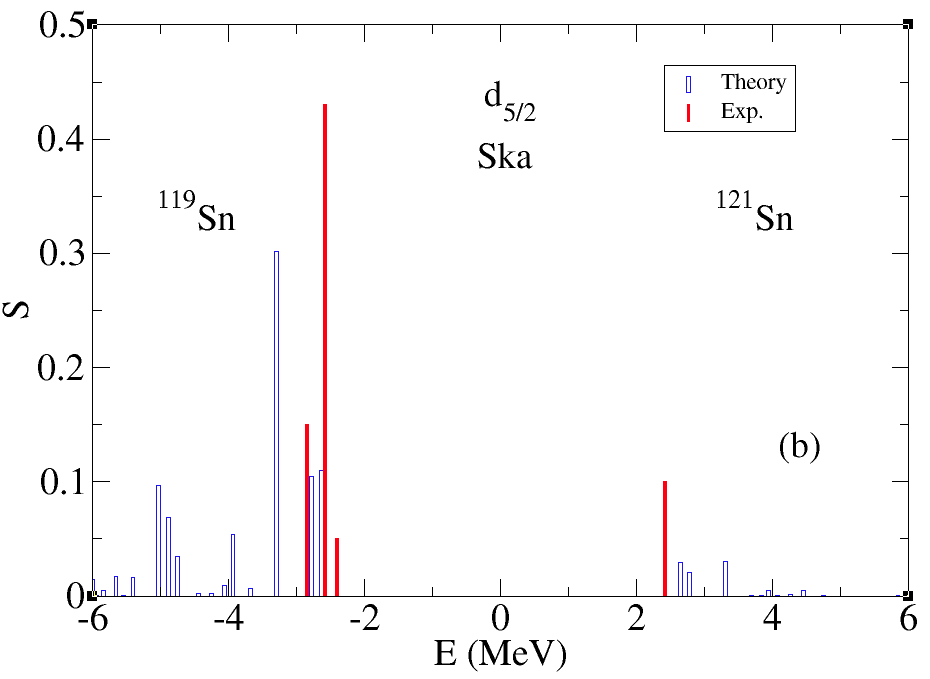}
\end{center}
\caption{ The theoretical strength function calculated for the $d_{5/2}$ orbital 
is compared to the spectroscopic factors associated with experimental levels detected in one-neutron transfer reactions.
We compare results obtained with a  mean field produced with
the SGII interaction (a) and with the SKa interaction (b)
(cf. Fig. \ref{fig:strength_d52} for the corresponding 
calculation with the SLy4 mean field).} 
\label{fig:strength_d52_sgii}
\end{figure*}

\begin{figure*}[b!]
\begin{center}
\includegraphics[width=0.7\textwidth]{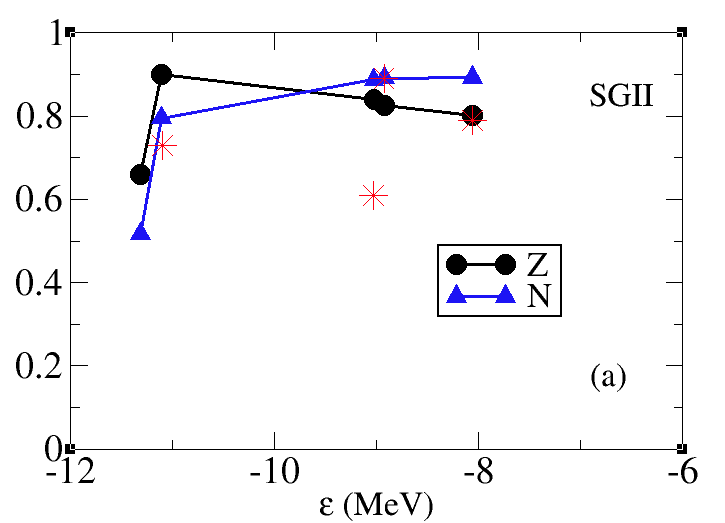}
\includegraphics[width=0.7\textwidth]{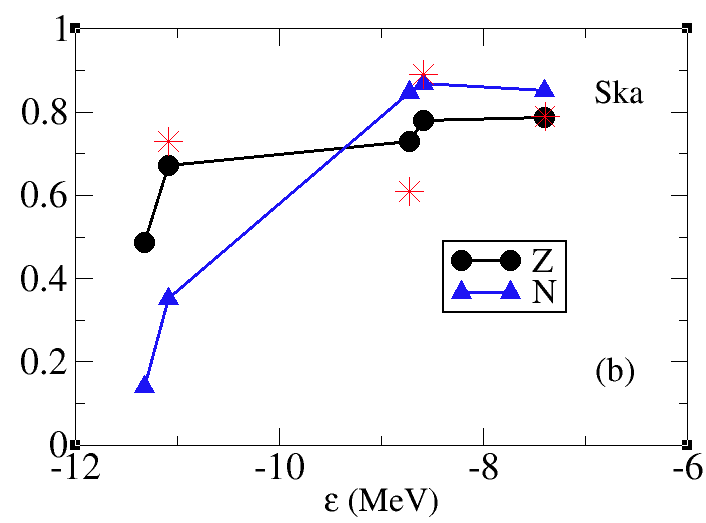}
\end{center}
\caption{ Comparison of the $N-$ and $Z-$ factors associated with the lowest quasiparticle peaks
in the  Nambu-Gor'kov calculation shown in Fig. \ref{fig:renogap_sgii}.
We compare results obtained with a  mean field produced with
the SGII interaction (a) and with the SKa interaction (b) 
(cf. Fig. \ref{fig:zeta_sly4} for the 
corresponden calculation with the SLy4 mean field). Also shown by stars are the values of the experimental
quasiparticle strength \cite{exp_transf2}, except for the $d_{5/2}$ orbital which shows a pronounced fragmentation.}
\label{fig:zeta_sgii}
\end{figure*}

\begin{figure*}[ht!]
\begin{center}
\includegraphics[width=0.7\textwidth]{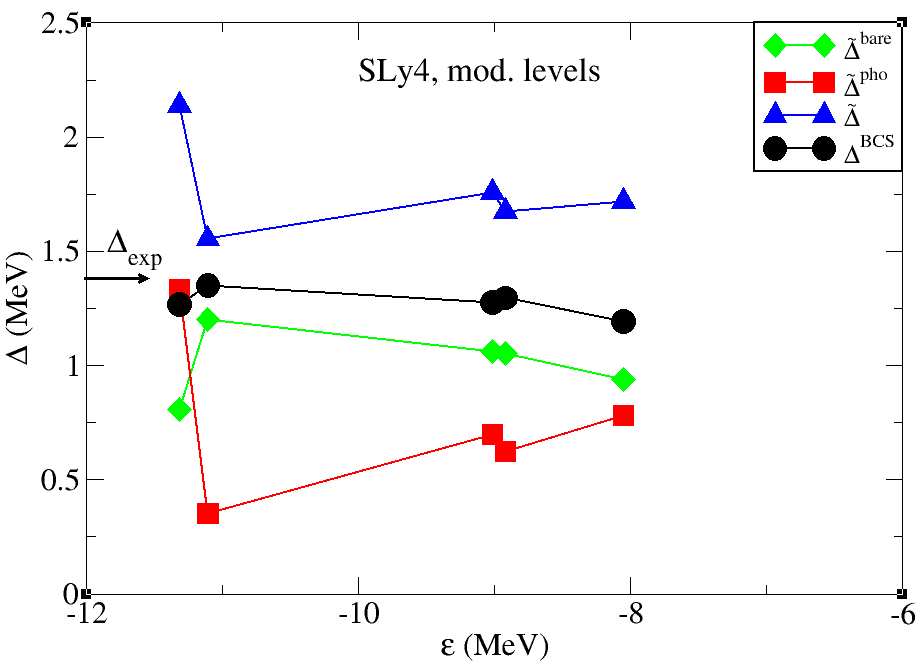}
\end{center}
\caption{ The same as in Fig. \ref{fig:renogap_sgii}, for the pairing gaps calculated
with the SGII mean field, but substituting the single-particle energies of the five valence orbitals 
with the energies calculated in the SLy4 mean field.}
\label{fig:renogap_hybrid}
\end{figure*}

The low-lying spectra are shown in Fig. \ref{fig:renospectra_sgii}, while the $Z-$ and $N-$factors of the 
lowest fragments are reported in Fig. \ref{fig:zeta_sgii}. Also in this case, the action of the PVC is similar
to that already discussed in the case of the  SLy4 interaction  (cf. Fig. \ref{fig:renospectra_sly4}).
However, the agreement with experiment is better than obtained previously.  This is due to the initial position 
of the $d_{5/2}$ and $g_{7/2}$ orbitals, which are closer to each other; furthermore, the $d_{5/2}$ single-particle
calculated with SGII lies at a distance of about 2.2 MeV from the Fermi energy, instead of 2.8 MeV as in the
case of SLy4: this leads to a good description of the fragmentation of this orbital, as shown in Fig. \ref{fig:strength_d52_sgii},
that in the case of SLy4 was obtained only shifting the energy of this level  (cf. Figs. \ref{fig:strength_d52}
and  \ref{fig:hindsight}). 
The quality of the spectrum obtained with SKa is almost as good as with SGII: however, the orbital $g_{7/2}$ 
lies more distant from the Fermi energy  and becomes fragmented, contrary to experiment (cf. Fig. \ref{fig:zeta_sgii}). 
It is interesting to notice that taking the SLy4 mean field, and changing the energies of the five valence orbitals
with those associated with the SGII interaction, one obtains practically the same spectrum and the same $Z-$ and $N-$ factors
as with SGII.  This in spite of the fact that $\Delta^{BCS}$ becomes equal to   about 1.25 MeV, to be compared with 1.1 MeV
(SLy4) and 1.6 MeV (SGII). At the same time,  the average value of $\tilde \Delta$ is equal to about 1.7  MeV, to be compared
with 1.6 MeV (SLy4) and 2.1 MeV (SGII) (cf. Fig. \ref{fig:renogap_hybrid}).  These 
results show that the low-energy spectrum is determined by the position of the  valence orbitals, while 
the absolute value of the BCS gap also depends on the effective mass associated with distant levels.

We can conclude  that renormalization effects 
are  similar for the three mean fields we have considered.
The comparison  with the odd-even
mass difference favours forces having low effective mass (SLy4 or SKa). The PVC  
improves the agreement of the spectral properties with experiment.
However, the quality of this agreement depends  on the specific position of the mean 
field single-particle levels close to the Fermi energy.
The value of the effective mass  far from the Fermi energy   determines the magnitude of the final gap 
$\tilde \Delta$ by shifting the value of $\Delta^{BCS}$, while  the value of $\Delta^{pho}$ and the properties of the low-lying
spectrum are not very sensitive to it. 


\clearpage

\subsection{\label{barepair} Bare pairing interaction}

\begin{figure*}[b!]
\begin{center}
\includegraphics[width=0.7\textwidth]{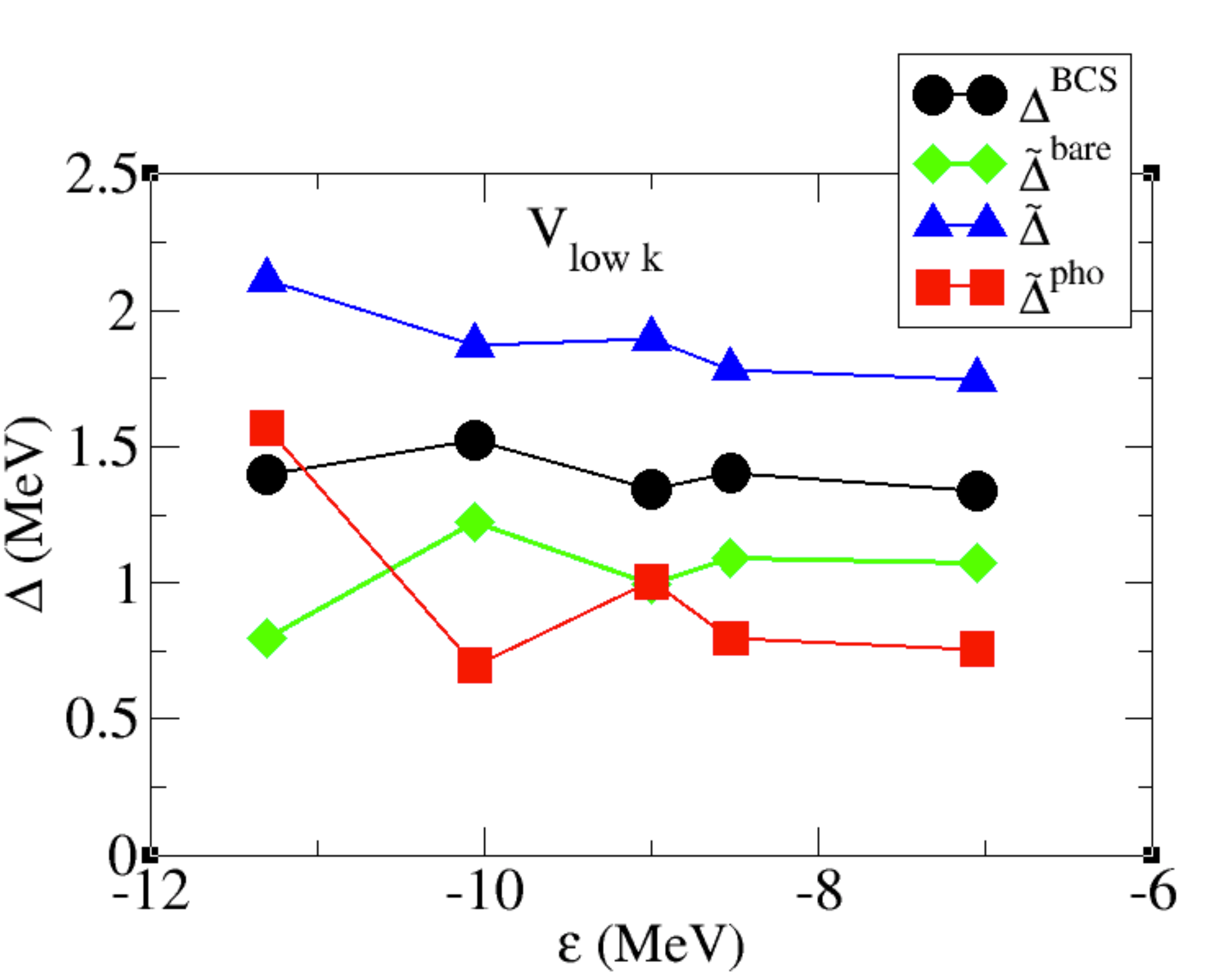}
\end{center}
\caption
{The state-dependent paring gap $\Delta^{BCS}$ calculated
 using  the $V_{{\rm low} k}$ potential with a cutoff
$\Lambda=4 $ fm$^{-1}$ as pairing force is compared to
the renormalized gap $\tilde \Delta$ (cf. Eq. (\ref{gap_prior}) )
obtained solving the Nambu-Gor'kov equations. The symbols refer to the position of the various valence 
orbitals in the SLy4 HF potential. We also show the decomposition of $\tilde \Delta$
into the bare and phonon contributions $\tilde \Delta^{bare}$ and $\tilde \Delta^{pho}$. 
We thank A. Pastore for providing us with the HFB calculation  with the $V_{{\rm low} k}$ potential. } 
\label{fig:renogap_sly4_vlowk}
\end{figure*}

The calculations reported in the  main text have been carried out adopting  $v_{14}$ as the bare pairing force.
However, the $V_{{\rm low} k} $ version of the Argonne potential has been used by several groups and here we show the
results using  this bare pairing  force in our Nambu-Gor'kov formalism. 
The corresponding bare and renormalized gap are shown in Fig. \ref{fig:renogap_sly4_vlowk}. The average value of the 
bare gap is equal to about 1.4 MeV, in agreement with ref.  \cite{hebeler}, to be compared with the value 1.1 MeV
obtained with the Argonne interaction.   
The effect of the renormalization processes increases  the gap on average by about 
500 keV, similar to the case of $v_{14}$ (600 keV, cf. Fig. \ref{fig:renogap_sly4}). 
Also the values of the phonon induced  component $\Delta^{pho}$ are  quite similar.  The results
of the  calculations  with $v_{14}$ and $V_{{\rm low} k}$ turn out to be  quite similar to 
those obtained in Section \ref{results} using the monopole force respectively with coupling constants $G_0 = 0.22 $  and $G_0 = 0.25 $ 
MeV (cf. Fig. \ref{fig:deltavsg}).

\clearpage

\subsection{\label{qrpa} QRPA}

\begin{figure*}[b!]
\begin{center}
\includegraphics[width=0.7\textwidth]{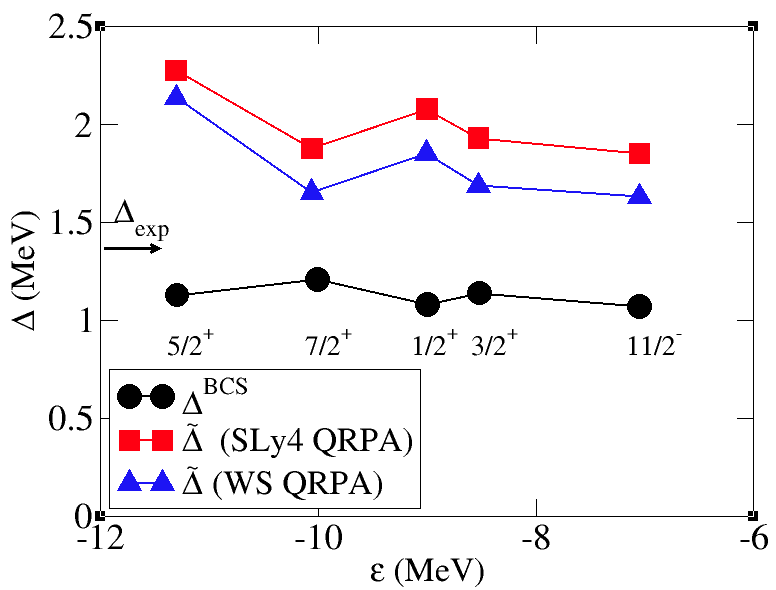}
\end{center}
\caption{ The renormalized pairing gaps $\tilde \Delta$, calculated using the levels of a Woods-Saxon potential 
to compute  the QRPA spectrum and the  PVC vertices, are shown by triangles (cf. Fig. \ref{fig:renogap_sly4})  
and are compared to the results obtained using  SLy4 single-particle levels (squares).
We also show by dots the pairing gap $\Delta^{BCS}$ calculated in BCS. } 
\label{fig:sly4_ff}
\end{figure*}

We recall (cf. Section \ref{formalism}) that the  results shown in the main text 
are based on effective PVC vertices calculated  with single-particle levels and wavefunction 
associated with a  Woods-Saxon potential,
with an associated effective mass $m^* =  m$. Another reasonable choice, often adopted in the literature, would be to use
instead the same single-particle levels used in the HF+BCS calculation. Using this prescription and 
readjusting  the coupling constants $\chi_{\lambda}$ of the QRPA calculation so as to reproduce the experimental properties
of the low-lying vibrational states, we obtain the renormalized pairing gaps shown in Fig. \ref{fig:sly4_ff}.
It is seen that the difference is not critical, our preferred choice leading  to gaps which are smaller by about 10\%.

\clearpage

\subsection {\label{bubble} Bubble approximation}

\begin{figure*}[t!]
\begin{center}
\includegraphics[width=0.5\textwidth]{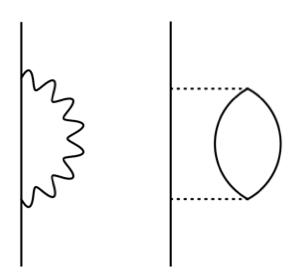}\\
\end{center}
\caption{ While in the rest of the work we have considered the effect of the coupling betwene quaisparticle 
and collective QRPA phonons (left), in this section we shall consider the coupling to uncorrelated 
excitations (right).}
\label{fig:bubble}
\end{figure*}

\begin{figure*}[t!]
\begin{center}
\includegraphics[width=0.7\textwidth]{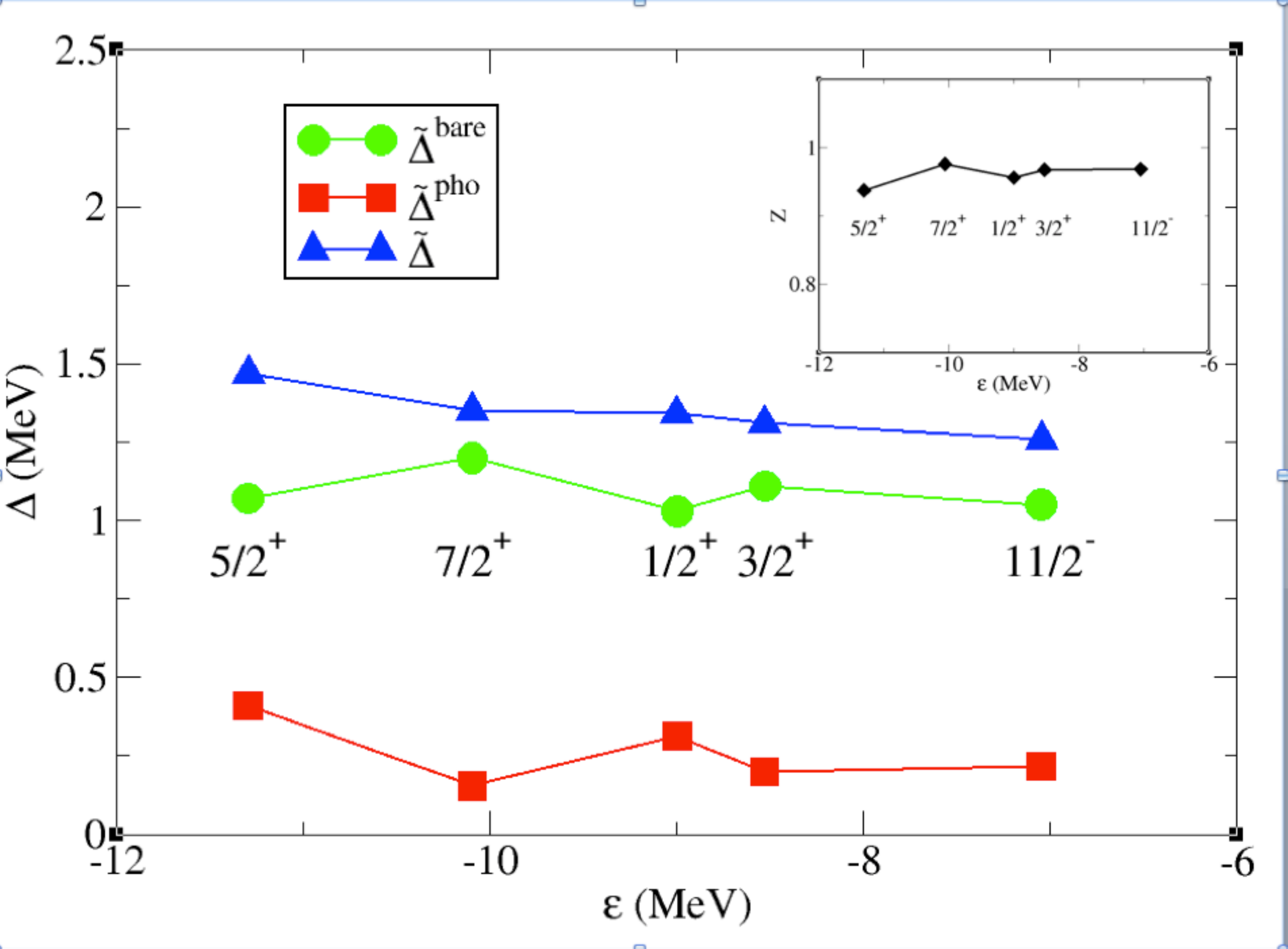}\\
\end{center}
\caption{ Pairing gaps associated with the main quasiparticle peaks obtained 
in the one-bubble approximation. We show the total renormalized gap $\tilde\Delta$ and its decomposition into
the bare and the phonon contributions. In the inset, the $Z-$values are also shown.} 
\label{fig:bubblegaps}
\end{figure*}





It is interesting to assess the importance of the collectivity of QRPA modes in the pairing calculation.
This can be done by  comparing our previous results with calculations performed coupling the quasiparticles with unperturbed p-h excitations (cf. Fig.  \ref{fig:bubble}).
In Fig. \ref{fig:bubblegaps} we show the pairing gap and the $Z-$ factors calculated 
at the p-h level:  they  are very close to 1, indicating that 
the renormalization processes are much less important as compared to the calculations with collective modes.    
Correspondingly, the phonon contribution to the gaps is drastically diminished
(compare with Figs. \ref{fig:renogap_sly4} and \ref{fig:zeta_sly4} ).
Also the quasiparticle spectrum becomes very close to that obtained at the HF+BCS level.  
On the other hand,  the total gap is reduced by only 10\% as compared to the QRPA result: this happens because, even if 
the phonon-induced interaction is considerably weaker, the
product $\tilde\Delta = Z \Delta^{BCS} +\tilde\Delta^{pho} $   is much less affected, because 
the $Z-$factor is correspondingly increased. Schematically, while in the QRPA case one finds $Z \approx 0.75 $ and
$\tilde\Delta = 0.75 \times 1.1 +0.8 $ = 1.6 MeV, in the present case    $Z \approx 0.95 $ and 
$\tilde\Delta = 0.95 \times 1.1 +0.3 $ = 1.4 MeV.
One then concludes that, although the value 
of the gap is similar using the QRPA or the unperturbed response, the physical picture is quite different. 
This is related to the fact that  the contribution of the phonon interaction to the total gap is equal to about 50\% 
considering the coupling to  QRPA modes,  and to about  20\% coupling to unperturbed p-h excitations.
Furthermore, we have performed a calculation correcting for the Pauli principle, namely removing the contribution from
a bubble when the particle or the hole coincides with the intermediate state $\{nlj\}$, with the proper
angular momentum factor $1/(2j+1)$ (cf. \cite{broglia_brink}, Appendix F). The pairing gap is reduced by less than
5\%. This represents an upper limit for role of  the Pauli correction in the calculations based on collective phonons
reported in the main text.

\end{document}